\documentclass[11pt]{article}
\usepackage[hmargin={3.0cm,3.0cm},vmargin={3.0cm,3.0cm}]{geometry}

\usepackage{amsmath,amssymb}
\usepackage{amsfonts}
\usepackage{mathrsfs}
\usepackage{dsfont}

\usepackage{epic}
\usepackage{eepic}
\usepackage{graphicx}

\usepackage{pgf}
\usepackage{tikz}
\usetikzlibrary{arrows.meta}

\usepackage{amsmath}
\usepackage{mathtools}

\usepackage{bm}

\usepackage[skip=2pt,font=small]{caption}
\usepackage{hyperref}
\hypersetup{
colorlinks,
linkcolor={blue!60!black},
citecolor={blue!90!black},
urlcolor={blue!90!black}}

\usepackage{footnote}

\usepackage{cite}
\usepackage{enumerate}
\usepackage[curve]{xy}

\usepackage[titletoc,title]{appendix}

\usepackage{amsthm}

\theoremstyle{definition}

\theoremstyle{remark}

\newcommand{\ds}{\displaystyle}

\renewcommand{\author}[1]{\large\rm #1\\ \bigskip}
\newcommand{\address}[1]{{\normalsize\it #1\\}\bigskip}
\renewcommand{\title}[1]{\bigskip\bigskip\Large\bf #1\bigskip\bigskip\\}

\newcommand{\Bigpsi}[3]{\phantom{\Psi}_2 \kern -.05em
\Psi_2\left(\genfrac{}{}{0pt}{}{#1}{#2}\biggl|#3\right)}

\newcommand{\ihat}{\mathbf{e}_i}
\newcommand{\jhat}{\mathbf{e}_j}
\newcommand{\khat}{\mathbf{e}_k}

\newcommand{\bea}{\begin{eqnarray}}
\newcommand{\eea}{\end{eqnarray}}

\newcommand{\n}{{\boldsymbol{n}}}

\newcommand{\ii}{\mathsf{i}}
\newcommand{\oW}{\overline{W}}
\renewcommand{\r}{{\mathsf r}}
\newcommand{\lag}{{\mathcal L}}

\newcommand{\V}{\mathbb{V}}

\newcommand{\q}{{\mathsf q}}
\newcommand{\p}{{\mathsf p}}

\newcommand{\bW}{\mathbb{V}}

\newcommand{\Gl}{{\mathscr G}}
\renewcommand{\L}{{\mathscr L}}

\newcommand{\ovphi}{{\overline{\varphi}}}
\newcommand{\olam}{{\overline{\Lambda}}}

\newcommand{\bp}{{\mathbf \p}}
\newcommand{\bq}{{\mathbf \q}}
\newcommand{\br}{{\mathbf \r}}

\def\EXP{\textrm{{\large e}}}

\newcommand{\bu}{{\boldsymbol{u}}}
\newcommand{\bv}{{\boldsymbol{v}}}
\newcommand{\x}{{\boldsymbol{x}}}
\newcommand{\y}{{\boldsymbol{y}}}
\newcommand{\al}{{\bm{\alpha}}}
\newcommand{\bt}{{\bm{\beta}}}
\newcommand{\gm}{{\bm{\gamma}}}
\newcommand{\w}{{\boldsymbol{w}}}

\newcommand{\xv}{{x}}

\newcommand{\qsym}{Q}
\newcommand{\qeq}{\mathbf{\qsym}}

\newcommand{\sn}{\hspace{0.05cm}\textrm{sn}}

\newcounter{app}
\newcounter{sapp}[app]

\begin{document}

\vglue 2cm

\begin{center}

\title{Extended Z-invariance for integrable vector and face models and multi-component integrable quad equations}
\author{Andrew P.~Kels}
\address{SISSA, Via Bonomea 265, 34136 Trieste, Italy}

\end{center}

\begin{abstract}

In a previous paper \cite{Kels:2017fyt}, the author has established an extension of the Z-invariance property for integrable edge-interaction models of statistical mechanics, that satisfy the star-triangle relation (STR) form of the Yang-Baxter equation (YBE).  In the present paper, an analogous extended Z-invariance property is shown to also hold for integrable vector models and interaction-round-a-face (IRF) models of statistical mechanics respectively.  
As for the previous case of the STR, the Z-invariance property is shown through the use of local cubic-type deformations of a 2-dimensional surface associated to the models, which allow an extension of the models onto a subset of next nearest neighbour vertices of $\mathbb{Z}^3$, while leaving the partition functions invariant.  These deformations are permitted as a consequence of the respective YBE's satisfied by the models.   The quasi-classical limit is also considered, and it is shown that an analogous Z-invariance property holds for the variational formulation of classical discrete Laplace equations which arise in this limit.  From this limit, new integrable 3D-consistent multi-component quad equations are proposed, which are constructed from a degeneration of the equations of motion for IRF Boltzmann weights.


\end{abstract}

\tableofcontents




\section{Introduction}

In a recent paper \cite{Kels:2017fyt}, the author has introduced an extension of Baxter's Z-invariance property \cite{Baxter:1978xr,Baxter:1986df} for exactly solved models of statistical mechanics, that satisfy the star-triangle relation form of the Yang-Baxter equation (YBE) \cite{Baxter:1982zz,PerkYBEs}. 
These lattice models were reformulated to lie on a two-dimensional surface of elementary squares, where each elementary square is associated to a Boltzmann weight of the model, and the partition function was shown to be invariant under ``cubic flips'' of this surface, as a consequence of the star-triangle relation.  These deformations were used to construct a model defined on a planar graph with edges connecting next-nearest-neighbour vertices in $\mathbb{Z}^3$, and whose partition function is equivalent to the usual square lattice model, up to some extra factors entering the star-triangle relation.  This is an extension of the usual formulation of Z-invariance, because rapidity lines which form closed directed loops are required in the definition of the deformed model, whereas traditionally such lines are not permitted.  In the quasi-classical limit, the resulting system of classical discrete Laplace equations were also shown to satisfy an analogous classical Z-invariance property, which is closely related to a closure property for Lagrangian multiforms \cite{LobbNijhoff} for systems of ABS equations \cite{ABS}.

In this paper, the results of \cite{Kels:2017fyt} are formulated for the cases of both vertex models and interaction-round-a-face (IRF) models of statistical mechanics \cite{Baxter:1982zz}.   The vertex and IRF models will first be constructed from edge-interaction models which satisfy another fundamental identity of statistical mechanics, known as the star-star relation \cite{Bazhanov:1990qk,Bazhanov:1992jqa,Baxter:1997tn,Bazhanov:2011mz,Bazhanov:2013bh,Yamazaki:2013nra,Gahramanov:2017idz,Kels:2017vbc}.  The latter relation implies the existence of a Yang-Baxter equation for associated vertex and IRF models, which in turn implies that transfer matrices commute in both of the latter formulations \cite{Baxter:1978xr,Baxter:1982zz,Baxter:1997tn}.  This approach allows the extended Z-invariance for both the vertex and IRF models to be formulated together, and means that the results of this paper may be applied to both vertex and IRF models which are derived from an edge-interaction model, and also pure vertex and IRF models which are formulated independently of an edge-interaction model.

To show the extended Z-invariance for both the vertex and IRF models, the respective models are first associated to an underlying surface made up of configurations of four squares arranged as in the diagram below, for the vertex and IRF models respectively.  Such groups of four squares are central objects for this paper, and are referred to as {\it elementary four-squares}.  Each elementary four-square is associated to a Boltzmann weight for the vertex or IRF model respectively, and the usual vertex or IRF models can be obtained by translating the respective four-squares in two orthogonal directions in the plane.  These four-squares may also be used as the building blocks of a more general two-dimensional surface, not restricted to lie in a plane, which contains a vertex or IRF model having variables on a subset of next nearest neighbour vertices of the lattice $\mathbb{Z}^3$.  The partition functions of these latter models, can be seen to be equivalent to the original respective models in the plane, up to simple factors coming from the expressions for the Yang-Baxter equations.  This is the extended Z-invariance property.  Each of the required deformations that are needed to show Z-invariance of the vertex and IRF models respectively, are pictured in Appendix \ref{app:IRF}.  These deformations are the analogues of the deformations which were previously used to show extended Z-invariance of  edge-interaction models, pictured in the Appendix of \cite{Kels:2017fyt}.

\begin{figure}[htb!]
\centering
\begin{tikzpicture}[scale=1.5]
\draw[-,gray] (-0.8,-0.8)--(0.8,-0.8)--(0.8,0.8)--(-0.8,0.8)--(-0.8,-0.8);
\draw[-,gray] (0,-0.8)--(0,0.8);\draw[-,gray] (-0.8,0)--(0.8,0);
\draw[-,very thick] (-0.8,0)--(0,0.8)--(0.8,0)--(0,-0.8)--(-0.8,0);
\draw[->,thick,dotted,black] (-0.4,-1)--(-0.4,1);\draw[->,dashed,black] (0.4,-1)--(0.4,1);
\draw[->,dashed,black] (-1,-0.4)--(1,-0.4);
\draw[->,thick,dotted,black] (-1,0.4)--(1,0.4);
\fill[black!] (-0.4,-1.4) circle (0.01pt);
\fill[black!] (-1,-0.4) circle (0.01pt);
\fill[black!] (0.4,-1.4) circle (0.01pt);
\fill[black!] (-0.95,0.4) circle (0.01pt);
\filldraw[fill=white,draw=black] (-0.8,-0.8) circle (1.5pt);
\filldraw[fill=white,draw=black] (0.8,-0.8) circle (1.5pt);
\filldraw[fill=white,draw=black] (0.8,0.8) circle (1.5pt);
\filldraw[fill=white,draw=black] (-0.8,0.8) circle (1.5pt);
\filldraw[fill=white,draw=black] (0,0) circle (1.5pt);
\filldraw[fill=black,draw=black] (0,-0.8) circle (1.5pt);
\filldraw[fill=black,draw=black] (0,0.8) circle (1.5pt);
\filldraw[fill=black,draw=black] (0.8,0) circle (1.5pt);
\filldraw[fill=black,draw=black] (-0.8,0) circle (1.5pt);

\fill (0,-1.3) circle(0.01pt)
node[below=0.05pt]{\color{black} Vertex};

\begin{scope}[xshift=120pt]
\draw[-,gray] (-0.8,-0.8)--(0.8,-0.8)--(0.8,0.8)--(-0.8,0.8)--(-0.8,-0.8);
\draw[-,gray] (0,-0.8)--(0,0.8);\draw[-,gray] (-0.8,0)--(0.8,0);
\draw[-,very thick] (-0.8,-0.8)--(0.8,0.8);\draw[-,very thick] (-0.8,0.8)--(0.8,-0.8);
\draw[->,dashed,black] (-0.4,-1)--(-0.4,1);\draw[->,thick,dotted,black] (0.4,-1)--(0.4,1);
\draw[->,dashed,black] (-1,-0.4)--(1,-0.4);
\draw[->,thick,dotted,black] (-1,0.4)--(1,0.4);
\fill[black!] (-0.4,-1.4) circle (0.01pt);
\fill[black!] (-1,-0.4) circle (0.01pt);
\fill[black!] (0.4,-1.4) circle (0.01pt);
\fill[black!] (-0.95,0.4) circle (0.01pt);
\filldraw[fill=black,draw=black] (-0.8,-0.8) circle (1.5pt);
\filldraw[fill=black,draw=black] (0.8,-0.8) circle (1.5pt);
\filldraw[fill=black,draw=black] (0.8,0.8) circle (1.5pt);
\filldraw[fill=black,draw=black] (-0.8,0.8) circle (1.5pt);
\filldraw[fill=black,draw=black] (0,0) circle (1.5pt);
\filldraw[fill=white,draw=black] (0,-0.8) circle (1.5pt);
\filldraw[fill=white,draw=black] (0,0.8) circle (1.5pt);
\filldraw[fill=white,draw=black] (0.8,0) circle (1.5pt);
\filldraw[fill=white,draw=black] (-0.8,0) circle (1.5pt);

\fill (0,-1.3) circle(0.01pt)
node[below=0.05pt]{\color{black} Interaction-round-a-face (IRF)};

\end{scope}

\end{tikzpicture}
\end{figure}

The quasi-classical limit of the IRF model is also considered in this paper.  This is an important limit that connects  integrable models of statistical mechanics \cite{Bazhanov:2007mh,Bazhanov:2010kz,Bazhanov:2016ajm,Kels:2017vbc,Kels:2018xge}, with discrete integrable systems that satisfy an integrability condition known as 3D-consistency \cite{nijhoff_walker_2001,BobSurQuadGraphs,ABS}.  Through this connection, the Yang-Baxter equation itself may be interpreted as a quantum counterpart of a discrete integrable equation, where the latter equation is identified as the equation of the saddle-point of the YBE. 
In this limit, it will be seen that the partition function of the IRF model reduces to an action functional for a system of classical discrete Laplace equations \cite{MR2467378,BG11}.  It is shown that this action functional also satisfies a classical Z-invariance property, analogous to the extended Z-invariance property that was described for the statistical mechanical model.  In this classical limit, the property of Z-invariance is a consequence of a classical counterpart of the Yang-Baxter equation.  This classical Yang-Baxter equation may also be interpreted as a local closure property \cite{LobbNijhoff} of the variational Laplace system, for which the action functional is invariant under the local cubic deformations of elementary four-squares, of the type pictured in Appendix \ref{app:IRF}.  

Some explicit multi-component classical equations are also considered, that were previously obtained from the quasi-classical limit of IRF Boltzmann weights by Bazhanov and Sergeev \cite{Bazhanov:2011mz}.  It is shown that an algebraic degeneration of the scalar case of these variational equations, results in a 3D-consistent linear quad equation that was previously studied by Atkinson \cite{Atkinson08,Atkinson09}.  Based on the form of these scalar equations, some new $n$-component 3D-consistent quad equations are proposed, for $n=1,2,\ldots$, which are different from the original multi-component variational equations that came from the Yang-Baxter equation.  The 3D-consistency of these $n$-component equations can be checked using numerical computations for small $n$ ($n\leq 10$), but the 3D-consistency appears difficult to prove directly using algebraic methods, due to the lengthy expressions that arise for $n\geq2$.  It is expected that the further study of the integrable properties of these equations will be an important problem.

This paper is organised as follows.  In Section \ref{sec:ssr}, the integrable vertex and IRF models are constructed from an edge-interaction model which satisfies the star-star relation.   The vertex and IRF models are then reformulated on two-dimensional surfaces of elementary four-squares, and it is seen that the respective partition functions remain invariant under cubic-type deformations of the surface, which is the extended Z-invariance property.  In Section \ref{sec:qcl}, the quasi-classical limit is considered.  It is shown that the action functional for the system of discrete Laplace equations  that arise in this limit, is also Z-invariant under the same deformations as for the lattice model.  This is a classical manifestation of the extended Z-invariance property, which is shown to be related to a local closure relation for the action functional.  Finally, in Section \ref{sec:Mcomp} new types of $n$-component 3D-consistent quad equations are constructed, based on the multi-component equations that are obtained in the quasi-classical limit of a continuous spin IRF model. 

\section{Star-star relation and Z-invariance of vertex and IRF models} \label{sec:ssr}

\subsection{Square lattice model}

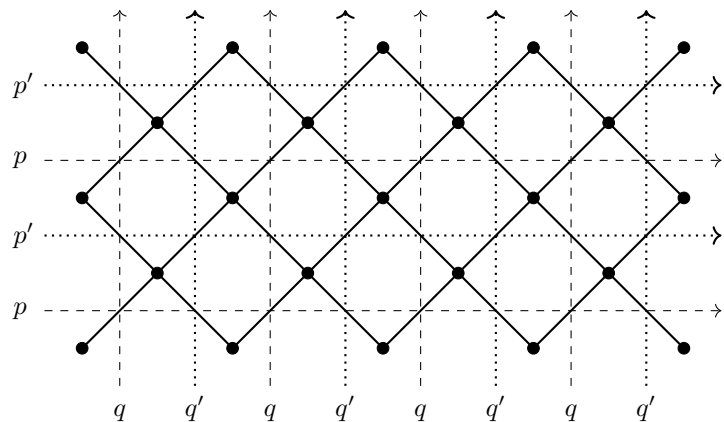
\begin{figure}[htb!]
\centering
\begin{tikzpicture}[scale=1]

\draw[-,thick] (-0.5,-0.5)--(3.5,3.5);
\draw[-,thick] (-0.5,3.5)--(3.5,-0.5);
\draw[-,thick] (-0.5,1.5)--(1.5,3.5)--(3.5,1.5)--(1.5,-0.5)--(-0.5,1.5);
\draw[-,thick] (-4.5,-0.5)--(-0.5,3.5);
\draw[-,thick] (-4.5,3.5)--(-0.5,-0.5);
\draw[-,thick] (-4.5,1.5)--(-2.5,3.5)--(-0.5,1.5)--(-2.5,-0.5)--(-4.5,1.5);
\fill[white!] (0,4.5) circle (0.1pt);
\foreach \x in {-4,-2,...,2}{
\draw[->,dashed] (\x,-1) -- (\x,4);
\fill[black!] (\x,-1) circle (0.08pt)
node[below=3.1pt]{\color{black}\small $q$};}
\foreach \x in {-3,-1,...,3}{
\draw[->,thick,dotted] (\x,-1) -- (\x,4);
\fill[black!] (\x,-1) circle (0.08pt)
node[below=0.05pt]{\color{black}\small $q'$};}
\foreach \y in {1,3}{
\draw[->,thick,dotted] (-5,\y) -- (4,\y);
\fill[black!] (-5,\y) circle (0.08pt)
node[left=0.05pt]{\color{black}\small $p'$};}
\foreach \y in {0,2}{
\draw[->,dashed] (-5,\y) -- (4,\y);
\fill[black!] (-5,\y) circle (0.08pt)
node[left=2.9pt]{\color{black}\small $p$};}

\foreach \y in {-0.5,1.5,3.5}{
\foreach \x in {-4.5,-2.5,...,3.5}{
\filldraw[fill=black,draw=black] (\x,\y) circle (2.2pt);}}
\foreach \y in {0.5,2.5}{
\foreach \x in {-3.5,-1.5,...,2.5}{
\filldraw[fill=black,draw=black] (\x,\y) circle (2.2pt);}}

\end{tikzpicture}

\caption{The square lattice $L$, and its medial rapidity graph, made up of the directed dashed and dotted lines.}
\label{fig-lattice}
\end{figure}

In this section, an edge interaction model will be defined on a square lattice, denoted by $L$, as pictured in Figure \ref{fig-lattice}.   The set of vertices of $L$ will be denoted $V(L)$, and the set of edges of $L$ by $E(L)$.  In Figure \ref{fig-lattice}, vertices $i\in V(L)$, are represented by solid (black) circles, and two nearest-neighbour vertices $i,j\in V(L)$ are connected by an edge $(ij)\in E(L)$.

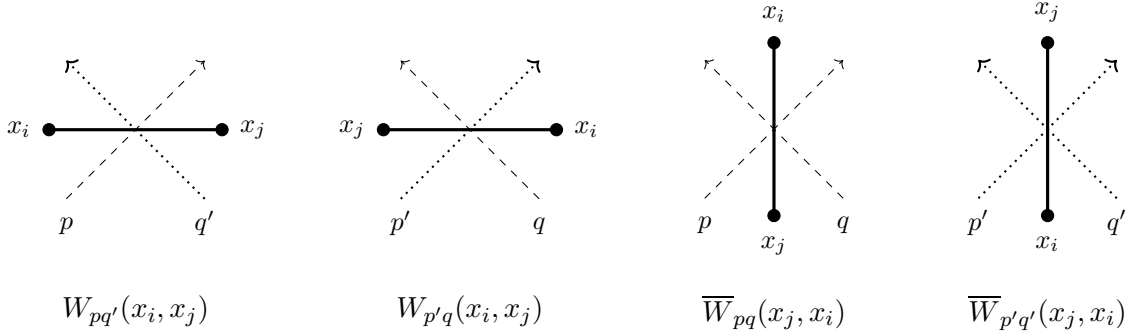
\begin{figure}[htb!]
\centering
\begin{tikzpicture}[scale=2.3]
\draw[-,very thick] (-0.5,2)--(0.5,2);
\draw[->,thick,dotted] (0.4,1.6)--(-0.4,2.4);
\fill[black!] (0.4,1.6) circle (0.01pt)
node[below=0.5pt]{\color{black}\small $q'$};
\draw[->,dashed] (-0.4,1.6)--(0.4,2.4);
\fill[black!] (-0.4,1.6) circle (0.01pt)
node[below=3.1pt]{\color{black}\small $p$};
\filldraw[fill=black,draw=black] (-0.5,2) circle (1.0pt)
node[left=3pt]{\color{black}\small $\xv_i$};
\filldraw[fill=black,draw=black] (0.5,2) circle (1.0pt)
node[right=3pt]{\color{black}\small $\xv_j$};

\fill (0,1.1) circle(0.01pt)
node[below=0.05pt]{\color{black} $ W_{pq'}(\xv_i,\xv_j)$};

\begin{scope}[xshift=55pt]
\draw[-,very thick] (-0.5,2)--(0.5,2);
\draw[->,dashed] (0.4,1.6)--(-0.4,2.4);
\fill[black!] (0.4,1.6) circle (0.01pt)
node[below=3.3pt]{\color{black}\small $q$};
\draw[->,thick,dotted] (-0.4,1.6)--(0.4,2.4);
\fill[black!] (-0.4,1.6) circle (0.01pt)
node[below=0.5pt]{\color{black}\small $p'$};
\filldraw[fill=black,draw=black] (-0.5,2) circle (1.0pt)
node[left=3pt]{\color{black}\small $\xv_j$};
\filldraw[fill=black,draw=black] (0.5,2) circle (1.0pt)
node[right=3pt]{\color{black}\small $\xv_i$};

\fill (0,1.1) circle(0.01pt)
node[below=0.05pt]{\color{black} $ W_{p'q}(\xv_i,\xv_j)$};
\end{scope}

\begin{scope}[xshift=105pt,yshift=57pt]
\draw[-,very thick] (0,-0.5)--(0,0.5);
\draw[->,dashed] (-0.4,-0.4)--(0.4,0.4);
\fill[black!] (-0.4,-0.4) circle (0.01pt)
node[below=3.2pt]{\color{black}\small $p$};
\draw[->,dashed] (0.4,-0.4)--(-0.4,0.4);
\fill[black!] (0.4,-0.4) circle (0.01pt)
node[below=3.2pt]{\color{black}\small $q$};
\filldraw[fill=black,draw=black] (0,-0.5) circle (1.0pt)
node[below=4pt]{\color{black}\small $\xv_j$};
\filldraw[fill=black,draw=black] (0,0.5) circle (1.0pt)
node[above=4pt]{\color{black}\small $\xv_i$};

\fill (0,-0.9) circle(0.01pt)
node[below=0.05pt]{\color{black} $\oW_{pq}(\xv_j,\xv_i)$};
\end{scope}

\begin{scope}[xshift=150pt,yshift=57pt]
\draw[-,very thick] (0,-0.5)--(0,0.5);
\draw[->,thick,dotted] (-0.4,-0.4)--(0.4,0.4);
\fill[black!] (-0.4,-0.4) circle (0.01pt)
node[below=0.5pt]{\color{black}\small $p'$};
\draw[->,thick,dotted] (0.4,-0.4)--(-0.4,0.4);
\fill[black!] (0.4,-0.4) circle (0.01pt)
node[below=0.5pt]{\color{black}\small $q'$};
\filldraw[fill=black,draw=black] (0,-0.5) circle (1.0pt)
node[below=4pt]{\color{black}\small $\xv_i$};
\filldraw[fill=black,draw=black] (0,0.5) circle (1.0pt)
node[above=4pt]{\color{black}\small $\xv_j$};

\fill (0,-0.9) circle(0.01pt)
node[below=0.05pt]{\color{black} $\oW_{p'q'}(\xv_j,\xv_i)$};
\end{scope}

\end{tikzpicture}
\caption{From left to right, the four different types of edges belonging to $E^{(1)}(L)$, $E^{(2)}(L)$,$E^{(3)}(L)$, $E^{(4)}(L)$ respectively, and the conventions used for the two types of Boltzmann weights $ W$ and $\oW$ that are associated to them.}
\label{fig-crosses}
\end{figure}

The pairs of directed alternating dashed and dotted horizontal and vertical lines form the rapidity graph.  The lines of the rapidity graph cross edges $(ij)\in E(L)$ of $L$ at 45 degree angles.  There are pairs of variables $\bp=\{p,p'\}$, and $\bq=\{q,q'\}$, associated to horizontal and vertical rapidity lines respectively.  For a latter pair, the primed rapidity variable is represented by a dotted line, and the non primed rapidity variable is represented by a dashed line.  From Figure \ref{fig-lattice}, it is clear that the intersections of different rapidity lines distinguish four types of edges in the lattice graphically.  The sets of these four types of edges of $L$, will be denoted by $E^{(1)}(L), E^{(2)}(L), E^{(3)}(L), E^{(4)}(L)$, according to Figure \ref{fig-crosses}.

The spin variables $\xv_i$ are assigned to the vertices $i\in V(L)$.  The spin variables typically take values in a subset of either the integers or reals.  The statistical mechanical model involves nearest-neighbour interactions between pairs of spins $\xv_i,\xv_j$, associated to two vertices $i,j\in V(L)$ connected by an edge $(ij)\in E(L)$.  The interaction is represented by Boltzmann weights, denoted $ W_{pq}(\xv_i,\xv_j)$ and $\oW_{pq}(\xv_i,\xv_j)$, which are associated to the different edges of $L$ according to Figure \ref{fig-crosses}.

The Boltzmann weights are assumed here to satisfy the following inversion relations
\begin{align}
\label{invrels2}
\begin{gathered}
\ds W_{pq'}(\xv_1,\xv_2)\, W_{q'p}(\xv_2,\xv_1)=1, \\[0.3cm]
\ds\sum_{\xv_0}\,\oW_{pq}(\xv_1,\xv_0)\,\oW_{qp}(\xv_0,\xv_2)=\delta_{\xv_1,\xv_2}.
\end{gathered}
\end{align}

The partition function for the edge-interaction model, is given explicitly in terms Boltzmann weights on edges shown in Figure \ref{fig-crosses}, as
\begin{align}
\label{Zdefm}
Z_0=\hspace{-0.1cm}\sum_{{\mathbf{x}}}\prod_{(ij)\in E^{(1)}(L)}\!\!\!\!\! W_{pq'}(\xv_i,\xv_j)\!\!\!\!\!\prod_{(ij)\in E^{(2)}(L)}\!\!\!\!\! W_{p'q}(\xv_i,\xv_j)\!\!\!\!\!\prod_{(ij)\in E^{(3)}(L)}\!\!\!\!\!\oW_{pq}(\xv_i,\xv_j)\!\!\!\!\!\prod_{(ij)\in E^{(4)}(L)}\!\!\!\!\!\oW_{p'q'}(\xv_i,\xv_j),
\end{align}
where each product is taken over the respective sets of edges in $E(L)$, the sum $\sum_{\mathbf{x}}$ is taken over all configurations of spins $x_i$ associated to interior vertices $i\in V_{int}(L)$, and the boundary spins are kept fixed.  Note that the expression \eqref{Zdefm} is for integer valued models; for real valued models, the sum in \eqref{Zdefm} should simply be replaced by an integral over all interior spin configurations.


\subsection{Formulation as a vertex or an IRF model} \label{sec:IRF}

The edge-interaction model defined on the square lattice in Figure \ref{fig-lattice}, can be reformulated as either a vertex model, or as an interaction-round-a-face (IRF) model  \cite{Bazhanov:2011mz}.  This is done by forming the appropriate Boltzmann weights for the latter models, from different combinations of the four edge Boltzmann weights of Figure \ref{fig-crosses}.  

\subsubsection{IRF formulation}
\begin{figure}[htb!]
\centering
\begin{tikzpicture}[scale=1.2]

\draw[-,very thick] (-1,-1)--(1,1);
\draw[-,very thick] (1,-1)--(-1,1);

\filldraw[fill=black!,draw=black!] (-1,-1) circle (2pt)
node[below=1.5pt]{\color{black}\small $\xv_c$};
\filldraw[fill=black!,draw=black!] (1,-1) circle (2pt)
node[below=1.5pt]{\color{black}\small $\xv_d$};
\filldraw[fill=black!,draw=black!] (-1,1) circle (2pt)
node[above=1.5pt]{\color{black}\small $\xv_a$};
\filldraw[fill=black!,draw=black!] (1,1) circle (2pt)
node[above=1.5pt]{\color{black}\small $\xv_b$};

\filldraw[fill=black!,draw=black!] (0,0) circle (2pt)
node[below=5pt]{\color{black}\small $\xv_i$};

\draw[->,thick,dotted] (-1.2,0.5) -- (1.2,0.5);
\draw[black!] (-1.2,0.5) circle (0.01pt)
node[left=1.5pt]{\color{black}\small $p'$};
\draw[->,dashed] (-1.2,-0.5) -- (1.2,-0.5);
\draw[black!] (-1.3,-0.5) circle (0.01pt)
node[left=1.5pt]{\color{black}\small $p$};
\draw[->,dashed] (-0.5,-1.2) -- (-0.5,1.2);
\draw[black!] (-0.5,-1.27) circle (0.01pt)
node[below=1.5pt]{\color{black}\small $q$};
\draw[->,thick,dotted] (0.5,-1.2) -- (0.5,1.2);
\draw[black!] (0.5,-1.2) circle (0.01pt)
node[below=1.5pt]{\color{black}\small $q'$};

\draw[black!] (0,-1.8) circle (0.01pt)
node[below=0.1pt]{\color{black} $\bW^{(1)}_{\bp\bq}(\xv_a,\xv_b,\xv_c,\xv_d)$};

\begin{scope}[xshift=150pt]

\draw[-,very thick] (-1,-1)--(1,1);
\draw[-,very thick] (1,-1)--(-1,1);

\filldraw[fill=black!,draw=black!] (-1,-1) circle (2pt)
node[below=1.5pt]{\color{black}\small $\xv_c$};
\filldraw[fill=black!,draw=black!] (1,-1) circle (2pt)
node[below=1.5pt]{\color{black}\small $\xv_d$};
\filldraw[fill=black!,draw=black!] (-1,1) circle (2pt)
node[above=1.5pt]{\color{black}\small $\xv_a$};
\filldraw[fill=black!,draw=black!] (1,1) circle (2pt)
node[above=1.5pt]{\color{black}\small $\xv_b$};

\filldraw[fill=black!,draw=black!] (0,0) circle (2pt)
node[below=5pt]{\color{black}\small $\xv_j$};

\draw[->,dashed] (-1.2,0.5) -- (1.2,0.5);
\draw[black!] (-1.3,0.5) circle (0.01pt)
node[left=1.5pt]{\color{black}\small $p$};
\draw[->,thick,dotted] (-1.2,-0.5) -- (1.2,-0.5);
\draw[black!] (-1.2,-0.5) circle (0.01pt)
node[left=1.5pt]{\color{black}\small $p'$};
\draw[->,thick,dotted] (-0.5,-1.2) -- (-0.5,1.2);
\draw[black!] (-0.5,-1.2) circle (0.01pt)
node[below=1.5pt]{\color{black}\small $q'$};
\draw[->,dashed] (0.5,-1.2) -- (0.5,1.2);
\draw[black!] (0.5,-1.27) circle (0.01pt)
node[below=1.5pt]{\color{black}\small $q$};

\draw[black!] (0,-1.8) circle (0.01pt)
node[below=0.1pt]{\color{black} $\bW^{(2)}_{\bp\bq}(\xv_a,\xv_b,\xv_c,\xv_d)$};

\end{scope}
\end{tikzpicture}

\caption{A four-edge star centered at a vertex $i\in V^{(1)}(L)$ (left), and a four-edge star centered at a vertex $j \in V^{(2)}(L)$ (right), and their associated IRF Boltzmann weights \eqref{V1} and \eqref{V2} respectively.}
\label{fig-IRF}
\end{figure}
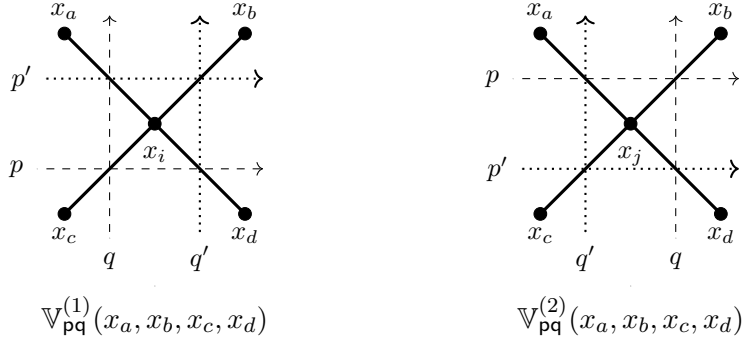

First the formulation of the IRF model will be given, since the star-star relation is expressed in terms of the IRF Boltzmann weights.  Note that the square lattice $L$ is bipartite, and the set of its vertices $V(L)$ may be split into two disjoint subsets, denoted by $V^{(1)}(L)$, and $V^{(2)}(L)$, such that edges $(ij)\in E(L)$, always connect a vertex $i\in V^{(1)}(L)$, with a vertex $j\in V^{(2)}(L)$.  In Figure \ref{fig-lattice}, two types of four-edge stars with a common vertex $i\in V^{(1)}(L)$, or $j\in V^{(2)}(L)$ respectively, can be distinguished by the intersections of rapidity lines on the associated edges, as is depicted in Figure \ref{fig-IRF}.  These two types of four-edge stars, respectively centered at vertices $i\in V^{(1)}(L)$ and $j\in V^{(2)}(L)$, are associated with the following IRF Boltzmann weights,
\begin{align}
\label{V1}
\ds \bW_{\bp\bq}^{(1)}(\xv_a,\xv_b,\xv_c,\xv_d)=\ds\sum_{x_i}\, \oW_{pq}(\xv_c,\xv_i)\,\oW_{p'q'}(\xv_b,\xv_i)\, W_{p'q}(\xv_i,\xv_a)\, W_{pq'}(\xv_i,\xv_d),
\end{align}
and
\begin{align}
\label{V2}
\ds \bW_{\bp\bq}^{(2)}(\xv_a,\xv_b,\xv_c,\xv_d)=\sum_{x_j}\,\oW_{pq}(\xv_j,\xv_b)\,  \oW_{p'q'}(\xv_j,\xv_c)\, W_{p'q}(\xv_d,\xv_j)\, W_{pq'}(\xv_a,\xv_j),
\end{align}
as indicated in Figure \ref{fig-IRF}.  As in the preceding section, the sums in \eqref{V1} and \eqref{V2}, are taken over the set of values of the interior spins $x_i$, and $x_j$, respectively, while the exterior boundary spins $\xv_a,\xv_b,\xv_c,\xv_d$ take some fixed values.

The model on the square lattice $L$ pictured in Figure \ref{fig-lattice}, may be formed with periodic translations of either of the stars pictured in Figure \ref{fig-IRF}, and the partition function \eqref{Zdefm} may be written in terms of either of the Boltzmann weights $\V^{(1)}_{\bp\bq}$, or $\V^{(2)}_{\bp\bq}$, respectively as
\begin{align}
\label{IRFZ1}
\ds Z_0=\ds\sum_{\mathbf{x}^{(2)}}\prod_{i\in V^{(1)}(L)}\V^{(1)}_{\bp\bq}(\xv_a,\xv_b,\xv_c,\xv_d),
\end{align}
or
\begin{align}
\label{IRFZ2}
Z_0=\ds\sum_{\mathbf{x}^{(1)}}\prod_{j\in V^{(2)}(L)}\V^{(2)}_{\bp\bq}(\xv_a,\xv_b,\xv_c,\xv_d).
\end{align}
The sum $\sum_{\x^{(2)}}$ in \eqref{IRFZ1}, represents $\sum_{\xv_1}\sum_{\xv_2},\ldots,\sum_{\xv_m}$, where $\xv_1,\xv_2,\ldots,\xv_m$, are spins assigned to vertices $i_1,i_2,\ldots,i_m\in V^{(2)}(L)\cap V_{int}(L)$.  The product in \eqref{IRFZ1}, is a product of all Boltzmann weights $\V^{(1)}_{\bp\bq}$, centered at vertices $i\in V^{(1)}(L)$.  The sum and product in the expression \eqref{IRFZ2} should be interpreted analogously.  The different expressions for the partition functions \eqref{Zdefm}, \eqref{IRFZ1}, and \eqref{IRFZ2}, are equivalent up to the boundary contributions.

The Boltzmann weights $\bW^{(1)}_{\bp\bq}$, and $\bW^{(2)}_{\bp\bq}$, are assumed here to satisfy the following {\it star-star relation},
\begin{align}
\label{star-star}
\ds W_{q'q}(\xv_d,\xv_c)\, W_{qq'}(\xv_a,\xv_b) \,\bW_{\bp\bq}^{(1)}(\xv_a,\xv_b,\xv_c,\xv_d)= W_{p'p}(\xv_c,\xv_a)\, W_{pp'}(\xv_b,\xv_d)\,\bW_{\bp\bq}^{(2)}(\xv_a,\xv_b,\xv_c,\xv_d),
\end{align}
which has a graphical representation shown in Figure \ref{star-star}.

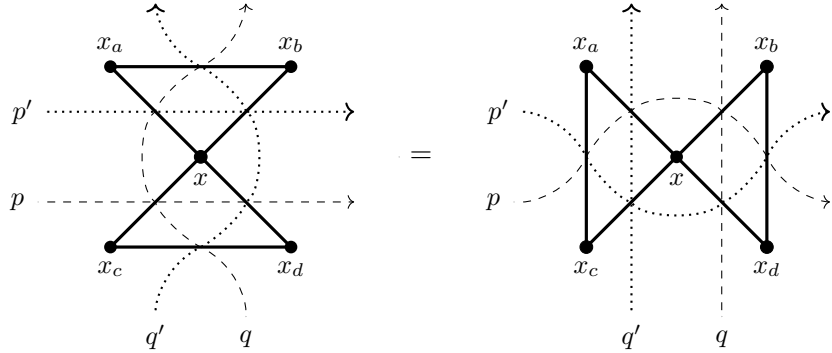
\begin{figure}[htb]
\centering
\begin{tikzpicture}[scale=1.2]

\draw[-,very thick] (-1,-1)--(1,1);
\draw[-,very thick] (1,-1)--(-1,1);
\draw[-,very thick] (-1,1)--(1,1);
\draw[-,very thick] (-1,-1)--(1,-1);

\fill[black!] (-1,-1) circle (2pt)
node[below=1.5pt]{\color{black}\small $\xv_c$};
\fill[black!] (1,-1) circle (2pt)
node[below=1.5pt]{\color{black}\small $\xv_d$};
\fill[black!] (-1,1) circle (2pt)
node[above=1.5pt]{\color{black}\small $\xv_a$};
\fill[black!] (1,1) circle (2pt)
node[above=1.5pt]{\color{black}\small $\xv_b$};

\filldraw[fill=black!,draw=black!] (0,0) circle (2pt)
node[below=2.5pt]{\color{black}\small $\xv$};

\draw[->,thick,dotted] (-1.7,0.5) -- (1.7,0.5);
\draw[black!] (-1.7,0.5) circle (0.01pt)
node[left=1.5pt]{\color{black}\small $p'$};
\draw[->,dashed] (-1.7,-0.5) -- (1.7,-0.5);
\draw[black!] (-1.8,-0.5) circle (0.01pt)
node[left=1.5pt]{\color{black}\small $p$};
\draw[->,thick,dotted] (-0.5,-1.7) .. controls (-0.5,-1.6) and (-0.4,-1.2) .. (0,-1) .. controls (0.2,-0.9) and (0.5,-0.6) .. (0.5,-0.5) .. controls (0.7,-0.4) and (0.7,0.4) .. (0.5,0.5) .. controls (0.5,0.6) and (0.2,0.9) .. (0,1) .. controls (-0.4,1.2) and (-0.5,1.6) .. (-0.5,1.7);
\draw[black!] (-0.5,-1.7) circle (0.01pt)
node[below=1.5pt]{\color{black}\small $q'$};
\draw[->,dashed] (0.5,-1.7) .. controls (0.5,-1.6) and (0.4,-1.2) .. (0,-1) .. controls (-0.2,-0.9) and (-0.5,-0.6) .. (-0.5,-0.5) .. controls (-0.7,-0.4) and (-0.7,0.4) .. (-0.5,0.5) .. controls (-0.5,0.6) and (-0.2,0.9) .. (0,1) .. controls (0.4,1.2) and (0.5,1.6) ..  (0.5,1.7);
\draw[black!] (0.5,-1.77) circle (0.01pt)
node[below=1.5pt]{\color{black}\small $q$};


\draw[black!] (2.2,0) circle (0.01pt)
node[right=0.1pt]{\color{black}=};

\begin{scope}[xshift=150pt]

\draw[-,very thick] (-1,-1)--(1,1);
\draw[-,very thick] (1,-1)--(-1,1);
\draw[-,very thick] (-1,-1)--(-1,1);
\draw[-,very thick] (1,-1)--(1,1);

\filldraw[fill=black!,draw=black!] (-1,-1) circle (2pt)
node[below=1.5pt]{\color{black}\small $\xv_c$};
\filldraw[fill=black!,draw=black!] (1,-1) circle (2pt)
node[below=1.5pt]{\color{black}\small $\xv_d$};
\filldraw[fill=black!,draw=black!] (-1,1) circle (2pt)
node[above=1.5pt]{\color{black}\small $\xv_a$};
\filldraw[fill=black!,draw=black!] (1,1) circle (2pt)
node[above=1.5pt]{\color{black}\small $\xv_b$};

\fill[black!] (0,0) circle (2pt)
node[below=2.5pt]{\color{black}\small $\xv$};

\draw[->,thick,dotted] (-1.7,0.5) .. controls (-1.6,0.5) and (-1.2,0.4) .. (-1,0) .. controls (-0.9,-0.2) and (-0.6,-0.5) .. (-0.5,-0.5) .. controls (-0.4,-0.7) and (0.4,-0.7) .. (0.5,-0.5) .. controls (0.6,-0.5) and (0.9,-0.2) .. (1,0) .. controls (1.2,0.4) and (1.6,0.5) .. (1.7,0.5);
\draw[black!] (-1.7,0.5) circle (0.01pt)
node[left=1.5pt]{\color{black}\small $p'$};
\draw[->,dashed] (-1.7,-0.5) .. controls (-1.6,-0.5) and (-1.2,-0.4) .. (-1,0) .. controls (-0.9,0.2) and (-0.6,0.5) .. (-0.5,0.5) .. controls (-0.4,0.7) and (0.4,0.7) .. (0.5,0.5) .. controls (0.6,0.5) and (0.9,0.2) .. (1,0) .. controls (1.2,-0.4) and (1.6,-0.5) .. (1.7,-0.5);
\draw[black!] (-1.8,-0.5) circle (0.01pt)
node[left=1.5pt]{\color{black}\small $p$};
\draw[->,thick,dotted] (-0.5,-1.7) -- (-0.5,1.7);
\draw[black!] (-0.5,-1.7) circle (0.01pt)
node[below=1.5pt]{\color{black}\small $q'$};
\draw[->,dashed] (0.5,-1.7) -- (0.5,1.7);
\draw[black!] (0.5,-1.77) circle (0.01pt)
node[below=1.5pt]{\color{black}\small $q$};


\end{scope}
\end{tikzpicture}

\caption{The star-star relation \eqref{star-star}.}
\label{ssfig}
\end{figure}

\begin{figure}[tbh!]
\centering
\begin{tikzpicture}[scale=1.3]

\begin{scope}[xshift=-180pt]


\draw[-,very thick] (2,0)--(0,0);\draw[-,very thick] (1,1.73)--(1,-1.73);
\draw[-,very thick] (1,1.73)--(-2,0);\draw[-,very thick] (-1,1.73)--(0,0);
\draw[-,very thick] (1,-1.73)--(-2,0);\draw[-,very thick] (-1,-1.73)--(0,0);

\filldraw[fill=black,draw=black] (-1,-1.73) circle (2.0pt)
node[below=1pt]{\small $\xv_e$};
\filldraw[fill=black,draw=black] (1,-1.73) circle (2.0pt)
node[below=1pt]{\small $\xv_d$};
\filldraw[fill=black,draw=black] (2,0) circle (2.0pt)
node[right=1pt]{\small $\xv_f$};
\filldraw[fill=black,draw=black] (1,1.73) circle (2.0pt)
node[above=1pt]{\small $\xv_b$};
\filldraw[fill=black,draw=black] (-1,1.73) circle (2.0pt)
node[above=1pt]{\small $\xv_a$};
\filldraw[fill=black,draw=black] (-2,0) circle (2.0pt)
node[left=1pt]{\small $\xv_c$};
\filldraw[fill=black,draw=black] (0,0) circle (2.0pt)
node[below=4pt]{\small $\xv$};
\filldraw[fill=black,draw=black] (-0.5,0.87) circle (2.0pt);
\filldraw[fill=black,draw=black] (-0.5,-0.87) circle (2.0pt);
\filldraw[fill=black,draw=black] (1,0) circle (2.0pt);

\draw[-,very thick] (-2,0.01)--(0,0.01);\draw[-,very thick] (-1,1.72)--(1,1.72);
\draw[-,very thick] (-2,-0.01)--(0,-0.01);\draw[-,very thick] (-1,1.74)--(1,1.74);

\draw[->,thick,dotted] (0.7,-2.08)--(0.5,-1.73)--(-0.25,-0.44) .. controls (-0.5,-0.22).. (-1,0) .. controls (-1.25,0.22) .. (-1.25,0.44) .. controls (-1.15,0.88) .. (-0.75,1.31) .. controls (-0.4,1.55) .. (0,1.73) .. controls (0.3,1.9) .. (0.5,2.08);
\draw[->,dashed] (-0.3,-2.08)--(-0.5,-1.73)--(-1.25,-0.43) .. controls (-1.25,-0.22).. (-1,0)--(-0.25,0.44) .. controls (0.25,0.88) .. (0.25,1.31) .. controls (0.15,1.52) .. (0,1.73) .. controls (-0.2,1.9) .. (-0.3,2.08);
\draw[->,thick,dotted] (-1.95,-0.44)--(-1.75,-0.44) .. controls (0.25,-0.44).. (0.25,-0.44) .. controls (0.25,-0.44) .. (1.25,1.31)--(1.45,1.66);
\draw[->,dashed] (-1.45,-1.31)--(-1.25,-1.31) .. controls (0.75,-1.31).. (0.75,-1.31) .. controls (0.75,-1.31) .. (1.75,0.44)--(1.95,0.81);
\draw[->,thick,dotted] (1.95,-0.81)--(1.75,-0.44) .. controls (0.75,1.31).. (0.75,1.31) .. controls (0.75,1.31) .. (-1.25,1.31)--(-1.45,1.31);
\draw[->,dashed] (1.45,-1.66)--(1.25,-1.31) .. controls (0.25,0.44).. (0.25,0.44) .. controls (0.25,0.44) .. (-1.75,0.44)--(-1.95,0.44);


\draw[black!] (-1.95,-0.44) circle (0.01pt)
node[left=1pt]{\color{black}\small $p'$};
\draw[black!] (-1.45,-1.31) circle (0.01pt)
node[left=1pt]{\color{black}\small $p$};
\draw[black!] (-0.3,-2.18) circle (0.01pt)
node[below=1pt]{\color{black}\small $q$};
\draw[black!] (0.7,-2.08) circle (0.01pt)
node[below=1pt]{\color{black}\small $q'$};
\draw[black!] (1.95,-0.81) circle (0.01pt)
node[right=1pt]{\color{black}\small $r'$};
\draw[black!] (1.45,-1.66) circle (0.01pt)
node[right=1pt]{\color{black}\small $r$};

\end{scope}


\draw[-,very thick] (-2,0)--(0,0);\draw[-,very thick] (-1,1.73)--(-1,-1.73);
\draw[-,very thick] (-1,1.73)--(2,0);\draw[-,very thick] (1,1.73)--(0,0);
\draw[-,very thick] (-1,-1.73)--(2,0);\draw[-,very thick] (1,-1.73)--(0,0);

\filldraw[fill=black,draw=black] (-1,-1.73) circle (2.0pt)
node[below=1pt]{\small $\xv_e$};
\filldraw[fill=black,draw=black] (1,-1.73) circle (2.0pt)
node[below=1pt]{\small $\xv_d$};
\filldraw[fill=black,draw=black] (2,0) circle (2.0pt)
node[right=1pt]{\small $\xv_f$};
\filldraw[fill=black,draw=black] (1,1.73) circle (2.0pt)
node[above=1pt]{\small $\xv_b$};
\filldraw[fill=black,draw=black] (-1,1.73) circle (2.0pt)
node[above=1pt]{\small $\xv_a$};
\filldraw[fill=black,draw=black] (-2,0) circle (2.0pt)
node[left=1pt]{\small $\xv_c$};
\filldraw[fill=black,draw=black] (0,0) circle (2.0pt)
node[below=4pt]{\small $\xv'$};
\filldraw[fill=black,draw=black] (0.5,0.87) circle (2.0pt);
\filldraw[fill=black,draw=black] (0.5,-0.87) circle (2.0pt);
\filldraw[fill=black,draw=black] (-1,0) circle (2.0pt);

\draw[-,very thick] (-1,-1.72)--(1,-1.72); \draw[-,very thick] (0,-0.01)--(2,-0.01);
\draw[-,very thick] (-1,-1.74)--(1,-1.74); \draw[-,very thick] (0,0.01)--(2,0.01);

\draw[->,dashed] (-0.7,-2.08) .. controls (-0.4,-1.9).. (0,-1.73) .. controls (0.5,-1.5) .. (0.75,-1.31) .. controls (1.1,-0.87) .. (1.25,-0.44) .. controls (1.25,-0.22).. (1,0) .. controls (0.5,0.22) .. (-0.5,1.73)--(-0.7,2.08);
\draw[->,thick,dotted] (0.3,-2.08)--(0,-1.73)--(-0.25,-1.31) .. controls (-0.15,-0.87) and (0,-0.65) .. (0.25,-0.44) .. controls (0.5,-0.22) .. (1,0) .. controls (1.5,0.22) .. (0.5,1.73)--(0.3,2.08);
\draw[->,thick,dotted] (1.95,-0.44)--(1.75,-0.44) .. controls (-0.25,-0.44).. (-0.25,-0.44) .. controls (-0.25,-0.44) .. (-1.25,1.31)--(-1.45,1.66);
\draw[->,dashed] (1.45,-1.31)--(1.25,-1.31) .. controls (-0.75,-1.31).. (-0.75,-1.31) .. controls (-0.75,-1.31) .. (-1.75,0.44)--(-1.95,0.81);
\draw[->,thick,dotted] (-1.95,-0.81)--(-1.75,-0.44) .. controls (-0.75,1.31).. (-0.75,1.31) .. controls (-0.75,1.31) .. (1.25,1.31)--(1.45,1.31);
\draw[->,dashed] (-1.45,-1.66)--(-1.25,-1.31) .. controls (-0.25,0.44).. (-0.25,0.44) .. controls (-0.25,0.44) .. (1.75,0.44)--(1.95,0.44);


\draw[black!] (1.95,-0.44) circle (0.01pt)
node[right=1pt]{\color{black}\small $r'$};
\draw[black!] (1.45,-1.31) circle (0.01pt)
node[right=1pt]{\color{black}\small $r$};
\draw[black!] (0.4,-2) circle (0.01pt)
node[below=1pt]{\color{black}\small $q'$};
\draw[black!] (-0.8,-2.08) circle (0.01pt)
node[below=1pt]{\color{black}\small $q$};
\draw[black!] (-1.95,-0.81) circle (0.01pt)
node[left=1pt]{\color{black}\small $p'$};
\draw[black!] (-1.45,-1.66) circle (0.01pt)
node[left=1pt]{\color{black}\small $p$};

\draw[black] (-2.9,0) circle (0.01pt)
node[left=1pt]{\color{black}\small $=$};

\end{tikzpicture}
\caption{Yang Baxter equation \eqref{YBE-IRF} for pairs of rapidity lines in the IRF formulation.}
\label{SSRYBE}
\end{figure}
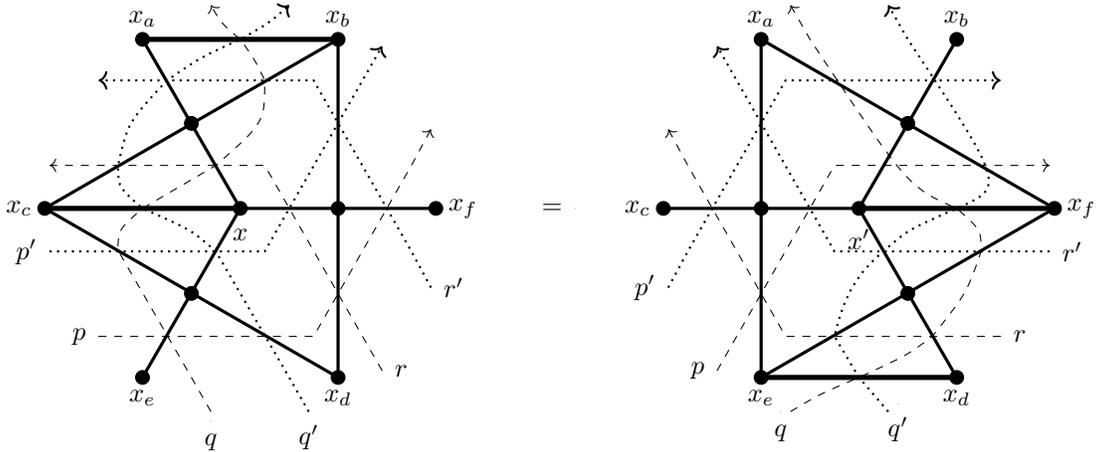

The star-star relation \eqref{star-star} is an important relation for models of statistical mechanics, particularly it implies the integrability of the associated  edge-interaction model through a Yang-Baxter equation defined for IRF weights.  The latter Yang-Baxter equation can be written in terms of either of the IRF Boltzmann weights \eqref{V1}, or \eqref{V2}, and it is given here in terms of \eqref{V1} as
\begin{align}
\label{YBE-IRF}
\begin{split}
\ds &\sum_x\, W_{qq'}(\xv_c,\xv)\, W_{q'q}(\xv_b,\xv_a)\,\bW^{(1)}_{\bp\bq}(\xv_c,\xv,\xv_e,\xv_d)\,\bW^{(1)}_{\bp\br}(\xv,\xv_b,\xv_d,\xv_f)\,\bW^{(1)}_{\bq\br}(\xv_c,\xv_a,\xv,\xv_b)\\
&=\ds \sum_{x'}\, W_{qq'}(\xv_e,\xv_d)\, W_{q'q}(\xv_f,\xv')\,\bW^{(1)}_{\bq\br}(\xv_e,\xv',\xv_d,\xv_f)\,\bW^{(1)}_{\bp\br}(\xv_c,\xv_a,\xv_e,\xv')\,\bW^{(1)}_{\bp\bq}(\xv_a,\xv_b,\xv',\xv_f),
\end{split}
\end{align}
with the graphical representation shown in Figure \ref{SSRYBE}.  In the expression for the Yang-Baxter equation \eqref{YBE-IRF}, spins $\xv_a,\xv_b,\xv_c,\xv_d,\xv_e,\xv_f,\xv_g$, take fixed values,  and the sums on the left and right hand sides, are taken over the sets of values of interior spins $\xv$, and $\xv'$, respectively.  The existence of the Yang-Baxter equation \eqref{YBE-IRF}, ensures that the transfer matrices of the IRF model defined with Boltzmann weights \eqref{V1} commute, and this implies that the transfer matrices for the original edge interaction model commute {in pairs} \cite{Baxter:1997tn}.  Note that the factors of Boltzmann weights $ W_{q'q},W_{qq'}$, appearing on both sides of the Yang-Baxter equation \eqref{YBE-IRF}, may be absorbed into a redefinition of the IRF weight $\bW^{(1)}_{\bp\bq}$.  However for the purpose  here of using the Yang-Baxter equation to show the extended Z-invariance, it is convenient to leave the edge Boltzmann weights as separate factors.

Note that the IRF Boltzmann weights \eqref{V1}, and \eqref{V2}, also satisfy the following inversion relation
\begin{align}
\label{IRFinv}
\sum_{\xv_0}\,\V^{(1)}_{\bp\bq}(\xv_a,\xv_0,\xv,\xv_d)\,\V^{(2)}_{\bq\bp}(\xv_a,\xv',\xv_0,\xv_d)=\sum_{\xv_0}\,\V^{(1)}_{\bp\bq}(\xv_a,\xv,\xv_0,\xv_d)\,\V^{(2)}_{\bq\bp}(\xv_a,\xv_0,\xv',\xv_d)=\delta_{\xv,\xv'},
\end{align}
where spins $\xv_a,\xv_d,\xv,\xv'$, take fixed values, and the sum is taken over the set of values for the spin $\xv_0$.  One way to obtain the relation \eqref{IRFinv}, is by expanding the IRF weights \eqref{V1}, \eqref{V2}, in terms of edge Boltzmann weights of Figure \ref{fig-crosses}, and then using the appropriate inversion relations given in \eqref{invrels2}.

\subsubsection{Vertex formulation}
\begin{figure}[htb!]
\centering
\begin{tikzpicture}[scale=1.5]
\draw[-,very thick] (-0.8,0)--(0,0.8)--(0.8,0)--(0,-0.8)--(-0.8,0);
\draw[->,thick,dotted,black] (-0.4,-1)--(-0.4,1);\draw[->,dashed,black] (0.4,-1)--(0.4,1);
\draw[->,dashed,black] (-1,-0.4)--(1,-0.4);
\draw[->,thick,dotted,black] (-1,0.4)--(1,0.4);
\fill[black!] (-0.4,-1.4) circle (0.01pt)
node[above=1.5pt]{\color{black}\small $q'$};
\fill[black!] (-1,-0.4) circle (0.01pt)
node[left=3.1pt]{\color{black}\small $p$};
\fill[black!] (0.4,-1.4) circle (0.01pt)
node[above=1.5pt]{\color{black}\small $q$};
\fill[black!] (-0.95,0.4) circle (0.01pt)
node[left=3.1pt]{\color{black}\small $p'$};
\filldraw[fill=black,draw=black] (0,-0.8) circle (1.5pt)
node[below=3pt]{\color{black}\small $\xv_j$};
\filldraw[fill=black,draw=black] (0,0.8) circle (1.5pt)
node[above=3pt]{\color{black}\small $\xv'_j$};
\filldraw[fill=black,draw=black] (0.8,0) circle (1.5pt)
node[right=3pt]{\color{black}\small $\xv'_i$};
\filldraw[fill=black,draw=black] (-0.8,0) circle (1.5pt)
node[left=3pt]{\color{black}\small $\xv_i$};

\fill (0,-1.3) circle(0.01pt)
node[below=0.05pt]{\color{black} $\left<\xv_i,\,\xv_j\,|\,\mathbb{R}_{\mathbf{p}\mathbf{q}}\,|\,\xv'_i,\,\xv'_j\right>$};

\end{tikzpicture}
\caption{Box configuration of edges and associated vertex Boltzmann weight \eqref{vertexBW}.}
\label{fig-vertex}
\end{figure}
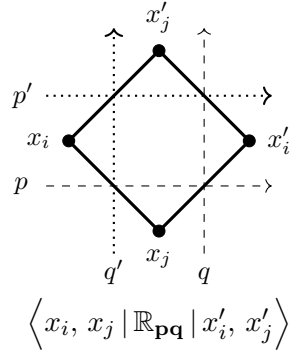

Next the vertex formulation will be given.  First note that the intersection of different rapidity lines on edges $(ij)\in E(L)$, in Figure \ref{fig-lattice}, distinguishes two types of faces of the lattice $L$.  Let $F^{(1)}(L)$ denote the set of faces of the lattice $L$ with the arrangement of edges as shown in Figure \ref{fig-vertex}, and $F^{(2)}(L)$ will denote the set of remaining faces such that $F^{(1)}(L)\cup F^{(2)}(L)=F(L)$.  Then the star-star relation \eqref{star-star} also implies a Yang-Baxter equation in the vertex formulation, where the following R-matrix
\begin{align}
\label{vertexBW}
\left<\xv_i,\,\xv_j\,|\,\mathbb{R}_{\mathbf{p}\mathbf{q}}\,|\,\xv'_i,\,\xv'_j\right>=\oW_{pq}(\xv_j,\xv'_i)\,\oW_{p'q'}(\xv'_j,\xv_i)\, W_{p'q}(\xv'_i,\xv'_j)\, W_{pq'}(\xv_i,\xv_j),
\end{align}
depicted in Figure \ref{fig-vertex}, corresponds to a vertex Boltzmann weight for four edges around a face\footnote{Also referred to simply as a ``box'' \cite{Bazhanov:1992jqa,Bazhanov:2011mz}.} $(i,j,i',j')\in F^{(1)}(L)$.   Here the four spins $\xv_i,\xv'_i$, and $\xv_j,\xv'_j$, are assigned to vertices $i,i'\in V^{(1)}(L)$, and $j,j'\in V^{(2)}(L)$ respectively.

The lattice in Figure \ref{fig-lattice}, may be reproduced with periodic translations of faces $(i,j,i',j')\in F^{(1)}(L)$, and the partition function \eqref{Zdefm} may be written in terms of the Boltzmann weight \eqref{vertexBW}, as
\begin{align}
\label{vertexZ}
Z_0=\sum_{\mathbf{x}}\prod_{(i,j,i',j')\in F^{(1)}(L)}\left<\xv_i,\,\xv_j\,|\,\mathbb{R}_{\mathbf{p}\mathbf{q}}\,|\,\xv'_i,\,\xv'_j\right>.
\end{align}
Here the sum has the same meaning as in \eqref{Zdefm}, the four spins $\xv_i,\xv'_i$, and $\xv_j,\xv'_j$, are once again assigned to the vertices $i,i'\in V^{(1)}(L)$, and $j,j'\in V^{(2)}(L)$ respectively, and the product is taken over faces $(i,j,i',l')\in F^{(1)}(L)$, while boundary spins are kept fixed.

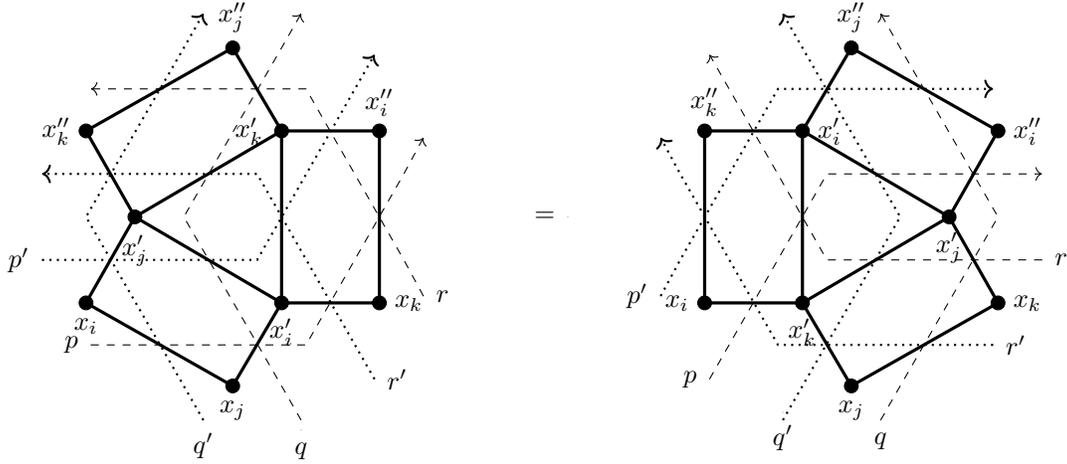
\begin{figure}[tbh]
\centering
\begin{tikzpicture}[scale=1.3]

\begin{scope}[xshift=-180pt]


\draw[-,very thick] (0,1.73)--(-1.5,0.88)--(-1,0)--(0.5,0.88)--(0,1.73);
\draw[-,very thick] (0,-1.73)--(-1.5,-0.88)--(-1,0)--(0.5,-0.88)--(0,-1.73);
\draw[-,very thick] (0.5,0.88)--(1.5,0.88)--(1.5,-0.88)--(0.5,-0.88)--(0.5,0.88);

\filldraw[fill=black,draw=black] (0,1.73) circle (2.0pt)
node[above=2pt]{\small $\xv''_j$};
\filldraw[fill=black,draw=black] (-1.5,0.88) circle (2.0pt)
node[left=2pt]{\small $\xv''_k$};
\filldraw[fill=black,draw=black] (-1.5,-0.88) circle (2.0pt)
node[below=2pt]{\small $\xv_i$};
\filldraw[fill=black,draw=black] (0,-1.73) circle (2.0pt)
node[below=2pt]{\small $\xv_j$};
\filldraw[fill=black,draw=black] (1.5,-0.88) circle (2.0pt)
node[right=2pt]{\small $\xv_k$};
\filldraw[fill=black,draw=black] (1.5,0.88) circle (2.0pt)
node[above=2pt]{\small $\xv''_i$};
\filldraw[fill=black,draw=black] (0.5,0.88) circle (2.0pt)
node[left=3pt]{\small $\xv'_k$};
\filldraw[fill=black,draw=black] (-1,0) circle (2.0pt)
node[below=3pt]{\small $\xv'_j$};
\filldraw[fill=black,draw=black] (0.5,-0.88) circle (2.0pt)
node[below=2pt]{\small $\xv'_i$};

\draw[->,dashed] (0.7,-2.08)--(0.5,-1.73) .. controls (-0.5,0).. (-0.5,0) .. controls (-0.5,0) .. (0.5,1.73)--(0.7,2.08);
\draw[->,thick,dotted] (-0.3,-2.08)--(-0.5,-1.73) .. controls (-1.5,0).. (-1.5,0) .. controls (-1.5,0) .. (-0.5,1.73)--(-0.3,2.08);
\draw[->,thick,dotted] (-1.95,-0.44)--(-1.75,-0.44) .. controls (0.25,-0.44).. (0.25,-0.44) .. controls (0.25,-0.44) .. (1.25,1.31)--(1.45,1.66);
\draw[->,dashed] (-1.45,-1.31)--(-1.25,-1.31) .. controls (0.75,-1.31).. (0.75,-1.31) .. controls (0.75,-1.31) .. (1.75,0.44)--(1.95,0.81);
\draw[->,dashed] (1.95,-0.81)--(1.75,-0.44) .. controls (0.75,1.31).. (0.75,1.31) .. controls (0.75,1.31) .. (-1.25,1.31)--(-1.45,1.31);
\draw[->,thick,dotted] (1.45,-1.66)--(1.25,-1.31) .. controls (0.25,0.44).. (0.25,0.44) .. controls (0.25,0.44) .. (-1.75,0.44)--(-1.95,0.44);


\draw[black!] (-1.95,-0.44) circle (0.01pt)
node[left=1pt]{\color{black}\small $p'$};
\draw[black!] (-1.45,-1.31) circle (0.01pt)
node[left=1pt]{\color{black}\small $p$};
\draw[black!] (-0.3,-2.08) circle (0.01pt)
node[below=1pt]{\color{black}\small $q'$};
\draw[black!] (0.7,-2.18) circle (0.01pt)
node[below=1pt]{\color{black}\small $q$};
\draw[black!] (1.95,-0.81) circle (0.01pt)
node[right=1pt]{\color{black}\small $r$};
\draw[black!] (1.45,-1.66) circle (0.01pt)
node[right=1pt]{\color{black}\small $r'$};

\end{scope}


\draw[-,very thick] (0,1.73)--(1.5,0.88)--(1,0)--(-0.5,0.88)--(0,1.73);
\draw[-,very thick] (0,-1.73)--(1.5,-0.88)--(1,0)--(-0.5,-0.88)--(0,-1.73);
\draw[-,very thick] (-0.5,0.88)--(-1.5,0.88)--(-1.5,-0.88)--(-0.5,-0.88)--(-0.5,0.88);

\filldraw[fill=black,draw=black] (0,1.73) circle (2.0pt)
node[above=2pt]{\small $\xv''_j$};
\filldraw[fill=black,draw=black] (1.5,0.88) circle (2.0pt)
node[right=2pt]{\small $\xv''_i$};
\filldraw[fill=black,draw=black] (1.5,-0.88) circle (2.0pt)
node[right=2pt]{\small $\xv_k$};
\filldraw[fill=black,draw=black] (0,-1.73) circle (2.0pt)
node[below=2pt]{\small $\xv_j$};
\filldraw[fill=black,draw=black] (-1.5,-0.88) circle (2.0pt)
node[left=2pt]{\small $\xv_i$};
\filldraw[fill=black,draw=black] (-1.5,0.88) circle (2.0pt)
node[above=2pt]{\small $\xv''_k$};
\filldraw[fill=black,draw=black] (-0.5,0.88) circle (2.0pt)
node[right=2pt]{\small $\xv'_i$};
\filldraw[fill=black,draw=black] (1,0) circle (2.0pt)
node[below=2pt]{\small $\xv'_j$};
\filldraw[fill=black,draw=black] (-0.5,-0.88) circle (2.0pt)
node[below=2pt]{\small $\xv'_k$};

\draw[->,thick,dotted] (-0.7,-2.08)--(-0.5,-1.73) .. controls (0.5,0).. (0.5,0) .. controls (0.5,0) .. (-0.5,1.73)--(-0.7,2.08);
\draw[->,dashed] (0.3,-2.08)--(0.5,-1.73) .. controls (1.5,0).. (1.5,0) .. controls (1.5,0) .. (0.5,1.73)--(0.3,2.08);
\draw[->,dashed] (1.95,-0.44)--(1.75,-0.44) .. controls (-0.25,-0.44).. (-0.25,-0.44) .. controls (-0.25,-0.44) .. (-1.25,1.31)--(-1.45,1.66);
\draw[->,thick,dotted] (1.45,-1.31)--(1.25,-1.31) .. controls (-0.75,-1.31).. (-0.75,-1.31) .. controls (-0.75,-1.31) .. (-1.75,0.44)--(-1.95,0.81);
\draw[->,thick,dotted] (-1.95,-0.81)--(-1.75,-0.44) .. controls (-0.75,1.31).. (-0.75,1.31) .. controls (-0.75,1.31) .. (1.25,1.31)--(1.45,1.31);
\draw[->,dashed] (-1.45,-1.66)--(-1.25,-1.31) .. controls (-0.25,0.44).. (-0.25,0.44) .. controls (-0.25,0.44) .. (1.75,0.44)--(1.95,0.44);


\draw[black!] (1.95,-0.44) circle (0.01pt)
node[right=1pt]{\color{black}\small $r$};
\draw[black!] (1.45,-1.31) circle (0.01pt)
node[right=1pt]{\color{black}\small $r'$};
\draw[black!] (0.3,-2.08) circle (0.01pt)
node[below=1pt]{\color{black}\small $q$};
\draw[black!] (-0.7,-2) circle (0.01pt)
node[below=1pt]{\color{black}\small $q'$};
\draw[black!] (-1.95,-0.81) circle (0.01pt)
node[left=1pt]{\color{black}\small $p'$};
\draw[black!] (-1.45,-1.66) circle (0.01pt)
node[left=1pt]{\color{black}\small $p$};

\draw[black!] (-2.9,0) circle (0.01pt)
node[left=1pt]{\color{black}\small $=$};

\end{tikzpicture}
\caption{Yang Baxter equation \eqref{YBE-vertex} for pairs of rapidity lines in the vertex formulation.}
\label{VYBE}
\end{figure}

The star-star relation \eqref{star-star}, implies the following Yang-Baxter equation 
\begin{align}
\label{YBE-vertex}
\begin{array}{c}
\ds\sum_{\xv'_i,\xv'_j,\xv'_k} \left<\xv_i,\,\xv_j\,|\,\mathbb{R}_{\mathbf{p}\mathbf{q}}\,|\,\xv'_i,\,\xv'_j\right>\,\left<\xv'_i,\,\xv_k\,|\,\mathbb{R}_{\mathbf{p}\mathbf{r}}\,|\,\xv''_i,\,\xv'_k\right>\,\left<\xv'_j,\,\xv'_k\,|\,\mathbb{R}_{\mathbf{q}\mathbf{r}}\,|\,\xv''_j,\,\xv''_k\right>\phantom{xxxx}\\[0.7cm]
\ds=\sum_{\xv'_i,\xv'_j,\xv'_k} \left<\xv_j,\,\xv_k\,|\,\mathbb{R}_{\mathbf{q}\mathbf{r}}\,|\,\xv'_j,\,\xv'_k\right>\,\left<\xv_i,\,\xv'_k\,|\,\mathbb{R}_{\mathbf{p}\mathbf{r}}\,|\,\xv'_i,\,\xv''_k\right>\,\left<\xv'_i,\,\xv'_j\,|\,\mathbb{R}_{\mathbf{p}\mathbf{q}}\,|\,\xv''_i,\,\xv''_j\right>,
\end{array}
\end{align}
depicted graphically in Figure \ref{VYBE}, for the vertex Boltzmann weights \eqref{vertexBW}.  In the expression \eqref{YBE-vertex}, the spins $\xv_i,\xv_j,\xv_k,\xv''_i,\xv''_j,\xv''_k$, are assigned fixed values, and the sums on both sides of \eqref{YBE-vertex} are taken over the set of values of the three interior spins $\xv'_i,\xv'_j,\xv'_k$.

Finally, the vertex Boltzmann weight \eqref{vertexBW}, also satisfies the following inversion relation
\begin{align}
\label{Vinv}
\sum_{\xv'_i,\xv'_j}\left<\xv_i,\xv_j\,|\,\mathbb{R}_{\bp\bq}\,|\,\xv'_i,\xv'_j\right>\,\left<\xv'_j,\xv'_i\,|\,\mathbb{R}_{\bq\bp}\,|\,\hat{\xv}_j,\hat{\xv}_i\right>=\delta_{\xv_i,\hat{\xv}_i}\,\delta_{\xv_j,\hat{\xv}_j},
\end{align}
where the spins $\xv_i,\xv_j,\hat{\xv}_i,\hat{\xv}_j$, take fixed values, and the sum is taken over the set of values of the spins $\xv'_i,\xv'_j$.  Similarly to the case for the IRF model, this inversion relation may be obtained by expanding the R-matrices \eqref{vertexBW} in terms of edge Boltzmann weights of Figure \ref{fig-crosses}, and then using the appropriate inversion relations given in \eqref{invrels2}.

\subsection{Extended Z-invariance for the integrable vertex and IRF models}\label{sec:z-invar2}

\subsubsection{Two-dimensional surface associated to the vertex and IRF models}\label{sec:sig}

Both the IRF and vertex models defined in the previous section, satisfy an {\it extended Z-invariance} property, as was established previously for edge-interaction models of statistical mechanics which satisfy the star-triangle relation \cite{Kels:2017fyt}.  To show the extended Z-invariance property, the models will be redefined on 2-dimensional surfaces made up of {\it elementary four-squares}, where each four-square is associated to a Boltzmann weight of the respective IRF or vertex models.  The surfaces may be deformed by using local cubic-type deformations that are pictured in Appendix \ref{app:IRF}, which are permitted as a consequence of the Yang-Baxter equations \eqref{YBE-IRF}, \eqref{YBE-vertex}, and inversion relations \eqref{IRFinv}, \eqref{Vinv}, satisfied by the respective models.  Then as was shown for the cases of the star-triangle relation \cite{Kels:2017fyt}, these local deformations are enough to show the property of extended Z-invariance.

\begin{figure}[htb]
\centering
\begin{tikzpicture}[scale=1.76]

\draw[black!] (0.5,0) circle (0.01pt);

\foreach \x in {0.5,2.5,...,4.5}{
\draw[->,black,dashed] (\x-0.15,-0.3) -- (\x+1.15,2.3);
\fill[black!] (\x-0.15,-0.3) circle (0.1pt)
node[below=0.05pt]{\color{black}\small $q$};}

\foreach \x in {1.5,3.5,...,5.5}{
\draw[->,thick,black,dotted] (\x-0.15,-0.3) -- (\x+1.15,2.3);
\fill[black!] (\x-0.15,-0.25) circle (0.1pt)
node[below=0.05pt]{\color{black}\small $q'$};}

\foreach \y in {0.25,1.25}{
\draw[->,black,dashed] (0.5*\y-0.4,\y) -- (0.5*\y+6+0.4,\y);
\fill[black!] (0.5*\y-0.4,\y) circle (0.1pt)
node[left=0.05pt]{\color{black}\small $p$};}

\foreach \y in {0.75,1.75}{
\draw[->,thick,black,dotted] (0.5*\y-0.4,\y) -- (0.5*\y+6+0.4,\y);
\fill[black!] (0.5*\y-0.35,\y) circle (0.1pt)
node[left=0.05pt]{\color{black}\small $p'$};}

\draw[-,thin,gray] (0,0)--(6,0)--(7,2)--(1,2)--(0,0);
\foreach \x in {1,2,...,5}
\draw[-,thin,gray] (\x,0)--(\x+1,2);

\foreach \y in {0.5,1,1.5}
\draw[-,thin,gray] (0.5*\y,\y)--(6+0.5*\y,\y);

\foreach \x in {0,2,4}{
\filldraw[draw=black,fill=black] (\x+1,2) circle (1.1pt);
\filldraw[draw=black,fill=white] (\x+2,2) circle (1.1pt);
\foreach \y in {0,1}{
\filldraw[draw=black,fill=black] (\x+\y*0.5,\y) circle (1.1pt);
\filldraw[draw=black,fill=white] (\x+\y*0.5+1,\y) circle (1.1pt);
\filldraw[draw=black,fill=white] (\x+0.25+\y*0.5,\y+0.5) circle (1.1pt);
\filldraw[draw=black,fill=black] (\x+0.25+\y*0.5+1,\y+0.5) circle (1.1pt);
}}
\foreach \y in {0,1}{
\filldraw[draw=black,fill=black] (6+\y*0.5,\y) circle (1.1pt);
\filldraw[draw=black,fill=white] (6+\y*0.5+0.25,\y+0.5) circle (1.1pt);
}
\filldraw[draw=black,fill=black] (7,2) circle (1.1pt);

\draw[-, thick] (0.5,1)--(3,2);\draw[-, thick] (0,0)--(5,2);\draw[-, thick] (2,0)--(7,2);\draw[-, thick] (4,0)--(6.5,1);\draw[-, thick] (2,0)--(0.5,1);\draw[-, thick] (4,0)--(1,2);\draw[-, thick] (6,0)--(3,2);\draw[-, thick] (6.5,1)--(5,2);

\fill[black] (7.86,0) circle (0.01pt)
node[below=1.5pt]{\color{black} $+\ihat$};
\fill[black] (7.66,0.33) circle (0.01pt)
node[right=1.5pt]{\color{black} $+\jhat$};
\fill[black] (7.5,0.75*0.5) circle (0.01pt)
node[left=1.5pt]{\color{black} $+\khat$};
\draw[->, thick] (7.5,0)--(8.25,0);\draw[->, thick] (7.5,0)--(7.83,0.67);\draw[->, thick] (7.5,0)--(7.5,0.75);

\end{tikzpicture}
\caption{An example of a surface $\sigma_0$, made up of faces and edges edges that connect the black and white nearest neighbour vertices of $\mathbb{Z}^2$.  The edges connecting next nearest neighbour vertices of $\sigma_0$, form the lattice of Figure \ref{fig-lattice}. This particular surface is naturally associated to an IRF model with 6 faces of the type that are shown on the right hand side of Figure \ref{e4s}.}
\label{Glattice}
\end{figure}
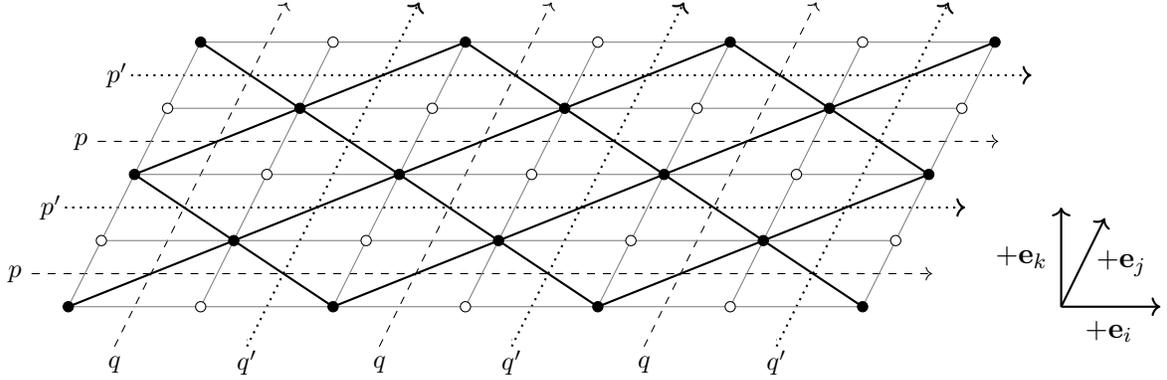

The first step will be to associate the models of the previous section to a two-dimensional surface, denoted by $\sigma_0$, which is made up of elementary squares lying in a plane, as shown in Figure \ref{Glattice}.  Let $V(\sigma_0)$, $E(\sigma_0)$, and $F(\sigma_0)$, denote respectively the sets of vertices, edges, and elementary squares (faces), of $\sigma_0$.  Vertices $i\in V(\sigma_0)$, are depicted as both the black and white vertices in Figure \ref{Glattice}.  Black next nearest neighbour vertices of $\sigma_0$ are connected by edges on diagonals of elementary squares; these diagonal edges do not belong to $E(\sigma_0)$, but together form a separate square lattice, $L$, equivalent to the lattice depicted in Figure \ref{fig-lattice}.


The directed rapidity graph, $\L$, is shown in Figure \ref{Glattice} consisting of the continuous directed dotted and dashed lines, that cross edges $(ij)\in E(\sigma_0)$ perpendicularly, and cross edges $(ij)\in E(L)$ at 45 degree angles.  Alternating rapidity lines that are aligned in the $+\ihat$ direction, are labelled by one of the variables $p,p'$, and alternating rapidity lines that are aligned in the $+\jhat$ direction, are labelled by one of the variables $q,q'$.

\begin{figure}[htb!]
\centering
\begin{tikzpicture}[scale=1.5]
\draw[->,very thick] (-2.4,-0.2)--(-1.9,-0.2);
\fill[black!] (-2.15,-0.2) circle (0.01pt)
node[below=1pt]{\color{black}\small $+\ihat$};
\draw[->,very thick] (-2.4,-0.2)--(-2.4,0.3);
\fill[black!] (-2.4,0.05) circle (0.01pt)
node[left=1pt]{\color{black}\small $+\jhat$};
\draw[-,gray] (-0.8,-0.8)--(0.8,-0.8)--(0.8,0.8)--(-0.8,0.8)--(-0.8,-0.8);
\draw[-,gray] (0,-0.8)--(0,0.8);\draw[-,gray] (-0.8,0)--(0.8,0);
\draw[-,very thick] (-0.8,0)--(0,0.8)--(0.8,0)--(0,-0.8)--(-0.8,0);
\draw[->,thick,dotted,black] (-0.4,-1)--(-0.4,1);\draw[->,dashed,black] (0.4,-1)--(0.4,1);
\draw[->,dashed,black] (-1,-0.4)--(1,-0.4);
\draw[->,thick,dotted,black] (-1,0.4)--(1,0.4);
\fill[black!] (-0.4,-1.4) circle (0.01pt)
node[above=1.5pt]{\color{black}\small $q'$};
\fill[black!] (-1,-0.4) circle (0.01pt)
node[left=3.1pt]{\color{black}\small $p$};
\fill[black!] (0.4,-1.4) circle (0.01pt)
node[above=1.5pt]{\color{black}\small $q$};
\fill[black!] (-0.95,0.4) circle (0.01pt)
node[left=3.1pt]{\color{black}\small $p'$};
\filldraw[fill=white,draw=black] (-0.8,-0.8) circle (1.5pt);
\filldraw[fill=white,draw=black] (0.8,-0.8) circle (1.5pt);
\filldraw[fill=white,draw=black] (0.8,0.8) circle (1.5pt);
\filldraw[fill=white,draw=black] (-0.8,0.8) circle (1.5pt);
\filldraw[fill=white,draw=black] (0,0) circle (1.5pt);
\filldraw[fill=black,draw=black] (0,-0.8) circle (1.5pt)
node[below=3pt]{\color{black}\small $\xv_j$};
\filldraw[fill=black,draw=black] (0,0.8) circle (1.5pt)
node[above=3pt]{\color{black}\small $\xv'_j$};
\filldraw[fill=black,draw=black] (0.8,0) circle (1.5pt)
node[right=3pt]{\color{black}\small $\xv'_i$};
\filldraw[fill=black,draw=black] (-0.8,0) circle (1.5pt)
node[left=3pt]{\color{black}\small $\xv_i$};

\fill (0,-1.3) circle(0.01pt)
node[below=0.05pt]{\color{black} $\left<\xv_i,\,\xv_j\,|\,\mathbb{R}_{\mathbf{p}\mathbf{q}}\,|\,\xv'_i,\,\xv'_j\right>$};

\begin{scope}[xshift=100pt]
\draw[->,very thick,white!] (2.3,-0.8)--(2.4,-0.8);
\draw[-,gray] (-0.8,-0.8)--(0.8,-0.8)--(0.8,0.8)--(-0.8,0.8)--(-0.8,-0.8);
\draw[-,gray] (0,-0.8)--(0,0.8);\draw[-,gray] (-0.8,0)--(0.8,0);
\draw[-,very thick] (-0.8,-0.8)--(0.8,0.8);\draw[-,very thick] (-0.8,0.8)--(0.8,-0.8);
\draw[->,dashed,black] (-0.4,-1)--(-0.4,1);\draw[->,thick,dotted,black] (0.4,-1)--(0.4,1);
\draw[->,dashed,black] (-1,-0.4)--(1,-0.4);
\draw[->,thick,dotted,black] (-1,0.4)--(1,0.4);
\fill[black!] (-0.4,-1.4) circle (0.01pt)
node[above=1.5pt]{\color{black}\small $q$};
\fill[black!] (-1,-0.4) circle (0.01pt)
node[left=3.1pt]{\color{black}\small $p$};
\fill[black!] (0.4,-1.4) circle (0.01pt)
node[above=1.5pt]{\color{black}\small $q'$};
\fill[black!] (-0.95,0.4) circle (0.01pt)
node[left=3.1pt]{\color{black}\small $p'$};
\filldraw[fill=black,draw=black] (-0.8,-0.8) circle (1.5pt)
node[below=3pt]{\color{black}\small $\xv_c$};
\filldraw[fill=black,draw=black] (0.8,-0.8) circle (1.5pt)
node[below=3pt]{\color{black}\small $\xv_d$};
\filldraw[fill=black,draw=black] (0.8,0.8) circle (1.5pt)
node[above=3pt]{\color{black}\small $\xv_b$};
\filldraw[fill=black,draw=black] (-0.8,0.8) circle (1.5pt)
node[above=3pt]{\color{black}\small $\xv_a$};
\filldraw[fill=black,draw=black] (0,0) circle (1.5pt);
\filldraw[fill=white,draw=black] (0,-0.8) circle (1.5pt);
\filldraw[fill=white,draw=black] (0,0.8) circle (1.5pt);
\filldraw[fill=white,draw=black] (0.8,0) circle (1.5pt);
\filldraw[fill=white,draw=black] (-0.8,0) circle (1.5pt);

\fill (0,-1.3) circle(0.01pt)
node[below=0.05pt]{\color{black} $\bW^{(1)}_{pq}(\xv_a,\xv_b,\xv_c,\xv_d)$};

\end{scope}

\end{tikzpicture}
\caption{Elementary four-squares $\sigma^4_{ij}(\mathbf{n})$, for the vertex formulation (left), and for the IRF formulation (right), and respective Boltzmann weights \eqref{vertexBW}, and \eqref{V1}.}
\label{e4s}
\end{figure}
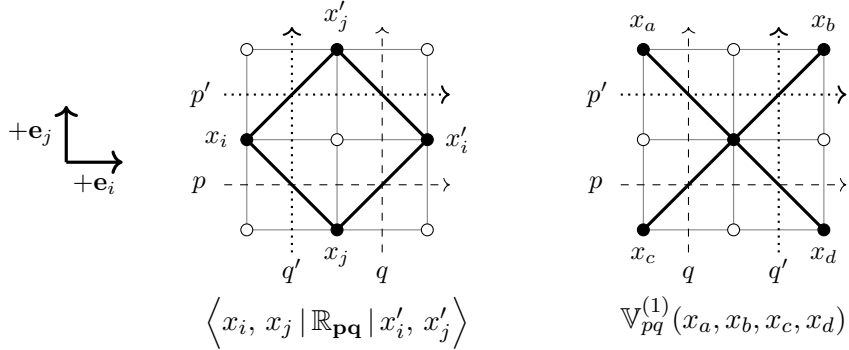

The IRF and vertex Boltzmann weights are naturally associated to the respective groups of four elementary squares shown in Figure \ref{e4s}.  These groups of four elementary squares will be referred to in the following by {\it elementary four-squares}, and are denoted by the notation $\sigma^4_{ij}$, where the $i$ and $j$ indices refer to two orthogonal lattice directions, as pictured in Figure \ref{Glattice}.  An elementary four-square is specified as $\sigma^4_{ij}(\mathbf{n})=(\mathbf{n},\mathbf{n}+2\ihat,\mathbf{n}+2(\ihat+\jhat),\mathbf{n}+2\jhat)$, where $\mathbf{n}$ is the coordinate of the vertex, and $\ihat,\jhat$ are orthogonal lattice directions.  Each elementary four-square is associated to a Boltzmann weight for either the vertex or IRF models, and these four-squares will be used in the following subsections as the building blocks of a general two-dimensional surface $\sigma$, with vertices in a subset of the cubic lattice $\mathbb{Z}^3$, on which the vertex or IRF models are defined.

To define the models of statistical mechanics on $\sigma_0$, spin variables $x_i$, are assigned only to black vertices $i\in V(L)$, while white vertices don't play any role for the definition of the vertex or IRF models.  Spins on the boundary of $\sigma_0$ are prescribed some fixed values. Note also that this boundary will differ depending on whether we are dealing with a vertex or IRF model.  For example, the surface $\sigma_0$ pictured in Figure \ref{Glattice} is naturally associated to an IRF model with six faces of the type shown in Figure \ref{e4s}.  The Boltzmann weights for the respective models are assigned according to Figure \ref{e4s}, and the partition functions are given by \eqref{IRFZ1}, and \eqref{vertexZ}, for the IRF and vertex models respectively.  This defines the IRF and vertex models on $\sigma_0$.


\subsubsection{Deformed two-dimensional surface for vertex model}\label{sec:vertexsigma}

Next the vertex and IRF models will be associated to a more general type of two-dimensional surface, denoted by $\sigma$, which is made up of elementary four-squares which are not restricted to lie in a plane.  First the vertex model will be considered, because this model turns out to have a simpler description on $\sigma$ than the IRF model.

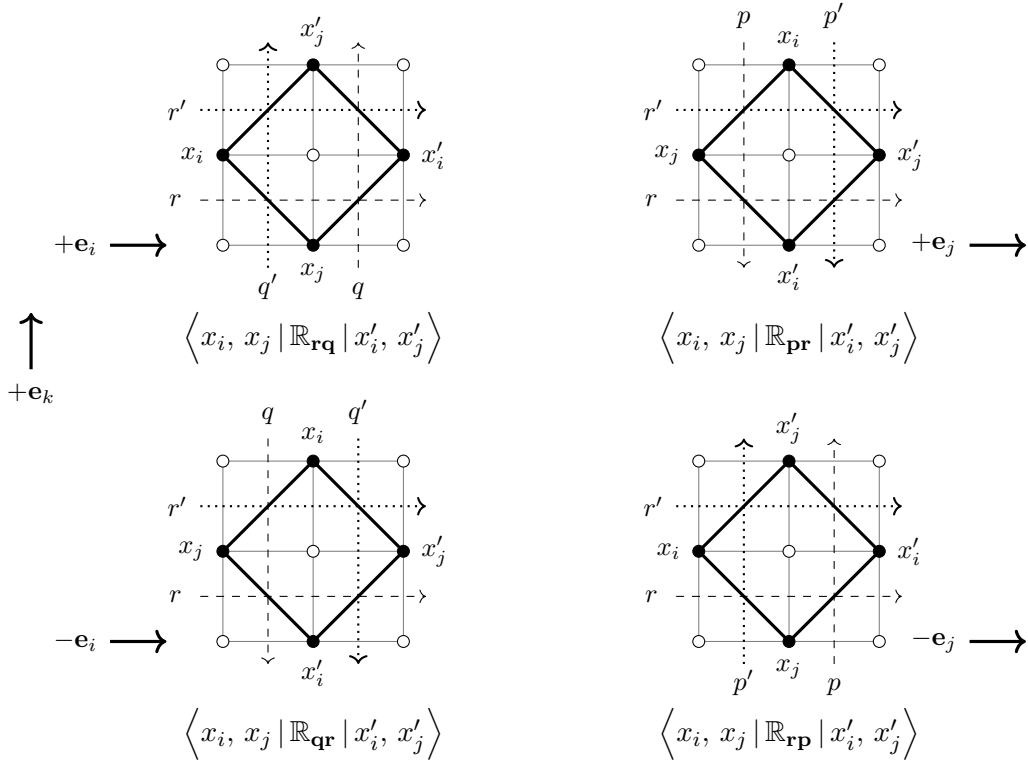
\begin{figure}[htb!]
\centering
\begin{tikzpicture}[scale=1.5]
\draw[->,very thick] (-2.5,-1.9)--(-2.5,-1.4);
\fill[black!] (-2.5,-1.9) circle (0.01pt)
node[below=1.5pt]{\color{black}\small $+\khat$};
\draw[->,very thick] (-1.8,-0.8)--(-1.3,-0.8);
\fill[black!] (-1.8,-0.8) circle (0.01pt)
node[left=1pt]{\color{black}\small $+\ihat$};
\draw[-,gray] (-0.8,-0.8)--(0.8,-0.8)--(0.8,0.8)--(-0.8,0.8)--(-0.8,-0.8);
\draw[-,gray] (0,-0.8)--(0,0.8);\draw[-,gray] (-0.8,0)--(0.8,0);
\draw[-,very thick] (-0.8,0)--(0,0.8)--(0.8,0)--(0,-0.8)--(-0.8,0);
\draw[->,thick,dotted,black] (-0.4,-1)--(-0.4,1);\draw[->,dashed,black] (0.4,-1)--(0.4,1);
\draw[->,dashed,black!30!black] (-1,-0.4)--(1,-0.4);
\draw[->,thick,dotted,black!30!black] (-1,0.4)--(1,0.4);
\fill[black!] (-0.4,-1.4) circle (0.01pt)
node[above=1.5pt]{\color{black}\small $q'$};
\fill[black!] (-1,-0.4) circle (0.01pt)
node[left=3.1pt]{\color{black}\small $r$};
\fill[black!] (0.4,-1.4) circle (0.01pt)
node[above=1.5pt]{\color{black}\small $q$};
\fill[black!] (-0.95,0.4) circle (0.01pt)
node[left=3.1pt]{\color{black}\small $r'$};
\filldraw[fill=white,draw=black] (-0.8,-0.8) circle (1.5pt);
\filldraw[fill=white,draw=black] (0.8,-0.8) circle (1.5pt);
\filldraw[fill=white,draw=black] (0.8,0.8) circle (1.5pt);
\filldraw[fill=white,draw=black] (-0.8,0.8) circle (1.5pt);
\filldraw[fill=white,draw=black] (0,0) circle (1.5pt);
\filldraw[fill=black,draw=black] (0,-0.8) circle (1.5pt)
node[below=3pt]{\color{black}\small $\xv_j$};
\filldraw[fill=black,draw=black] (0,0.8) circle (1.5pt)
node[above=3pt]{\color{black}\small $\xv'_j$};
\filldraw[fill=black,draw=black] (0.8,0) circle (1.5pt)
node[right=3pt]{\color{black}\small $\xv'_i$};
\filldraw[fill=black,draw=black] (-0.8,0) circle (1.5pt)
node[left=3pt]{\color{black}\small $\xv_i$};

\fill (0,-1.3) circle(0.01pt)
node[below=0.05pt]{\color{black} $\left<\xv_i,\,\xv_j\,|\,\mathbb{R}_{\mathbf{r}\mathbf{q}}\,|\,\xv'_i,\,\xv'_j\right>$};

\begin{scope}[xshift=120pt]
\draw[->,very thick] (1.6,-0.8)--(2.1,-0.8);
\fill[black!] (1.6,-0.8) circle (0.01pt)
node[left=1pt]{\color{black}\small $+\jhat$};
\draw[-,gray] (-0.8,-0.8)--(0.8,-0.8)--(0.8,0.8)--(-0.8,0.8)--(-0.8,-0.8);
\draw[-,gray] (0,-0.8)--(0,0.8);\draw[-,gray] (-0.8,0)--(0.8,0);
\draw[-,very thick] (-0.8,0)--(0,0.8)--(0.8,0)--(0,-0.8)--(-0.8,0);
\draw[->,dashed,black] (-0.4,1)--(-0.4,-1);\draw[->,thick,dotted,black] (0.4,1)--(0.4,-1);
\draw[->,dashed,black!30!black] (-1,-0.4)--(1,-0.4);
\draw[->,thick,dotted,black!30!black] (-1,0.4)--(1,0.4);
\fill[black!] (-0.4,1) circle (0.01pt)
node[above=1.5pt]{\color{black}\small $p$};
\fill[black!] (-1,-0.4) circle (0.01pt)
node[left=3.1pt]{\color{black}\small $r$};
\fill[black!] (0.4,1) circle (0.01pt)
node[above=1.5pt]{\color{black}\small $p'$};
\fill[black!] (-0.95,0.4) circle (0.01pt)
node[left=3.1pt]{\color{black}\small $r'$};
\filldraw[fill=white,draw=black] (-0.8,-0.8) circle (1.5pt);
\filldraw[fill=white,draw=black] (0.8,-0.8) circle (1.5pt);
\filldraw[fill=white,draw=black] (0.8,0.8) circle (1.5pt);
\filldraw[fill=white,draw=black] (-0.8,0.8) circle (1.5pt);
\filldraw[fill=white,draw=black] (0,0) circle (1.5pt);
\filldraw[fill=black,draw=black] (0,-0.8) circle (1.5pt)
node[below=3pt]{\color{black}\small $\xv'_i$};
\filldraw[fill=black,draw=black] (0,0.8) circle (1.5pt)
node[above=3pt]{\color{black}\small $\xv_i$};
\filldraw[fill=black,draw=black] (0.8,0) circle (1.5pt)
node[right=3pt]{\color{black}\small $\xv'_j$};
\filldraw[fill=black,draw=black] (-0.8,0) circle (1.5pt)
node[left=3pt]{\color{black}\small $\xv_j$};

\fill (0,-1.3) circle(0.01pt)
node[below=0.05pt]{\color{black} $\left<\xv_i,\,\xv_j\,|\,\mathbb{R}_{\mathbf{p}\mathbf{r}}\,|\,\xv'_i,\,\xv'_j\right>$};

\end{scope}

\begin{scope}[yshift=-100pt]
\draw[->,very thick] (-1.8,-0.8)--(-1.3,-0.8);
\fill[black!] (-1.8,-0.8) circle (0.01pt)
node[left=1pt]{\color{black}\small $-\ihat$};
\draw[-,gray] (-0.8,-0.8)--(0.8,-0.8)--(0.8,0.8)--(-0.8,0.8)--(-0.8,-0.8);
\draw[-,gray] (0,-0.8)--(0,0.8);\draw[-,gray] (-0.8,0)--(0.8,0);
\draw[-,very thick] (-0.8,0)--(0,0.8)--(0.8,0)--(0,-0.8)--(-0.8,0);
\draw[->,dashed,black] (-0.4,1)--(-0.4,-1);\draw[->,thick,dotted,black] (0.4,1)--(0.4,-1);
\draw[->,dashed,black!30!black] (-1,-0.4)--(1,-0.4);
\draw[->,thick,dotted,black!30!black] (-1,0.4)--(1,0.4);
\fill[black!] (-0.4,1) circle (0.01pt)
node[above=1.5pt]{\color{black}\small $q$};
\fill[black!] (-1,-0.4) circle (0.01pt)
node[left=3.1pt]{\color{black}\small $r$};
\fill[black!] (0.4,1) circle (0.01pt)
node[above=1.5pt]{\color{black}\small $q'$};
\fill[black!] (-0.95,0.4) circle (0.01pt)
node[left=3.1pt]{\color{black}\small $r'$};
\filldraw[fill=white,draw=black] (-0.8,-0.8) circle (1.5pt);
\filldraw[fill=white,draw=black] (0.8,-0.8) circle (1.5pt);
\filldraw[fill=white,draw=black] (0.8,0.8) circle (1.5pt);
\filldraw[fill=white,draw=black] (-0.8,0.8) circle (1.5pt);
\filldraw[fill=white,draw=black] (0,0) circle (1.5pt);
\filldraw[fill=black,draw=black] (0,-0.8) circle (1.5pt)
node[below=3pt]{\color{black}\small $\xv'_i$};
\filldraw[fill=black,draw=black] (0,0.8) circle (1.5pt)
node[above=3pt]{\color{black}\small $\xv_i$};
\filldraw[fill=black,draw=black] (0.8,0) circle (1.5pt)
node[right=3pt]{\color{black}\small $\xv'_j$};
\filldraw[fill=black,draw=black] (-0.8,0) circle (1.5pt)
node[left=3pt]{\color{black}\small $\xv_j$};

\fill (0,-1.3) circle(0.01pt)
node[below=0.05pt]{\color{black} $\left<\xv_i,\,\xv_j\,|\,\mathbb{R}_{\mathbf{q}\mathbf{r}}\,|\,\xv'_i,\,\xv'_j\right>$};

\end{scope}

\begin{scope}[xshift=120pt,yshift=-100pt]
\draw[->,very thick] (1.6,-0.8)--(2.1,-0.8);
\fill[black!] (1.6,-0.8) circle (0.01pt)
node[left=1pt]{\color{black}\small $-\jhat$};
\fill[black!] (2,0) circle (0.01pt)
node[below=1.5pt]{\color{white}\small $+\khat$};
\draw[-,gray] (-0.8,-0.8)--(0.8,-0.8)--(0.8,0.8)--(-0.8,0.8)--(-0.8,-0.8);
\draw[-,gray] (0,-0.8)--(0,0.8);\draw[-,gray] (-0.8,0)--(0.8,0);
\draw[-,very thick] (-0.8,0)--(0,0.8)--(0.8,0)--(0,-0.8)--(-0.8,0);
\draw[->,thick,dotted,black] (-0.4,-1)--(-0.4,1);\draw[->,dashed,black] (0.4,-1)--(0.4,1);
\draw[->,dashed,black!30!black] (-1,-0.4)--(1,-0.4);
\draw[->,thick,dotted,black!30!black] (-1,0.4)--(1,0.4);
\fill[black!] (-0.4,-1.4) circle (0.01pt)
node[above=1.5pt]{\color{black}\small $p'$};
\fill[black!] (-1,-0.4) circle (0.01pt)
node[left=3.1pt]{\color{black}\small $r$};
\fill[black!] (0.4,-1.4) circle (0.01pt)
node[above=1.5pt]{\color{black}\small $p$};
\fill[black!] (-0.95,0.4) circle (0.01pt)
node[left=3.1pt]{\color{black}\small $r'$};
\filldraw[fill=white,draw=black] (-0.8,-0.8) circle (1.5pt);
\filldraw[fill=white,draw=black] (0.8,-0.8) circle (1.5pt);
\filldraw[fill=white,draw=black] (0.8,0.8) circle (1.5pt);
\filldraw[fill=white,draw=black] (-0.8,0.8) circle (1.5pt);
\filldraw[fill=white,draw=black] (0,0) circle (1.5pt);
\filldraw[fill=black,draw=black] (0,-0.8) circle (1.5pt)
node[below=3pt]{\color{black}\small $\xv_j$};
\filldraw[fill=black,draw=black] (0,0.8) circle (1.5pt)
node[above=3pt]{\color{black}\small $\xv'_j$};
\filldraw[fill=black,draw=black] (0.8,0) circle (1.5pt)
node[right=3pt]{\color{black}\small $\xv'_i$};
\filldraw[fill=black,draw=black] (-0.8,0) circle (1.5pt)
node[left=3pt]{\color{black}\small $\xv_i$};

\fill (0,-1.3) circle(0.01pt)
node[below=0.05pt]{\color{black} $\left<\xv_i,\,\xv_j\,|\,\mathbb{R}_{\mathbf{r}\mathbf{p}}\,|\,\xv'_i,\,\xv'_j\right>$};

\end{scope}

\end{tikzpicture}
\caption{In clockwise order starting from the top right, the Boltzmann weight \eqref{vertexBW} that is assigned to elementary four-squares $\sigma^4_{jk}(\n)$, $\sigma^4_{kj}(\n)$, $\sigma^4_{ki}(\n)$, $\sigma^4_{ik}(\n)$, respectively, in the vertex formulation.}
\label{4VBW}
\end{figure}

A straightforward way to obtain the desired surface $\sigma$, is to shift each elementary four-square $\sigma^4_{ij}(\n)$ of $\sigma_0$, that contains the edge configuration on the left hand side of Figure \ref{e4s}, by $\n\pm2k_\n\khat$, for some integers $k_\n\in\mathbb{Z}$.  The elementary four-squares $\sigma^4_{ik},\sigma^4_{jk},\sigma^4_{ki},\sigma^4_{kj}$, depicted in Figure \ref{4VBW}, are then added where required, to form the simply connected surface $\sigma$.  Directed rapidity lines labelled $p,p'$, $q,q'$, and $r,r'$, are assigned to the different elementary squares, according to Figures \ref{4VBW}.  The type of deformations that arise from this process are shown in Appendix \ref{app:IRF}.  

More specifically, the elementary four-squares $\sigma^4_{ij},\sigma^4_{ik},\sigma^4_{ki},\sigma^4_{jk},\sigma^4_{kj}$, are used as building blocks of a surface $\sigma$, as follows.  Let $F^4(\sigma_0)=\{\sigma^4_{ij}(n_{i_1},n_{j_1},0)),\sigma^4_{ij}(n_{i_2},n_{j_2},0)),\ldots,\}$, be the set of elementary four-squares of some flat surface $\sigma_0$, as was defined in the previous subsection.  Then for an admissible set of integers $n_{k_1},n_{k_2},\ldots$, the surface $\sigma$ is defined to be the unique surface that consists of the following set of elementary four-squares
\begin{align}
F^4(\sigma)=\{\sigma^4_{ij}(n_{i_1},n_{j_1},n_{k_1})),\sigma^4_{ij}(n_{i_2},n_{j_2},n_{k_2})),\ldots,\}\cup F^4_k,
\end{align}
where $F^4_k$ contains elementary four-squares of the type $\sigma^4_{ik},\sigma^4_{jk},\sigma^4_{ki},\sigma^4_{kj}$, chosen such that $\sigma$ has the same boundary as $\sigma_0$, $\sigma$ is simply connected and oriented, and the following additional {\it corner condition} is satisfied:
\\[0.2cm]
\textbf{Corner condition.} \textit{{For any $\n$, $\sigma$ cannot contain pairs of elementary four-squares $\sigma^4_{ki}(\n),\sigma^4_{kj}(\n+2\ihat)$, that are associated to pairs of positively oriented rapidity lines $r,r'$. Similarly, for any $\n$, $\sigma$ cannot contain pairs of elementary four-squares $\sigma^4_{ki}(\n+2\jhat),\sigma^4_{kj}(\n)$, that are associated to pairs of negatively oriented rapidity lines $r,r'$.}}
\\[0.2cm]
The integers $n_{k_1},n_{k_2},\ldots$, should be chosen so that the above properties can be satisfied.    The above corner condition is required, due to there being no Yang-Baxter equation satisfied where pairs of parallel rapidity lines form closed directed loops, and the corner condition ensures that problematic corners with this associated rapidity configuration are avoided.

\begin{figure}[htb]
\centering
\begin{tikzpicture}[scale=1.76]

\filldraw[fill=black,draw=black] (0.25,0.0) circle (1.1pt);
\filldraw[fill=black,draw=black] (1.25,0.0) circle (1.1pt);
\draw[-,thick] (0.0,-0.5)--(0.125,0.25)--(0.25,0.0)--(0.75,0.5)--(1.25,0.0)--(1.75,0.5)--(2.25,0.0);
\draw[-,thick,dotted] (2.5,0.5-0.25)--(0.25,0.5-0.25)--(0.0,0.0-0.25);
\filldraw[fill=white,draw=white] (0.1,0.0)--(1.4,0.0)--(1.4,-0.7)--(0.0,-0.7)--(0.0,0.0);
\filldraw[fill=white,draw=white] (2.0,0.0)--(2.25,0.5)--(2.75,0.5)--(2.5,0.0)--(2.0,0.0);



%

\draw[-,thin,dashed] (0.125/2-0.2,0.125)--(0.125/2,0.125)--(0.125/2,0.0);\draw[->,thin,dashed] (2.0+ 0.125/2,0.125)--(6.0+0.125/2+0.2,0.125);
\draw[-,thick,dotted] (0.25-0.125/2-0.2,0.5-0.125)--(0.25-0.125/2,0.5-0.125)--(0.25-0.125/2,0.0);\draw[->,thick,dotted] (2.25- 0.125/2,0.5-0.125)--(6.25-0.125/2+0.2,0.5-0.125);
\draw[->,thin,dashed] (0.25+0.125/2-0.2,0.625)--(0.25+6.0+0.125/2+0.2,0.625);
\draw[->,thin,dashed] (0.50+0.125/2-0.2,1.125)--(0.50+6.0+0.125/2+0.2,1.125);
\draw[->,thin,dashed] (0.75+0.125/2-0.2,1.625)--(0.75+6.0+0.125/2+0.2,1.625);
\draw[->,thick,dotted] (0.50-0.125/2-0.2,1.0-0.125)--(0.50+6.0-0.125/2+0.2,1.0-0.125);
\draw[->,thick,dotted] (0.75-0.125/2-0.2,1.5-0.125)--(0.75+6.0-0.125/2+0.2,1.5-0.125);
\draw[->,thick,dotted] (1.0 -0.125/2-0.2,2.0-0.125)--(1.0 +6.0-0.125/2+0.2,2.0-0.125);

\draw[-,thick,dotted] (0.25-0.1,0.0-0.2)--(0.25,0.0);\draw[->,thick,dotted] (0.5,0.0)--(0.5,0.5)--(1.25+0.1,2.0+0.2);
\draw[-,thick,dotted] (1.25-0.1,0.0-0.2)--(1.25,0.0);\draw[->,thick,dotted] (1.5,0.0)--(1.5,0.5)--(2.25+0.1,2.0+0.2);
\draw[-,thin,dashed] (0.75-0.1,0.0-0.2)--(0.75,0.0);\draw[->,thin,dashed] (1.0,0.0)--(1.0,0.5)--(1.75+0.1,2.0+0.2);
\draw[-,thin,dashed] (1.75-0.1,0.0-0.2)--(1.75,0.0);\draw[->,thin,dashed] (2.0,0.0)--(2.0,0.5)--(2.75+0.1,2.0+0.2);
\draw[->,thick,dotted] (2.25-0.1,0.0-0.2)--(3.25+0.1,2.0+0.2);
\draw[->,thick,dotted] (3.25-0.1,0.0-0.2)--(4.25+0.1,2.0+0.2);
\draw[->,thick,dotted] (4.25-0.1,0.0-0.2)--(5.25+0.1,2.0+0.2);
\draw[->,thick,dotted] (5.25-0.1,0.0-0.2)--(6.25+0.1,2.0+0.2);
\draw[->,thin,dashed] (2.75-0.1,0.0-0.2)--(3.75+0.1,2.0+0.2);
\draw[->,thin,dashed] (3.75-0.1,0.0-0.2)--(4.75+0.1,2.0+0.2);
\draw[->,thin,dashed] (4.75-0.1,0.0-0.2)--(5.75+0.1,2.0+0.2);
\draw[->,thin,dashed] (5.75-0.1,0.0-0.2)--(6.75+0.1,2.0+0.2);

\draw[-,thin,gray] (0,0)--(6,0)--(6.25,0.5)--(3.25,0.5);\draw[-,thin,gray] (4,2)--(1,2)--(0,0);
\draw[-,thin,gray] (2,0)--(3,2);

\foreach \y in {0.5,1,1.5}
\draw[-,thin,gray] (0.5*\y,\y)--(3+0.5*\y,\y);

\draw[-,thin,gray] (3,0)--(3.25,0.5);\draw[-,thin,gray] (4,0)--(4.25,0.5);\draw[-,thin,gray] (5,0)--(5.25,0.5);

\foreach \x in {0,1,...,5}{
\filldraw[draw=black,fill=black] (\x+1.5,2) circle (1.1pt);
\foreach \y in {0,1,...,3}{
\filldraw[draw=black,fill=black] (\x+\y*0.5/2+0.5,\y/2) circle (1.1pt);
\filldraw[draw=black,fill=black] (\x+\y*0.5/2+0.25/2,\y/2+0.25) circle (1.1pt);
}
}
\foreach \y in {0,1,...,3}{
\filldraw[draw=black,fill=black] (6+\y*0.5/2+0.25/2,\y/2+0.25) circle (1.1pt);
}


\begin{scope}[xshift=0.5cm]
\draw[-,thick] (-0.5+0.25*7/2,1.75)--(1,2);
\draw[-,thick] (-0.5+0.25*5/2,1.25)--(2,2);
\draw[-,thick] (-0.5+0.25*3/2,0.75)--(3,2);
\draw[-,thick] (0.25,0.5)--(4,2);
\draw[-,thick] (1.25,0.5)--(5,2);
\draw[-,thick] (1.5+0.25/2,0.25)--(6,2);
\draw[-,thick] (2,0)--(5.5+0.25*7/2,0.25*7);
\draw[-,thick] (3,0)--(5.5+0.25*5/2,0.25*5);
\draw[-,thick] (4,0)--(5.5+0.25*3/2,0.25*3);
\draw[-,thick] (5,0)--(5.5+0.25*1/2,0.25*1);

\draw[-,thick] (-0.5+0.25*3/2,0.75)--(0.25,0.5);
\draw[-,thick] (-0.5+0.25*5/2,1.25)--(1.25,0.5);
\draw[-,thick] (-0.5+0.25*7/2,1.75)--(3,0);
\draw[-,thick] (1,2)--(4,0);
\draw[-,thick] (2,2)--(5,0);
\draw[-,thick] (3,2)--(5.5+0.25*1/2,0.25*1);
\draw[-,thick] (4,2)--(5.5+0.25*3/2,0.25*3);
\draw[-,thick] (5,2)--(5.5+0.25*5/2,0.25*5);
\draw[-,thick] (6,2)--(5.5+0.25*7/2,0.25*7);

\draw[-,thick] (1.5+0.25/2,0.25)--(2,0);
\end{scope}

\filldraw[-,thin,fill=white,draw=gray] (3.25,0.5)--(6.25,0.5)--(7,2)--(7,3)--(4,3)--(3.25,1.5)--(3.25,0.5);
\filldraw[fill=black,draw=black] (4.75,2.0) circle (1.1pt);
\draw[-,thick] (4.25+0.25,1.5)--(4.5+0.25/2,2.25)--(4.75,2.0)--(5.0+0.25,2.5)--(5.75,2.0);
\draw[-,thick,dotted] (5.75,2.25)--(4.75,2.25)--(4.5,1.75);
\filldraw[,thin,fill=white,draw=white] (3.75+0.25,2.0)--(5.25+0.25,2.0)--(5.75+0.25/2,2.75)--(5.75+0.25/2+0.5,2.75)--(5.75,1.55)--(3.75+0.25,1.5)--(3.75+0.25,2.0);

\foreach \x in {4.25,5.25,6.25}
\draw[-,thin,gray] (\x,1.5)--(\x+0.75,3);
\foreach \y in {0,0.5,1}
\draw[-,thin,gray] (3.25+\y*0.5,\y+1.5)--(6.25+\y*0.5,\y+1.5)--(6.25+\y*0.5,\y+0.5);
\draw[-,thin,gray] (4.25,0.5)--(4.25,1.5);\draw[-,thin,gray] (5.25,0.5)--(5.25,1.5);
\draw[-,thin,gray] (4.75,2.5)--(4.75,2);
\foreach \x in {3.25,4.25,5.25,6.25} {
\filldraw[fill=black,draw=black] (\x,1.0) circle (1.1pt);
\foreach \y in {0,0.5,1} {
\filldraw[fill=black,draw=black] (\x+\y*0.5+0.25/2,\y+1.75) circle (1.1pt);
}
}
\foreach \x in {3.25,4.25,5.25} {
\filldraw[fill=black,draw=black] (\x+0.5,0.5) circle (1.1pt);
\foreach \y in {0,0.5,1,1.5} {
\filldraw[fill=black,draw=black] (\x+0.5+\y*0.5,1.5+\y) circle (1.1pt);
}
}
\foreach \y in {0.75,1.25,1.75} {
\filldraw[fill=black,draw=black] (6.25+0.25/2+\y/2-0.75/2,\y) circle (1.1pt);
\filldraw[fill=black,draw=black] (6.25+0.5/2+\y/2-0.75/2,\y+0.5+0.25) circle (1.1pt);
}

\draw[-,thick] (6.25,1.0)--(6.25+0.25/2,1.75)--(6.25+0.25,1.5)--(6.5+0.25/2,2.25)--(6.75,2.0)--(6.75+0.25/2,2.75)--(7.0,2.5);
\draw[-,thick] (6.25,1.0)--(6.25+0.25/2,0.75)--(6.25+0.25,1.5)--(6.5+0.25/2,1.25)--(6.75,2.0)--(6.75+0.25/2,1.75)--(7.0,2.5);

\draw[-,thick] (6.25+0.25/2,1.75)--(5.5+0.25,1.5)--(5.25+0.25/2,1.75)--(4.5+0.25,1.5)--(4.25+0.25/2,1.75)--(3.5+0.25,1.5)--(3.25+0.25/2,1.75);
\draw[-,thick] (6.25+0.25/2,1.75)--(5.75+0.25,2.0)--(5.25+0.25/2,1.75)--(4.75+0.25,2.0)--(4.25+0.25/2,1.75)--(3.75+0.25,2.0)--(3.25+0.25/2,1.75);

\draw[-,thick] (6.5+0.25/2,2.25)--(5.75+0.25,2.0)--(5.5+0.25/2,2.25);
\draw[-,thick] (4.5+0.25/2,2.25)--(3.75+0.25,2.0)--(3.5+0.25/2,2.25);
\draw[-,thick] (6.5+0.25/2,2.25)--(6.0+0.25,2.5)--(5.5+0.25/2,2.25);
\draw[-,thick] (4.5+0.25/2,2.25)--(4.0+0.25,2.5)--(3.5+0.25/2,2.25);

\draw[-,thick] (6.75+0.25/2,2.75)--(6.0+0.25,2.5)--(5.75+0.25/2,2.75)--(5.0+0.25,2.5)--(4.75+0.25/2,2.75)--(4.0+0.25,2.5)--(3.75+0.25/2,2.75);
\draw[-,thick] (6.75+0.25/2,2.75)--(6.25+0.25,3.0)--(5.75+0.25/2,2.75)--(5.25+0.25,3.0)--(4.75+0.25/2,2.75)--(4.25+0.25,3.0)--(3.75+0.25/2,2.75);

\draw[-,thick] (3.5+0.25,0.5)--(3.0+0.25,1.0);
\draw[-,thick] (4.5+0.25,0.5)--(3.5+0.25,1.5);
\draw[-,thick] (5.5+0.25,0.5)--(4.5+0.25,1.5);
\draw[-,thick] (6.0+0.25,1.0)--(5.5+0.25,1.5);

\draw[-,thick] (3.5+0.25,1.5)--(3.0+0.25,1.0);
\draw[-,thick] (3.5+0.25,0.5)--(4.5+0.25,1.5);
\draw[-,thick] (4.5+0.25,0.5)--(5.5+0.25,1.5);
\draw[-,thick] (5.5+0.25,0.5)--(6.0+0.25,1.0);

\draw[->,thin,dashed] (3.25,0.75)--(6.25,0.75)--(7.0,2.25);
\draw[->,thick,dotted] (3.25,1.25)--(6.25,1.25)--(7.0,2.75);

\draw[->,thick,dotted] (3.5,0.5)--(3.5,1.5)--(4.25,3.0);
\draw[-,thick,dotted] (4.5,0.5)--(4.5,1.5)--(4.75,2.0);\draw[->,thick,dotted] (5.0,2.0)--(5.0,2.5)--(5.25,3.0);
\draw[->,thick,dotted] (5.5,0.5)--(5.5,1.5)--(6.25,3.0);
\draw[->,thin,dashed] (4.0,0.5)--(4.0,1.5)--(4.75,3.0);
\draw[-,thin,dashed] (5.0,0.5)--(5.0,1.5)--(5.25,2.0);\draw[->,thin,dashed] (5.5,2.0)--(5.5,2.5)--(5.75,3.0);
\draw[->,thin,dashed] (6.0,0.5)--(6.0,1.5)--(6.75,3.0);

\draw[-,thick,dotted] (3.5 -0.125/2,2.0-0.125)--(6.5 -0.125/2,2.0-0.125)--(6.5 -0.125/2,1.0-0.125);
\draw[-,thick,dotted] (3.75-0.125/2,2.5-0.125)--(4.75-0.125/2,2.5-0.125)--(4.75-0.125/2,2,0);\draw[-,thick,dotted] (5.75-0.125/2,2.5-0.125)--(6.75-0.125/2,2.5-0.125)--(6.75-0.125/2,1.5-0.125);
\draw[-,thick,dotted] (4.0 -0.125/2,3.0-0.125)--(7.0 -0.125/2,3.0-0.125)--(7.0 -0.125/2,2.0-0.125);
\draw[-,thin,dashed] (3.25+0.125/2,1.5+0.125)--(6.25+0.125/2,1.5+0.125)--(6.25+0.125/2,0.5+0.125);
\draw[-,thin,dashed] (3.5 +0.125/2,2.0+0.125)--(4.5 +0.125/2,2.0+0.125)--(4.5 +0.125/2,2.0);\draw[-,thin,dashed] (5.5 +0.125/2,2.0+0.125)--(6.5 +0.125/2,2.0+0.125)--(6.5 +0.125/2,1.0+0.125);
\draw[-,thin,dashed] (3.75+0.125/2,2.5+0.125)--(6.75+0.125/2,2.5+0.125)--(6.75+0.125/2,1.5+0.125);

\filldraw[-,thin,fill=white,draw=gray] (1.25,0.5)--(2.25,0.5)--(2.5,1)--(2.5,3)--(1.5,3)--(1.25,2.5)--(1.25,0.5);
\draw[-,thin,gray] (2.25,0.5)--(2.25,2.5);\draw[-,thin,gray](1.25,2.5)--(2.25,2.5)--(2.5,3);\draw[-,thin,gray] (1.25,1.5)--(2.25,1.5)--(2.5,2);
\filldraw[fill=black,draw=black] (1.75,0.5) circle (1.1pt);
\filldraw[fill=black,draw=black] (1.75,1.5) circle (1.1pt);
\filldraw[fill=black,draw=black] (1.75,2.5) circle (1.1pt);
\filldraw[fill=black,draw=black] (2.25+0.25/2,0.75) circle (1.1pt);
\filldraw[fill=black,draw=black] (2.25+0.25/2,1.75) circle (1.1pt);
\filldraw[fill=black,draw=black] (2.25+0.25/2,2.75) circle (1.1pt);
\filldraw[fill=black,draw=black] (1.25,1.0) circle (1.1pt);
\filldraw[fill=black,draw=black] (1.25,2.0) circle (1.1pt);
\filldraw[fill=black,draw=black] (2.25,1.0) circle (1.1pt);
\filldraw[fill=black,draw=black] (2.25,2.0) circle (1.1pt);
\filldraw[fill=black,draw=black] (2.5,1.5) circle (1.1pt);
\filldraw[fill=black,draw=black] (2.5,2.5) circle (1.1pt);
\filldraw[fill=black,draw=black] (1.25+0.25/2,2.75) circle (1.1pt);
\filldraw[fill=black,draw=black] (2.0,3.0) circle (1.1pt);
\draw[-,thick] (1.75,0.5)--(1.25,1.0)--(1.75,1.5)--(1.25,2.0)--(1.75,2.5)--(1.25+0.25/2,2.75)--(2.0,3.0);
\draw[-,thick] (1.75,0.5)--(2.25,1.0)--(1.75,1.5)--(2.25,2.0)--(1.75,2.5)--(2.25+0.25/2,2.75)--(2.0,3.0);
\draw[-,thick] (2.25+0.25/2,0.75)--(2.25,1.0)--(2.25+0.25/2,1.75)--(2.25,2.0)--(2.25+0.25/2,2.75);
\draw[-,thick] (2.25+0.25/2,0.75)--(2.5,1.5)--(2.25+0.25/2,1.75)--(2.5,2.5)--(2.25+0.25/2,2.75);

\draw[-,thick,dotted] (1.25+3*0.125/2,2.5+3*0.5/4)--(2.25+3*0.125/2,2.5+3*0.5/4)--(2.25+3*0.125/2,0.5+3*0.5/4);
\draw[-,thin,dashed]  (1.25+1*0.125/2,2.5+1*0.5/4)--(2.25+1*0.125/2,2.5+1*0.5/4)--(2.25+1*0.125/2,0.5+1*0.5/4);
\draw[->,thick,dotted] (1.5,0.5)--(1.5,2.5)--(1.75,3.0);
\draw[->,thin,dashed] (2.0,0.5)--(2.0,2.5)--(2.25,3.0);

\draw[->,thin,dashed] (1.25,0.75)--(2.25,0.75)--(2.5,1.25);
\draw[->,thin,dashed] (1.25,1.75)--(2.25,1.75)--(2.5,2.25);
\draw[->,thick,dotted] (1.25,1.25)--(2.25,1.25)--(2.5,1.75);
\draw[->,thick,dotted] (1.25,2.25)--(2.25,2.25)--(2.5,2.75);



\filldraw[fill=white,draw=white] (1.125,0.25) circle (2.0pt);

\draw[-,thick,dotted] (1.1,0.25)--(1.3,0.25);

\draw[-,thin,gray] (0,0)--(0,-1)--(1,-1)--(1,0);\draw[-,thin,gray] (1,-1)--(2,-1)--(2,0);\draw[-,thin,gray] (2,-1)--(2.25,-0.5)--(2.25,0);\draw[-,thin,gray] (0.25,0.5)--(0.25,0);\draw[-,thin,gray] (1.25,0.5)--(1.25,0);

\fill[black] (7.86,0) circle (0.01pt)
node[below=1.5pt]{\color{black} $+\ihat$};
\fill[black] (7.66,0.33) circle (0.01pt)
node[right=1.5pt]{\color{black} $+\jhat$};
\fill[black] (7.5,0.75*0.5) circle (0.01pt)
node[left=1.5pt]{\color{black} $+\khat$};
\draw[->, thick] (7.5,0)--(8.25,0);\draw[->, thick] (7.5,0)--(7.83,0.67);\draw[->, thick] (7.5,0)--(7.5,0.75);

\end{tikzpicture}
\caption{An example of a vertex model on a surface $\sigma$, constructed from cubic-shaped deformations of a vertex model on a flat surface $\sigma_0$.  Boltzmann weights on elementary four-squares of $\sigma$, are shown in Figures \ref{e4s} and \eqref{4VBW}.  White vertices of $\sigma$ are not shown.}
\label{Glattice2}
\end{figure}
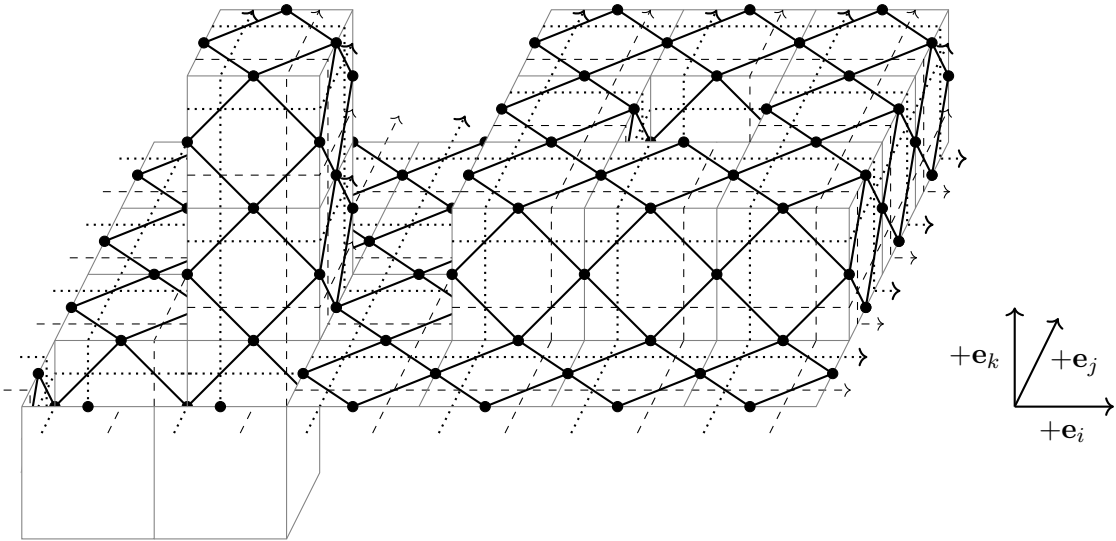

An example of such a surface $\sigma$, that arises from the deformation of a vertex model on $\sigma_0$ that satisfies the above conditions, is shown in Figure \ref{Glattice2}.   The to define the vertex model on $\sigma$, spin variables $\xv_i$, are assigned to vertices $i\in V(\Gl)$, where $\Gl$ is the graph that is formed by edges connecting black next nearest neighbour vertices of $V(\sigma)$.  Vertex Boltzmann weights \eqref{vertexBW} are assigned to each elementary square $\sigma^4_{ik}$, $\sigma^4_{jk}$, $\sigma^4_{ki}$, $\sigma^4_{kj}$, according to Figure \ref{4VBW}, and to $\sigma^4_{ij}$, according to the left hand side of Figure \ref{e4s}.  The partition function for the vertex model is given by the expression
\begin{align}
\label{zvertexsigma}
Z=\sum_{\mathbf{x}}\prod_{(i,j,i',j')\in F^{(4)}(\sigma)}\left<\xv_i,\,\xv_j\,|\,\mathbb{R}_{\mathbf{p}\mathbf{q}}\,|\,\xv'_i,\,\xv'_j\right>.
\end{align}
 This is similar to the expression for the partition function on $\sigma_0$ given in \eqref{vertexZ}, with the product over $F^{(1)}(L)$ being replaced with a product over elementary four-squares in $F^{(4)}(\sigma)$, and the sum being taken over all interior spins $\xv_1,\xv_2,\ldots,\xv_n$, assigned to vertices $i_1,i_2,\ldots,i_n\in V_{int}(\Gl)$.  This defines the vertex model on $\sigma$.

\subsubsection{Deformed two-dimensional surface for IRF model}

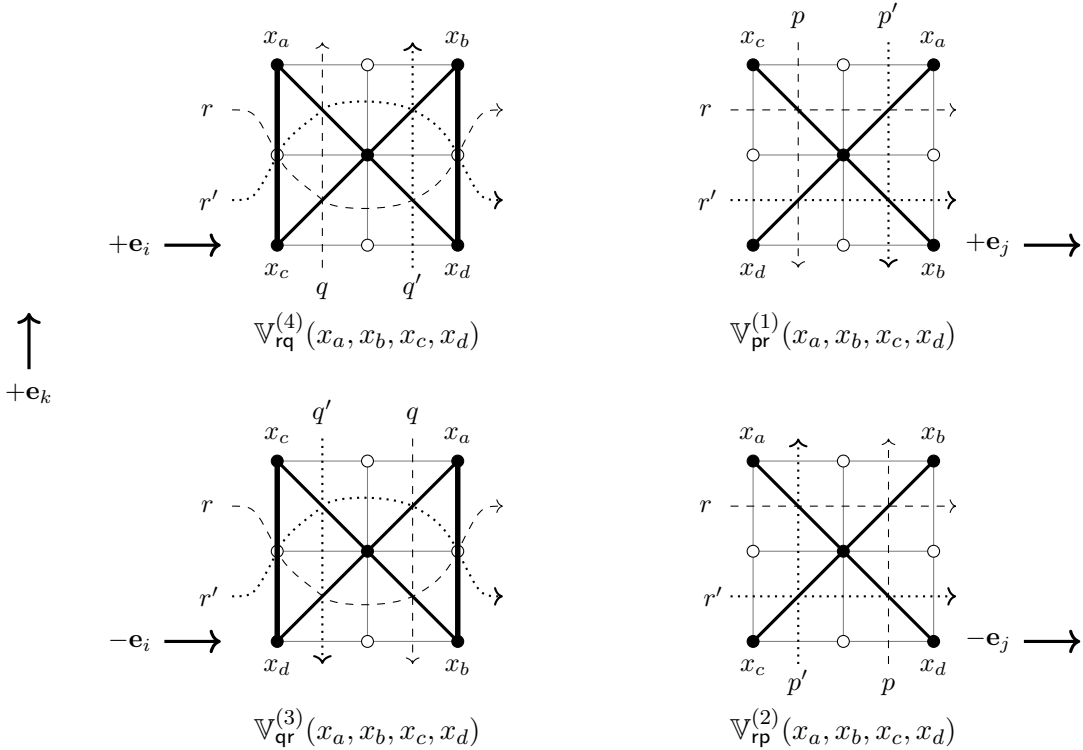
\begin{figure}[htb!]
\centering
\begin{tikzpicture}[scale=1.5]
\draw[->,very thick] (-3,-1.9)--(-3,-1.4);
\fill[black!] (-3,-1.9) circle (0.01pt)
node[below=1.5pt]{\color{black}\small $+\khat$};
\draw[->,very thick] (-1.8,-0.8)--(-1.3,-0.8);
\fill[black!] (-1.8,-0.8) circle (0.01pt)
node[left=1pt]{\color{black}\small $+\ihat$};
\draw[-,gray] (-0.8,-0.8)--(0.8,-0.8)--(0.8,0.8)--(-0.8,0.8)--(-0.8,-0.8);
\draw[-,gray] (0,-0.8)--(0,0.8);\draw[-,gray] (-0.8,0)--(0.8,0);
\draw[-,very thick] (-0.8,-0.8)--(0.8,0.8);\draw[-,very thick] (-0.8,0.8)--(0.8,-0.8);
\draw[->,dashed,black] (-0.4,-1)--(-0.4,1);\draw[->,thick,dotted,black] (0.4,-1)--(0.4,1);
\draw[->,dashed] (-1.2,0.4) .. controls (-1,0.4) .. (-0.8,0) .. controls (-0.7,-0.2) and (-0.5,-0.3) .. (-0.4,-0.4) .. controls (-0.3,-0.5) and (0.2,-0.5) .. (0.4,-0.4) .. controls (0.6,-0.3) and (0.7,-0.2) .. (0.8,0) .. controls (1,0.4) .. (1.2,0.4);
\draw[->,thick,dotted] (-1.2,-0.4) .. controls (-1,-0.4) .. (-0.8,0) .. controls (-0.7,0.2) and (-0.5,0.3) .. (-0.4,0.4) .. controls (-0.3,0.5) and (0.2,0.5) .. (0.4,0.4) .. controls (0.6,0.3) and (0.7,0.2) .. (0.8,0) .. controls (1,-0.4) .. (1.2,-0.4);
\fill[black!] (-0.4,-1.4) circle (0.01pt)
node[above=1.5pt]{\color{black}\small $q$};
\fill[black!] (-1.15,-0.4) circle (0.01pt)
node[left=3.1pt]{\color{black}\small $r'$};
\fill[black!] (0.4,-1.4) circle (0.01pt)
node[above=1.5pt]{\color{black}\small $q'$};
\fill[black!] (-1.2,0.4) circle (0.01pt)
node[left=3.1pt]{\color{black}\small $r$};
\filldraw[fill=black,draw=black] (-0.8,-0.8) circle (1.5pt)
node[below=3pt]{\color{black}\small $\xv_c$};
\filldraw[fill=black,draw=black] (0.8,-0.8) circle (1.5pt)
node[below=3pt]{\color{black}\small $\xv_d$};
\filldraw[fill=black,draw=black] (0.8,0.8) circle (1.5pt)
node[above=3pt]{\color{black}\small $\xv_b$};
\filldraw[fill=black,draw=black] (-0.8,0.8) circle (1.5pt)
node[above=3pt]{\color{black}\small $\xv_a$};
\filldraw[fill=black,draw=black] (0,0) circle (1.5pt);
\filldraw[fill=white,draw=black] (0,-0.8) circle (1.5pt);
\filldraw[fill=white,draw=black] (0,0.8) circle (1.5pt);
\filldraw[fill=white,draw=black] (0.8,0) circle (1.5pt);
\filldraw[fill=white,draw=black] (-0.8,0) circle (1.5pt);
\draw[-,very thick] (-0.81,-0.8)--(-0.81,0.8);\draw[-,very thick] (0.81,-0.8)--(0.81,0.8);
\draw[-,very thick] (-0.79,-0.8)--(-0.79,0.8);\draw[-,very thick] (0.79,-0.8)--(0.79,0.8);

\fill (0,-1.3) circle(0.01pt)
node[below=0.05pt]{\color{black} $\V_{\br\bq}^{(4)}(\xv_a,\xv_b,\xv_c,\xv_d)$};

\begin{scope}[xshift=120pt]
\draw[->,very thick] (1.6,-0.8)--(2.1,-0.8);
\fill[black!] (1.6,-0.8) circle (0.01pt)
node[left=1pt]{\color{black}\small $+\jhat$};
\draw[-,gray] (-0.8,-0.8)--(0.8,-0.8)--(0.8,0.8)--(-0.8,0.8)--(-0.8,-0.8);
\draw[-,gray] (0,-0.8)--(0,0.8);\draw[-,gray] (-0.8,0)--(0.8,0);
\draw[-,very thick] (-0.8,-0.8)--(0.8,0.8);\draw[-,very thick] (-0.8,0.8)--(0.8,-0.8);
\draw[->,dashed,black] (-0.4,1)--(-0.4,-1);\draw[->,thick,dotted,black] (0.4,1)--(0.4,-1);
\draw[->,thick,dotted,black!30!black] (-1,-0.4)--(1,-0.4);
\draw[->,dashed,black!30!black] (-1,0.4)--(1,0.4);
\fill[black!] (-0.4,1) circle (0.01pt)
node[above=1.5pt]{\color{black}\small $p$};
\fill[black!] (-0.95,-0.4) circle (0.01pt)
node[left=3.1pt]{\color{black}\small $r'$};
\fill[black!] (0.4,1) circle (0.01pt)
node[above=1.5pt]{\color{black}\small $p'$};
\fill[black!] (-1,0.4) circle (0.01pt)
node[left=3.1pt]{\color{black}\small $r$};
\filldraw[fill=black,draw=black] (-0.8,-0.8) circle (1.5pt)
node[below=3pt]{\color{black}\small $\xv_d$};
\filldraw[fill=black,draw=black] (0.8,-0.8) circle (1.5pt)
node[below=3pt]{\color{black}\small $\xv_b$};
\filldraw[fill=black,draw=black] (0.8,0.8) circle (1.5pt)
node[above=3pt]{\color{black}\small $\xv_a$};
\filldraw[fill=black,draw=black] (-0.8,0.8) circle (1.5pt)
node[above=3pt]{\color{black}\small $\xv_c$};
\filldraw[fill=black,draw=black] (0,0) circle (1.5pt);
\filldraw[fill=white,draw=black] (0,-0.8) circle (1.5pt);
\filldraw[fill=white,draw=black] (0,0.8) circle (1.5pt);
\filldraw[fill=white,draw=black] (0.8,0) circle (1.5pt);
\filldraw[fill=white,draw=black] (-0.8,0) circle (1.5pt);

\fill (0,-1.3) circle(0.01pt)
node[below=0.05pt]{\color{black} $\bW^{(1)}_{\bp\br}(\xv_a,\xv_b,\xv_c,\xv_d)$};

\end{scope}

\begin{scope}[yshift=-100pt]
\draw[->,very thick] (-1.8,-0.8)--(-1.3,-0.8);
\fill[black!] (-1.8,-0.8) circle (0.01pt)
node[left=1pt]{\color{black}\small $-\ihat$};
\draw[-,gray] (-0.8,-0.8)--(0.8,-0.8)--(0.8,0.8)--(-0.8,0.8)--(-0.8,-0.8);
\draw[-,gray] (0,-0.8)--(0,0.8);\draw[-,gray] (-0.8,0)--(0.8,0);
\draw[-,very thick] (-0.8,-0.8)--(0.8,0.8);\draw[-,very thick] (-0.8,0.8)--(0.8,-0.8);
\draw[->,thick,dotted,black] (-0.4,1)--(-0.4,-1);\draw[->,dashed,black] (0.4,1)--(0.4,-1);
\draw[->,dashed] (-1.2,0.4) .. controls (-1,0.4) .. (-0.8,0) .. controls (-0.7,-0.2) and (-0.5,-0.3) .. (-0.4,-0.4) .. controls (-0.3,-0.5) and (0.2,-0.5) .. (0.4,-0.4) .. controls (0.6,-0.3) and (0.7,-0.2) .. (0.8,0) .. controls (1,0.4) .. (1.2,0.4);
\draw[->,thick,dotted] (-1.2,-0.4) .. controls (-1,-0.4) .. (-0.8,0) .. controls (-0.7,0.2) and (-0.5,0.3) .. (-0.4,0.4) .. controls (-0.3,0.5) and (0.2,0.5) .. (0.4,0.4) .. controls (0.6,0.3) and (0.7,0.2) .. (0.8,0) .. controls (1,-0.4) .. (1.2,-0.4);
\fill[black!] (-0.4,1) circle (0.01pt)
node[above=1.5pt]{\color{black}\small $q'$};
\fill[black!] (-1.15,-0.4) circle (0.01pt)
node[left=3.1pt]{\color{black}\small $r'$};
\fill[black!] (0.4,1) circle (0.01pt)
node[above=1.5pt]{\color{black}\small $q$};
\fill[black!] (-1.2,0.4) circle (0.01pt)
node[left=3.1pt]{\color{black}\small $r$};
\filldraw[fill=black,draw=black] (-0.8,-0.8) circle (1.5pt)
node[below=3pt]{\color{black}\small $\xv_d$};
\filldraw[fill=black,draw=black] (0.8,-0.8) circle (1.5pt)
node[below=3pt]{\color{black}\small $\xv_b$};
\filldraw[fill=black,draw=black] (0.8,0.8) circle (1.5pt)
node[above=3pt]{\color{black}\small $\xv_a$};
\filldraw[fill=black,draw=black] (-0.8,0.8) circle (1.5pt)
node[above=3pt]{\color{black}\small $\xv_c$};
\filldraw[fill=black,draw=black] (0,0) circle (1.5pt);
\filldraw[fill=white,draw=black] (0,-0.8) circle (1.5pt);
\filldraw[fill=white,draw=black] (0,0.8) circle (1.5pt);
\filldraw[fill=white,draw=black] (0.8,0) circle (1.5pt);
\filldraw[fill=white,draw=black] (-0.8,0) circle (1.5pt);
\draw[-,very thick] (-0.81,-0.8)--(-0.81,0.8);\draw[-,very thick] (0.81,-0.8)--(0.81,0.8);
\draw[-,very thick] (-0.79,-0.8)--(-0.79,0.8);\draw[-,very thick] (0.79,-0.8)--(0.79,0.8);

\fill (0,-1.3) circle(0.01pt)
node[below=0.05pt]{\color{black} $\V_{\bq\br}^{(3)}(\xv_a,\xv_b,\xv_c,\xv_d)$};

\end{scope}

\begin{scope}[xshift=120pt,yshift=-100]
\draw[->,very thick] (1.6,-0.8)--(2.1,-0.8);
\fill[black!] (1.6,-0.8) circle (0.01pt)
node[left=1pt]{\color{black}\small $-\jhat$};
\fill[black!] (2,0) circle (0.01pt)
node[below=1.5pt]{\color{white}\small $+\khat$};
\draw[-,gray] (-0.8,-0.8)--(0.8,-0.8)--(0.8,0.8)--(-0.8,0.8)--(-0.8,-0.8);
\draw[-,gray] (0,-0.8)--(0,0.8);\draw[-,gray] (-0.8,0)--(0.8,0);
\draw[-,very thick] (-0.8,-0.8)--(0.8,0.8);\draw[-,very thick] (-0.8,0.8)--(0.8,-0.8);
\draw[->,thick,dotted,black] (-0.4,-1)--(-0.4,1);\draw[->,dashed,black] (0.4,-1)--(0.4,1);
\draw[->,thick,dotted,black!30!black] (-1,-0.4)--(1,-0.4);
\draw[->,dashed,black!30!black] (-1,0.4)--(1,0.4);
\fill[black!] (-0.4,-1.4) circle (0.01pt)
node[above=1.5pt]{\color{black}\small $p'$};
\fill[black!] (-0.9,-0.4) circle (0.01pt)
node[left=3.1pt]{\color{black}\small $r'$};
\fill[black!] (0.4,-1.4) circle (0.01pt)
node[above=1.5pt]{\color{black}\small $p$};
\fill[black!] (-1,0.4) circle (0.01pt)
node[left=3.1pt]{\color{black}\small $r$};
\filldraw[fill=black,draw=black] (-0.8,-0.8) circle (1.5pt)
node[below=3pt]{\color{black}\small $\xv_c$};
\filldraw[fill=black,draw=black] (0.8,-0.8) circle (1.5pt)
node[below=3pt]{\color{black}\small $\xv_d$};
\filldraw[fill=black,draw=black] (0.8,0.8) circle (1.5pt)
node[above=3pt]{\color{black}\small $\xv_b$};
\filldraw[fill=black,draw=black] (-0.8,0.8) circle (1.5pt)
node[above=3pt]{\color{black}\small $\xv_a$};
\filldraw[fill=black,draw=black] (0,0) circle (1.5pt);
\filldraw[fill=white,draw=black] (0,-0.8) circle (1.5pt);
\filldraw[fill=white,draw=black] (0,0.8) circle (1.5pt);
\filldraw[fill=white,draw=black] (0.8,0) circle (1.5pt);
\filldraw[fill=white,draw=black] (-0.8,0) circle (1.5pt);

\fill (0,-1.3) circle(0.01pt)
node[below=0.05pt]{\color{black} $\bW^{(2)}_{\br\bp}(\xv_a,\xv_b,\xv_c,\xv_d)$};

\end{scope}

\end{tikzpicture}
\caption{In clockwise order starting from the top right, the Boltzmann weights \eqref{V1}, \eqref{V2}, \eqref{V3}, \eqref{V4}, that are assigned to elementary four-squares $\sigma^4_{jk}(\n)$, $\sigma^4_{kj}(\n)$, $\sigma^4_{ki}(\n)$, $\sigma^4_{ik}(\n)$, respectively, in the IRF formulation.  Note that on the left hand side, certain edges of $\Gl$ are pictured to pass through white vertices of $\sigma$, however the white vertices play no role in the definition of the IRF model on $\Gl$.}
\label{4IRFBW}
\end{figure}

The IRF model on $\sigma$ may be defined in a similar fashion to the vertex model of the previous subsection.  To apply the construction of the previous subsection to the IRF model, the elementary four-squares for the vertex model in Figure \ref{4VBW}, and on the left hand side of Figure \ref{e4s}, should be replaced by elementary four-squares in Figure \ref{4IRFBW}, and on the right hand side of Figure \ref{e4s}, respectively.  

\begin{figure}[htb]
\centering
\begin{tikzpicture}[scale=1.76]

\draw[white!] (0.5,0) circle (0.01pt);

\draw[-,thick,dotted] (2.25,0.0)--(2.0,0.5-0.25)--(0.5,0.5-0.25)--(0.25,0.0)--(0.25-0.25/4,0.5-0.25-0.5/4)--(0.0,0.0-0.25);
\filldraw[fill=white,draw=white] (0.1,0.0)--(1.4,0.0)--(1.4,-0.7)--(0.0,-0.7)--(0.0,0.0);
\filldraw[fill=white,draw=white] (2.0,0.0)--(2.25,0.5)--(2.75,0.5)--(2.5,0.0)--(2.0,0.0);

%
%


\draw[-,thick] (0.75,0)--(1.25,0.5)--(1.75,0);
\draw[-,thick] (0.0,-1.0)--(0.25,0.5)--(1.25,-0.5)--(2.25,0.5);
\filldraw[fill=black,draw=black] (0.75,0.0) circle (1.1pt);
\filldraw[fill=black,draw=black] (1.75,0.0) circle (1.1pt);
\draw[-,thick] (0.75,0)--(1.25,0.5)--(1.75,0);
\draw[-,thick] (0.0,-1.0)--(0.25,0.5)--(1.25,-0.5)--(2.25,0.5);
\filldraw[-,thin,fill=white,draw=white] (0.0,0.0)--(2.0,0.0)--(2.0,-1.0)--(0.0,-1.0)--(0.0,0.0);

\draw[-,thin,gray] (0,0)--(6,0)--(6.25,0.5)--(3.25,0.5);\draw[-,thin,gray] (4,2)--(1,2)--(0,0);
\draw[-,thin,gray] (2,0)--(3,2);

\foreach \y in {0.5,1,1.5}
\draw[-,thin,gray] (0.5*\y,\y)--(3+0.5*\y,\y);

\draw[-,thin,gray] (3,0)--(3.25,0.5);\draw[-,thin,gray] (4,0)--(4.25,0.5);\draw[-,thin,gray] (5,0)--(5.25,0.5);

\foreach \x in {0,2,4}{
\filldraw[draw=black,fill=black] (\x+1,2) circle (1.1pt);
\filldraw[draw=black,fill=black] (\x+2,2) circle (1.1pt);
\foreach \y in {0,1}{
\filldraw[draw=black,fill=black] (\x+\y*0.5,\y) circle (1.1pt);
\filldraw[draw=black,fill=black] (\x+\y*0.5+1,\y) circle (1.1pt);
\filldraw[draw=black,fill=black] (\x+0.25+\y*0.5,\y+0.5) circle (1.1pt);
\filldraw[draw=black,fill=black] (\x+0.25+\y*0.5+1,\y+0.5) circle (1.1pt);
}
}
\foreach \y in {0,1}{
\filldraw[draw=black,fill=black] (6+\y*0.5,\y) circle (1.1pt);
\filldraw[draw=black,fill=black] (6+\y*0.5+0.25,\y+0.5) circle (1.1pt);
}

\foreach \x in {0,1,...,5}{
\foreach \y in {1,...,3}{
\filldraw[fill=black,draw=black] (\x+\y*0.25+0.5+0.25/2,\y*0.5+0.25) circle (1.1pt);
}
}

\foreach \x in {2,3,4,5}{
\filldraw[fill=black,draw=black] (\x+0.5+0.25/2,0.25) circle (1.1pt);
}

\draw[-,thick] (0.75,1.5)--(2,2);
\draw[-,thick] (0.5,1)--(3,2);
\draw[-,thick] (0.25,0.5)--(4,2);
\draw[-,thick] (1.25,0.5)--(5,2);
\draw[-,thick] (2.25,0.5)--(6,2);
\draw[-,thick] (2,0)--(7,2);
\draw[-,thick] (3,0)--(6.75,1.5);
\draw[-,thick] (4,0)--(6.5,1.0);
\draw[-,thick] (5,0)--(6.25,0.5);

\draw[-,thick] (0.5,1)--(1.25,0.5);
\draw[-,thick] (0.75,1.5)--(3,0);
\draw[-,thick] (1,2)--(4,0);
\draw[-,thick] (2,2)--(5,0);
\draw[-,thick] (3,2)--(6,0);
\draw[-,thick] (4,2)--(6.25,0.5);
\draw[-,thick] (5,2)--(6.5,1.0);
\draw[-,thick] (6,2)--(6.75,1.5);

\draw[-,very thick] (0.25,0.0)--(0.25,0.5);

\draw[-,thin,dashed] (0.125/2-0.2,0.125)--(0.125/2,0.125)--(0.125/2,0.0);\draw[->,thin,dashed] (2.0+ 0.125/2,0.125)--(6.0+0.125/2+0.2,0.125);
\draw[-,thick,dotted] (0.25-0.125/2-0.2,0.5-0.125)--(0.25-0.125/2,0.5-0.125)--(0.25-0.125/2,0.0);\draw[->,thick,dotted] (2.25- 0.125/2,0.5-0.125)--(6.25-0.125/2+0.2,0.5-0.125);
\draw[->,thin,dashed] (0.25+0.125/2-0.2,0.625)--(0.25+6.0+0.125/2+0.2,0.625);
\draw[->,thin,dashed] (0.50+0.125/2-0.2,1.125)--(0.50+6.0+0.125/2+0.2,1.125);
\draw[->,thin,dashed] (0.75+0.125/2-0.2,1.625)--(0.75+6.0+0.125/2+0.2,1.625);
\draw[->,thick,dotted] (0.50-0.125/2-0.2,1.0-0.125)--(0.50+6.0-0.125/2+0.2,1.0-0.125);
\draw[->,thick,dotted] (0.75-0.125/2-0.2,1.5-0.125)--(0.75+6.0-0.125/2+0.2,1.5-0.125);
\draw[->,thick,dotted] (1.0 -0.125/2-0.2,2.0-0.125)--(1.0 +6.0-0.125/2+0.2,2.0-0.125);

\draw[-,thin,dashed] (0.25-0.1,0.0-0.2)--(0.25,0.0);\draw[->,thin,dashed] (0.5,0.0)--(0.5,0.5)--(1.25+0.1,2.0+0.2);
\draw[-,thin,dashed] (1.25-0.1,0.0-0.2)--(1.25,0.0);\draw[->,thin,dashed] (1.5,0.0)--(1.5,0.5)--(2.25+0.1,2.0+0.2);
\draw[-,thick,dotted] (0.75-0.1,0.0-0.2)--(0.75,0.0);\draw[->,thick,dotted] (1.0,0.0)--(1.0,0.5)--(1.75+0.1,2.0+0.2);
\draw[-,thick,dotted] (1.75-0.1,0.0-0.2)--(1.75,0.0);\draw[->,thick,dotted] (2.0,0.0)--(2.0,0.5)--(2.75+0.1,2.0+0.2);
\draw[->,thin,dashed] (2.25-0.1,0.0-0.2)--(3.25+0.1,2.0+0.2);
\draw[->,thin,dashed] (3.25-0.1,0.0-0.2)--(4.25+0.1,2.0+0.2);
\draw[->,thin,dashed] (4.25-0.1,0.0-0.2)--(5.25+0.1,2.0+0.2);
\draw[->,thin,dashed] (5.25-0.1,0.0-0.2)--(6.25+0.1,2.0+0.2);
\draw[->,thick,dotted] (2.75-0.1,0.0-0.2)--(3.75+0.1,2.0+0.2);
\draw[->,thick,dotted] (3.75-0.1,0.0-0.2)--(4.75+0.1,2.0+0.2);
\draw[->,thick,dotted] (4.75-0.1,0.0-0.2)--(5.75+0.1,2.0+0.2);
\draw[->,thick,dotted] (5.75-0.1,0.0-0.2)--(6.75+0.1,2.0+0.2);


\filldraw[-,thin,fill=white,draw=gray] (3.25,0.5)--(6.25,0.5)--(7,2)--(7,3)--(4,3)--(3.25,1.5)--(3.25,0.5);
\filldraw[fill=black,draw=black] (5.75-0.5,2.0) circle (1.1pt);
\filldraw[-,thin,fill=white,draw=white] (5.75-0.6,2.0)--(5.75-0.4,2.0)--(5.75-0.4,1.9)--(5.75-0.6,1.9)--(5.75-0.6,2.0);

\draw[-,thick,dotted] (5.75,2.0)--(5.5,2.25)--(5.0,2.25)--(4.75,2.0)--(4.75-0.25/4,2.25-0.5/4)--(4.5,1.75);
\filldraw[,thin,fill=white,draw=white] (3.75+0.25,2.0)--(5.25+0.25,2.0)--(5.75+0.25/2,2.75)--(5.75+0.25/2+0.5,2.75)--(5.75,1.55)--(3.75+0.25,1.5)--(3.75+0.25,2.0);

\foreach \x in {4.25,5.25,6.25}
\draw[-,thin,gray] (\x,1.5)--(\x+0.75,3);
\foreach \y in {0,0.5,1}
\draw[-,thin,gray] (3.25+\y*0.5,\y+1.5)--(6.25+\y*0.5,\y+1.5)--(6.25+\y*0.5,\y+0.5);
\draw[-,thin,gray] (4.25,0.5)--(4.25,1.5);\draw[-,thin,gray] (5.25,0.5)--(5.25,1.5);
\draw[-,thin,gray] (4.75,2.5)--(4.75,2);
\foreach \x in {3.25,5.25} {
\filldraw[fill=black,draw=black] (\x,0.5) circle (1.1pt);
\filldraw[fill=black,draw=black] (\x+1,0.5) circle (1.1pt);
\foreach \y in {0,1} {
\filldraw[fill=black,draw=black] (\x+\y*0.5,\y+1.5) circle (1.1pt);
\filldraw[fill=black,draw=black] (\x+\y*0.5+1,\y+1.5) circle (1.1pt);
\filldraw[fill=black,draw=black] (\x+\y*0.5+0.25,\y+1.5+0.5) circle (1.1pt);
\filldraw[fill=black,draw=black] (\x+\y*0.5+1+0.25,\y+1.5+0.5) circle (1.1pt); }}
\filldraw[fill=black,draw=black] (6.5,1) circle (1.1pt);
\filldraw[fill=black,draw=black] (6.75,1.5) circle (1.1pt);
\filldraw[fill=black,draw=black] (7,2) circle (1.1pt);

\foreach \x in {3,4,5}{
\filldraw[fill=black,draw=black] (\x+0.75,1.0) circle (1.1pt);
\filldraw[fill=black,draw=black] (\x+0.75+0.25/2,1.75) circle (1.1pt);
\filldraw[fill=black,draw=black] (\x+0.75+0.25/2+0.5,2.75) circle (1.1pt);
}
\filldraw[fill=black,draw=black] (3.75+0.25/2+0.25,2.25) circle (1.1pt);
\filldraw[fill=black,draw=black] (5.75+0.25/2+0.25,2.25) circle (1.1pt);

\draw[-,thick] (3.25,0.5)--(4.25,1.5)--(5.25,0.5)--(6.25,1.5)--(6.5,1.0)--(6.75,2.5)--(7,2);
\draw[-,thick] (3.25,1.5)--(4.25,0.5)--(5.25,1.5)--(6.25,0.5)--(6.5,2)--(6.75,1.5)--(7,3);
\draw[-,thick] (3.25,1.5)--(4.5,2.0)--(5.25,1.5)--(6.5,2.0)--(5.75,2.5)--(7.0,3.0);
\draw[-,thick] (5.75,2.5)--(5.0,3.0)--(3.75,2.5)--(4.5,2.0);
\draw[-,thick] (6.25,1.5)--(5.5,2.0)--(4.25,1.5)--(3.5,2.0)--(4.75,2.5)--(4.0,3.0);
\draw[-,thick] (4.75,2.5)--(6.0,3.0)--(6.75,2.5)--(5.5,2.0);

\draw[-,very thick] (3.25,0.5)--(3.25,1.5);
\draw[-,very thick] (6.25,0.5)--(6.25,1.5);
\draw[-,very thick] (7.0,2.0)--(7.0,3.0);
\draw[-,very thick] (4.75,2.0)--(4.75,2.5);

\draw[-,thin,dashed] (3.25,1.0)--(3.5,0.75)--(6.0,0.75)--(6.25,1.0); \draw[->,thick,dotted] (6.25,1.0)--(6.25+0.25/4,0.75+0.5/4)--(7.0-0.25/4,2.25-0.5/4)--(7.0,2.5);
\draw[-,thick,dotted] (3.25,1.0)--(3.5,1.25)--(6.0,1.25)--(6.25,1.0); \draw[->,thin,dashed] (6.25,1.0)--(6.25+0.25/4,1.25+0.5/4)--(7.0-0.25/4,2.75-0.5/4)--(7.0,2.5);

\draw[->,thin,dashed] (3.5,0.5)--(3.5,1.5)--(4.25,3.0);
\draw[-,thin,dashed] (4.5,0.5)--(4.5,1.5)--(4.75,2.0);\draw[->,thin,dashed] (5.0,2.0)--(5.0,2.5)--(5.25,3.0);
\draw[->,thin,dashed] (5.5,0.5)--(5.5,1.5)--(6.25,3.0);
\draw[->,thick,dotted] (4.0,0.5)--(4.0,1.5)--(4.75,3.0);
\draw[-,thick,dotted] (5.0,0.5)--(5.0,1.5)--(5.25,2.0);\draw[->,thick,dotted] (5.5,2.0)--(5.5,2.5)--(5.75,3.0);
\draw[->,thick,dotted] (6.0,0.5)--(6.0,1.5)--(6.75,3.0);

\draw[-,thick,dotted] (3.5 -0.125/2,2.0-0.125)--(6.5 -0.125/2,2.0-0.125)--(6.5 -0.125/2,1.0-0.125);
\draw[-,thick,dotted] (3.75-0.125/2,2.5-0.125)--(4.75-0.125/2,2.5-0.125)--(4.75-0.125/2,2,0);\draw[-,thick,dotted] (5.75-0.125/2,2.5-0.125)--(6.75-0.125/2,2.5-0.125)--(6.75-0.125/2,1.5-0.125);
\draw[-,thick,dotted] (4.0 -0.125/2,3.0-0.125)--(7.0 -0.125/2,3.0-0.125)--(7.0 -0.125/2,2.0-0.125);
\draw[-,thin,dashed] (3.25+0.125/2,1.5+0.125)--(6.25+0.125/2,1.5+0.125)--(6.25+0.125/2,0.5+0.125);
\draw[-,thin,dashed] (3.5 +0.125/2,2.0+0.125)--(4.5 +0.125/2,2.0+0.125)--(4.5 +0.125/2,2.0);\draw[-,thin,dashed] (5.5 +0.125/2,2.0+0.125)--(6.5 +0.125/2,2.0+0.125)--(6.5 +0.125/2,1.0+0.125);
\draw[-,thin,dashed] (3.75+0.125/2,2.5+0.125)--(6.75+0.125/2,2.5+0.125)--(6.75+0.125/2,1.5+0.125);


\draw[-, thick] (4.5+0.25*2/3,2)--(4.75,2.5)--(5.75-0.5,2)--(5.75,2.5);


\filldraw[-,thin,fill=white,draw=gray] (1.25,0.5)--(2.25,0.5)--(2.5,1)--(2.5,3)--(1.5,3)--(1.25,2.5)--(1.25,0.5);
\draw[-,thin,gray] (2.25,0.5)--(2.25,2.5);\draw[-,thin,gray](1.25,2.5)--(2.25,2.5)--(2.5,3);\draw[-,thin,gray] (1.25,1.5)--(2.25,1.5)--(2.5,2);
\filldraw[fill=black,draw=black] (1.25,0.5) circle (1.1pt);
\filldraw[fill=black,draw=black] (2.25,0.5) circle (1.1pt);
\filldraw[fill=black,draw=black] (2.5,1) circle (1.1pt);
\filldraw[fill=black,draw=black] (1.25,1.5) circle (1.1pt);
\filldraw[fill=black,draw=black] (2.25,1.5) circle (1.1pt);
\filldraw[fill=black,draw=black] (2.5,2) circle (1.1pt);
\filldraw[fill=black,draw=black] (1.25,2.5) circle (1.1pt);
\filldraw[fill=black,draw=black] (2.25,2.5) circle (1.1pt);
\filldraw[fill=black,draw=black] (2.5,3) circle (1.1pt);
\filldraw[fill=black,draw=black] (1.5,3) circle (1.1pt);
\draw[-,thick] (1.25,0.5)--(2.25,1.5)--(2.5,1);
\draw[-,thick] (1.25,2.5)--(2.25,1.5)--(2.5,3)--(1.25,2.5);
\draw[-,thick] (1.5,3.0)--(2.25,2.5)--(1.25,1.5)--(2.25,0.5)--(2.5,2.0)--(2.25,2.5);

\draw[-,very thick] (1.25,0.5)--(1.25,2.5);
\draw[-,very thick] (2.25,0.5)--(2.25,2.5);
\draw[-,very thick] (2.5,1.0)--(2.5,3.0);

\filldraw[fill=black,draw=black] (1.75,2.0) circle (1.1pt);
\filldraw[fill=black,draw=black] (1.75,1.0) circle (1.1pt);
\filldraw[fill=black,draw=black] (2.25+0.25/2,0.5+1.5/2) circle (1.1pt);
\filldraw[fill=black,draw=black] (2.25+0.25/2,1.5+1.5/2) circle (1.1pt);
\filldraw[fill=black,draw=black] (2.0-0.25/2,2.75) circle (1.1pt);


\draw[-,thick,dotted] (1.25+3*0.125/2,2.5+3*0.5/4)--(2.25+3*0.125/2,2.5+3*0.5/4)--(2.25+3*0.125/2,0.5+3*0.5/4);
\draw[-,thin,dashed]  (1.25+1*0.125/2,2.5+1*0.5/4)--(2.25+1*0.125/2,2.5+1*0.5/4)--(2.25+1*0.125/2,0.5+1*0.5/4);
\draw[->,thin,dashed] (1.5,0.5)--(1.5,2.5)--(1.75,3.0);
\draw[->,thick,dotted] (2.0,0.5)--(2.0,2.5)--(2.25,3.0);

\draw[->,thin,dashed] (1.25,1.0)--(1.5,0.75)--(2.0,0.75)--(2.25,1.0)--(2.25+0.25/4,0.75+0.5/4)--(2.5-0.25/4,1.25-0.5/4)--(2.5,1.5);
\draw[->,thin,dashed] (1.25,2.0)--(1.5,1.75)--(2.0,1.75)--(2.25,2.0)--(2.25+0.25/4,1.75+0.5/4)--(2.5-0.25/4,2.25-0.5/4)--(2.5,2.5);
\draw[->,thick,dotted] (1.25,1.0)--(1.5,1.25)--(2.0,1.25)--(2.25,1.0)--(2.25+0.25/4,1.25+0.5/4)--(2.5-0.25/4,1.75-0.5/4)--(2.5,1.75);
\draw[->,thick,dotted] (1.25,2.0)--(1.5,2.25)--(2.0,2.25)--(2.25,2.0)--(2.25+0.25/4,2.25+0.5/4)--(2.5-0.25/4,2.75-0.5/4)--(2.5,2.75);

\draw[-,thin,gray] (0,0)--(0,-1)--(1,-1)--(1,0);\draw[-,thin,gray] (1,-1)--(2,-1)--(2,0);\draw[-,thin,gray] (2,-1)--(2.25,-0.5)--(2.25,0);\draw[-,thin,gray] (0.25,0.5)--(0.25,0);\draw[-,thin,gray] (1.25,0.5)--(1.25,0);
\filldraw[fill=black,draw=black] (0.25,0.5) circle (1.1pt);
\filldraw[fill=black,draw=black] (1,0) circle (1.1pt);




\fill[black] (7.86,0) circle (0.01pt)
node[below=1.5pt]{\color{black} $+\ihat$};
\fill[black] (7.66,0.33) circle (0.01pt)
node[right=1.5pt]{\color{black} $+\jhat$};
\fill[black] (7.5,0.75*0.5) circle (0.01pt)
node[left=1.5pt]{\color{black} $+\khat$};
\draw[->, thick] (7.5,0)--(8.25,0);\draw[->, thick] (7.5,0)--(7.83,0.67);\draw[->, thick] (7.5,0)--(7.5,0.75);

\end{tikzpicture}
\caption{An example of an IRF model on a surface $\sigma$, constructed from cubic-shaped deformations of an IRF model on a flat surface $\sigma_0$.  Boltzmann weights on elementary four-squares of $\sigma$ are shown in Figures \ref{e4s} and \ref{4IRFBW}.  White vertices of $\sigma$ are not shown.}
\label{Glattice3}
\end{figure}
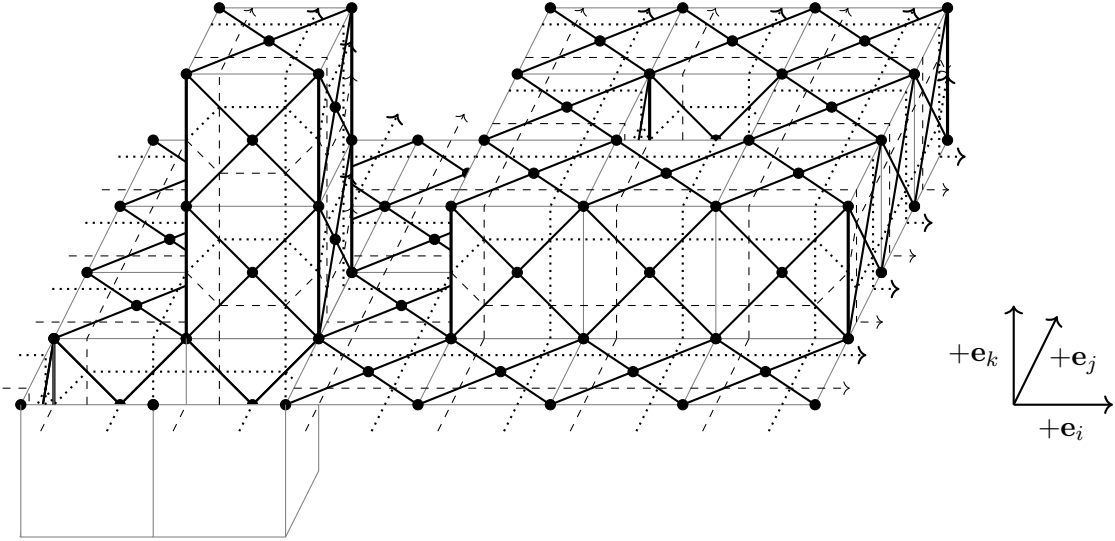

Figure \ref{Glattice3} shows an example of a surface $\sigma$, that can arise from a deformation of the IRF model on the flat surface $\sigma_0$, and satisfies the conditions given in the previous subsection.  Recall that in the previous subsection, the vertex model was defined on a graph $\Gl$ that was formed by connecting next nearest neighbour vertices of $\sigma$.  The IRF model is defined on a similar graph $\Gl$ here, but one obvious difference is the appearance of additional edges of $\Gl$, that connect two spins at either of the pairs of vertices $(\n,\n+\khat)$, or $(\n+\jhat,\n+\jhat+\khat)$, on elementary four-squares $\sigma^4_{ik}(\n)$, or $\sigma^4_{ki}(\n)$, respectively.  Such edges are required due to the form of the IRF Yang-Baxter equation \eqref{YBE-IRF} as seen in Figure \ref{SSRYBE}.  This means that two types of rapidity lines $r$, and $r'$, intersect with each other on the edges connecting these vertices, as seen in Figure \ref{4IRFBW}.  Due to the first inversion relation in \eqref{invrels2}, two of these types of edges cancel, when they connect the same two vertices that are common to either a pair of elementary four-squares $\sigma^4_{ik}(\n)$, $\sigma^4_{ik}(\n+\ihat)$, or a pair $\sigma^4_{ki}(\n)$, $\sigma^4_{ki}(\n+\ihat)$.  However if the respective pairs of vertices $(\n,\n+\khat)$, or $(\n+\jhat,\n+\jhat+\khat)$, of either elementary four-square $\sigma^4_{ik}(\n)$, or $\sigma^4_{ki}(\n)$, are common with any of the four-squares $\sigma^4_{jk}(\n+\ihat)$, $\sigma^4_{jk}(\n+\ihat-\jhat)$, $\sigma^4_{kj}(\n)$ or $\sigma^4_{kj}(\n-\jhat)$ (forming a 90 degree angle at the edge), then $\Gl$ will contain a single edge connecting these vertices.  Note that these edges of $\Gl$ are pictured in Figure \ref{4IRFBW} to pass through white vertices of $V(\sigma)$, but there are no white vertices in $V(\Gl)$.

The IRF model on $\Gl$, will depend on the two different types of IRF Boltzmann weights \eqref{V1}, and \eqref{V2}, and also on the edge Boltzmann weights given in Figure \ref{fig-crosses}.  Then there end up being a total of four different types of IRF Boltzmann weights, associated to five different types of elementary four-squares of $\sigma$, as shown in Figures \ref{e4s} and \ref{4IRFBW}.  In the latter figures, the IRF Boltzmann weights assigned to elementary four-squares $\sigma^4_{ki}$, and $\sigma^4_{ik}$, are given by
\begin{align}
\label{V3}
\V_{\bq\br}^{(3)}(\xv_a,\xv_b,\xv_c,\xv_d):= W_{rr'}(\xv_c,\xv_d)\, W_{r'r}(\xv_b,\xv_a)\,\V_{\bq\br}^{(2)}(\xv_a,\xv_b,\xv_c,\xv_d),
\end{align}
and
\begin{align}
\label{V4}
\V_{\br\bq}^{(4)}(\xv_a,\xv_b,\xv_c,\xv_d):= W_{rr'}(\xv_a,\xv_c)\, W_{r'r}(\xv_d,\xv_b)\,\V_{\br\bq}^{(1)}(\xv_a,\xv_b,\xv_c,\xv_d),
\end{align}
respectively.  Thus at least in terms of Boltzmann weights, the IRF model on $\sigma$ is somewhat more complicated than the vertex model on $\sigma$, since the latter model only had a dependence on the single expression for the Boltzmann weight given in \eqref{vertexBW}.  

Finally, the partition function for the IRF model on $\sigma$, is given by the expression
\begin{align}
\label{zIRFsigma}
\begin{array}{l}
\ds Z=\sum_{\x}\,\prod_{\sigma^4_{ij}}\,\V_{\bp\bq}^{(1)}(\xv_a,\xv_b,\xv_c,\xv_d)\,\prod_{\sigma^4_{jk}}\,\V_{\bp\br}^{(1)}(\xv_a,\xv_b,\xv_c,\xv_d)\,\prod_{\sigma^4_{kj}}\,\V_{\br\bp}^{(2)}(\xv_a,\xv_b,\xv_c,\xv_d) \\[0.7cm]
\ds\phantom{Z=\sum_{\x}\xv}\times\prod_{\sigma^4_{ki}}\,\V_{\bq\br}^{(3)}(\xv_a,\xv_b,\xv_c,\xv_d)\,\prod_{\sigma^4_{ik}}\,\V_{\br\bq}^{(4)}(\xv_a,\xv_b,\xv_c,\xv_d).
\end{array}
\end{align}
Here the products are taken over the Boltzmann weights associated to each of the five different types of elementary four-squares in $F^{(4)}(\sigma)$.  The sum is taken over all configurations of interior spins $\xv_1,\xv_2,\ldots,\xv_m$, that are assigned to vertices $i_1,i_2,\ldots,i_m\in V_{int}(G)$, where each vertex is common to at least two different elementary four-squares in $F^{(4)}(\sigma)$ (black corner vertices).  Boundary spins remain fixed.  Note that the spin interior to each elementary four-square in $F^{(4)}(\sigma)$ is already summed over in the definition of the IRF Boltzmann weights in \eqref{V1}, \eqref{V2}.  
This defines the IRF model on $\sigma$.

\subsubsection{Extended Z-invariance property}
The statement of extended Z-invariance for the vertex and IRF models on $\sigma$ is as follows:
\\[0.2cm]
\textbf{Extended Z-invariance.} \textit{{The expressions for the partition functions for the vertex model on $\sigma$ \eqref{zvertexsigma}, and the IRF model on $\sigma$ \eqref{zIRFsigma}, are equivalent to the respective expressions for the partition functions for the vertex model on $\sigma_0$ \eqref{vertexZ}, and the IRF model on $\sigma_0$ \eqref{IRFZ1}, up to some constant factors.}}
\\[0.2cm]
Note that the extended Z-invariance property also implies the equivalence of partition functions on two different surfaces $\sigma$, $\sigma'$, where both of the latter surfaces satisfy the conditions given in Section \ref{sec:vertexsigma}.  The arguments for extended Z-invariance of the vertex and IRF models, follow identically to the case of \cite{Kels:2017fyt}, with deformations of elementary squares in the latter case, being replaced by deformations of elementary four-squares that appear in Appendix \ref{app:IRF}.  Similarly to \cite{Kels:2017fyt}, the deformations described in Appendix \ref{app:IRF}, always involving shifting an elementary four-square $\sigma^4_{ij}(\n)$, to either $\sigma^4_{ij}(\n+\khat)$, or $\sigma^4_{ij}(\n-\khat)$, respectively, according to whether the associated pairs of rapidity lines $r,r'$, are positively or negatively oriented.  Then using these deformations, all elementary four-squares $\sigma^4_{ij}(\n)\in F^{(4)}(\sigma)$, can be shifted (in a particular order) until they have the same $\n$ coordinate as the boundary, at which stage $\sigma$ will coincide with the plane $\sigma_0$.  This implies that the partition functions on $\sigma$ and $\sigma_0$ are equivalent, up to some irrelevant constant factors.\footnote{For $N$-state discrete spin models the constant factor is simply given by a power of $N$.  As was noted in \cite{Kels:2017fyt}, there is some subtlety for continuous spin models where this constant becomes infinite (proportional to $\delta(0)$).  These cases then appear to require an appropriate regularisation in order to properly define the deformed partition functions. In any case, the observables of the deformed model will not be affected, since these are typically expressed in terms of derivatives of the partition function.}

The deformations in Appendix \ref{app:IRF} which are used to show the extended Z-invariance property, follow from the respective inversion relations \eqref{IRFinv}, and \eqref{Vinv}, and Yang-Baxter equations \eqref{YBE-IRF}, and \eqref{YBE-vertex}, satisfied by the integrable IRF and vertex models.  There are several important models of statistical mechanics which satisfy such relations \cite{Baxter:1972hz,Baxter:1982zz,ABF,Bazhanov:1990qk,Bazhanov:1992jqa,Bazhanov:2011mz}, but the explicit expressions for the Boltzmann weights were not considered here in order to remain as general and concise as possible.   Note that while the approach of this paper was to start with an edge-interaction model that was used to formulate the vertex and IRF models, the  extended Z-invariance property also applies to pure vertex and IRF models of statistical mechanics, whose Yang-Baxter equations are independent of the star-star relation.  In the latter cases, the Boltzmann weights that are associated to elementary four-squares\footnote{Since these models are not formed from edge Boltzmann weights, these would simply be elementary squares, rather than four-squares.} can simply be read off the respective expressions for the Yang-Baxter equations, similarly to the cases for this paper.
 
\begin{figure}[htb!]
\centering
\begin{tikzpicture}[scale=3.0]

\draw[-,very thick] (-0.6,1.4)--(0.6,1.4)--(0.6,2.6)--(-0.6,2.6)--(-0.6,1.4);
\draw[->,double,black] (0,1.2)--(0,2.8);
\draw[->,double,black] (-0.8,2)--(0.8,2);
\fill[black!] (0,1.2) circle (0.01pt)
node[below=1.5pt]{\color{black}\small $\mathbf{q}$};
\fill[black!] (-0.8,2) circle (0.01pt)
node[left=3.1pt]{\color{black}\small $\mathbf{p}$};
\filldraw[fill=black,draw=black] (-0.6,2.6) circle (0.8pt)
node[above=0.5pt]{\color{black}\small $x_a$};
\filldraw[fill=black,draw=black] (0.6,1.4) circle (0.8pt)
node[below=0.5pt]{\color{black}\small $x_d$};
\filldraw[fill=black,draw=black] (-0.6,1.4) circle (0.8pt)
node[below=0.5pt]{\color{black}\small $x_c$};
\filldraw[fill=black,draw=black] (0.6,2.6) circle (0.8pt)
node[above=0.5pt]{\color{black}\small $x_b$};

\draw[black] (1.55,2.1) circle (0.01pt)
node[below=1pt]{\color{black}\small $\times\left(\sum_{x_a}\delta_{x_a,x_a}\right)=$};

\begin{scope}[xshift=90]

\draw[-,very thick] (-0.6,1.4)--(0.6,1.4)--(0.6,2.6)--(-0.6,2.6)--(-0.6,1.4);
\draw[-,very thick] (-0.2,1.8)--(0.2,1.8)--(0.2,2.2)--(-0.2,2.2)--(-0.2,1.8);
\draw[-,very thick] (-0.6,1.4)--(-0.2,1.8);
\draw[-,very thick] (0.6,1.4)--(0.2,1.8);
\draw[-,very thick] (0.6,2.6)--(0.2,2.2);
\draw[-,very thick] (-0.6,2.6)--(-0.2,2.2);
\draw[->,double,black] (0,1.2)--(0,2.8);
\draw[->,double,black] (-0.8,2)--(0.8,2);
\draw[->,double,black] (-0.4,1.6)--(0.4,1.6)--(0.4,2.4)
node[above=0.5pt]{\small $\mathbf{r}$};
\draw[->,double,black] (0.4,2.4)--(-0.4,2.4)--(-0.4,1.6)
node[below=0.5pt]{\small $\mathbf{r}$};
\fill[black!] (0,1.2) circle (0.01pt)
node[below=1.5pt]{\color{black}\small $\mathbf{q}$};
\fill[black!] (-0.8,2) circle (0.01pt)
node[left=3.1pt]{\color{black}\small $\mathbf{p}$};
\filldraw[fill=black,draw=black] (-0.6,2.6) circle (0.8pt)
node[above=0.5pt]{\color{black}\small $x_a$};
\filldraw[fill=black,draw=black] (0.6,1.4) circle (0.8pt)
node[below=0.5pt]{\color{black}\small $x_d$};
\filldraw[fill=black,draw=black] (-0.6,1.4) circle (0.8pt)
node[below=0.5pt]{\color{black}\small $x_c$};
\filldraw[fill=black,draw=black] (0.6,2.6) circle (0.8pt)
node[above=0.5pt]{\color{black}\small $x_b$};
\filldraw[fill=black,draw=black] (-0.2,2.2) circle (0.8pt)
node[above=0.5pt]{\color{black}\small $x'_a$};
\filldraw[fill=black,draw=black] (0.2,1.8) circle (0.8pt)
node[below=0.5pt]{\color{black}\small $x'_d$};
\filldraw[fill=black,draw=black] (-0.2,1.8) circle (0.8pt)
node[below=0.5pt]{\color{black}\small $x'_c$};
\filldraw[fill=black,draw=black] (0.2,2.2) circle (0.8pt)
node[above=0.5pt]{\color{black}\small $x'_b$};

\end{scope}

\begin{scope}[scale=1.1,yshift=-60]

\draw[-,very thick] (0,2.6)--(-0.6,2.0)--(0,1.4)--(0.6,2.0)--(0,2.6);
\draw[->,double,black] (0,1.2)--(0,2.8);
\draw[->,double,black] (-0.8,2)--(0.8,2);
\fill[black!] (0,1.2) circle (0.01pt)
node[below=1.5pt]{\color{black}\small $\mathbf{q}$};
\fill[black!] (-0.8,2) circle (0.01pt)
node[left=3.1pt]{\color{black}\small $\mathbf{p}$};
\filldraw[fill=black,draw=black] (0,2.6) circle (0.6pt)
node[right=0.5pt]{\color{black}\small $x'_j$};
\filldraw[fill=black,draw=black] (0.6,2.0) circle (0.6pt)
node[below=0.5pt]{\color{black}\small $x'_i$};
\filldraw[fill=black,draw=black] (0,1.4) circle (0.6pt)
node[left=0.5pt]{\color{black}\small $x_j$};
\filldraw[fill=black,draw=black] (-0.6,2.0) circle (0.6pt)
node[above=0.5pt]{\color{black}\small $x_i$};

\draw[black] (1.35,2.1) circle (0.01pt)
node[below=1pt]{\color{black}\small $\times\left(\sum_{x'''_k}\delta_{x'''_k,x'''_k}\right)=$};

\begin{scope}[xshift=80]

\draw[-,very thick] (0,2.6)--(-0.6,2.0)--(0,1.4)--(0.6,2.0)--(0,2.6);
\draw[-,very thick] (0,2.2)--(-0.2,2.0)--(0,1.8)--(0.2,2.0)--(0,2.2);
\draw[-,very thick] (0.2,2.2)--(0.2,1.8)--(-0.2,1.8)--(-0.2,2.2)--(0.2,2.2);
\draw[-,very thick] (0,2.6)--(0.2,2.2)--(0.6,2.0)--(0.2,1.8)--(0,1.4)--(-0.2,1.8)--(-0.6,2.0)--(-0.2,2.2)--(0,2.6);
\draw[->,double,black] (0,1.2)--(0,2.8);
\draw[->,double,black] (-0.8,2)--(0.8,2);
\draw[->,double,black] (0,2.4)--(-0.4,2.0)--(0,1.6); \draw[->,double,black] (0,1.6)--(0.4,2.0)--(0,2.4);
\fill[black!] (0,1.2) circle (0.01pt)
node[below=1.5pt]{\color{black}\small $\mathbf{q}$};
\fill[black!] (-0.8,2) circle (0.01pt)
node[left=3.1pt]{\color{black}\small $\mathbf{p}$};
\filldraw[fill=black,draw=black] (0,2.6) circle (0.6pt)
node[right=0.5pt]{\color{black}\small $x'_j$};
\filldraw[fill=black,draw=black] (0.6,2.0) circle (0.6pt)
node[below=0.5pt]{\color{black}\small $x'_i$};
\filldraw[fill=black,draw=black] (0,1.4) circle (0.6pt)
node[left=0.5pt]{\color{black}\small $x_j$};
\filldraw[fill=black,draw=black] (-0.6,2.0) circle (0.6pt)
node[above=0.5pt]{\color{black}\small $x_i$};
\filldraw[fill=black,draw=black] (0,2.2) circle (0.6pt)
;
\filldraw[fill=black,draw=black] (0.2,2.0) circle (0.6pt)
;
\filldraw[fill=black,draw=black] (0,1.8) circle (0.6pt)
;
\filldraw[fill=black,draw=black] (-0.2,2.0) circle (0.6pt)
;
\filldraw[fill=black,draw=black] (0.2,2.2) circle (0.6pt)
;
\filldraw[fill=black,draw=black] (0.2,1.8) circle (0.6pt)
;
\filldraw[fill=black,draw=black] (-0.2,2.2) circle (0.6pt)
;
\filldraw[fill=black,draw=black] (-0.2,1.8) circle (0.6pt)
;

\end{scope}
\end{scope}

\end{tikzpicture}
\caption{A pair of deformations for IRF and vertex models, corresponding to Figures \ref{IRFcube} (top), and \ref{Vcube} (bottom), respectively.  Some graphical simplifications were used in order to present the above deformations in a clear manner.  The pairs of rapidity lines labelled $(p,p')$, $(q,q')$, or $(r,r')$, are represented by a single rapidity line labelled $\mathbf{p}$, $\mathbf{q}$, or $\mathbf{r}$, respectively.  For the vertex deformation, the labelling of vertices on the right hand side has not been shown due to space limitations.  Also the variables on vertices interior to elementary four-squares, and the twist of $\mathbf{r}$-type rapidity lines are not shown for the IRF deformation.  Note that the $r$-type rapidity lines form closed directed loops, both with themselves, and in combination with the $\mathbf{p}$- and $\mathbf{q}$-type rapidity lines.  These two deformations of the IRF and vertex Boltzmann weights, are central to the extended formulation of Z-invariance that has been presented in this paper.}
\label{newd}
\end{figure}
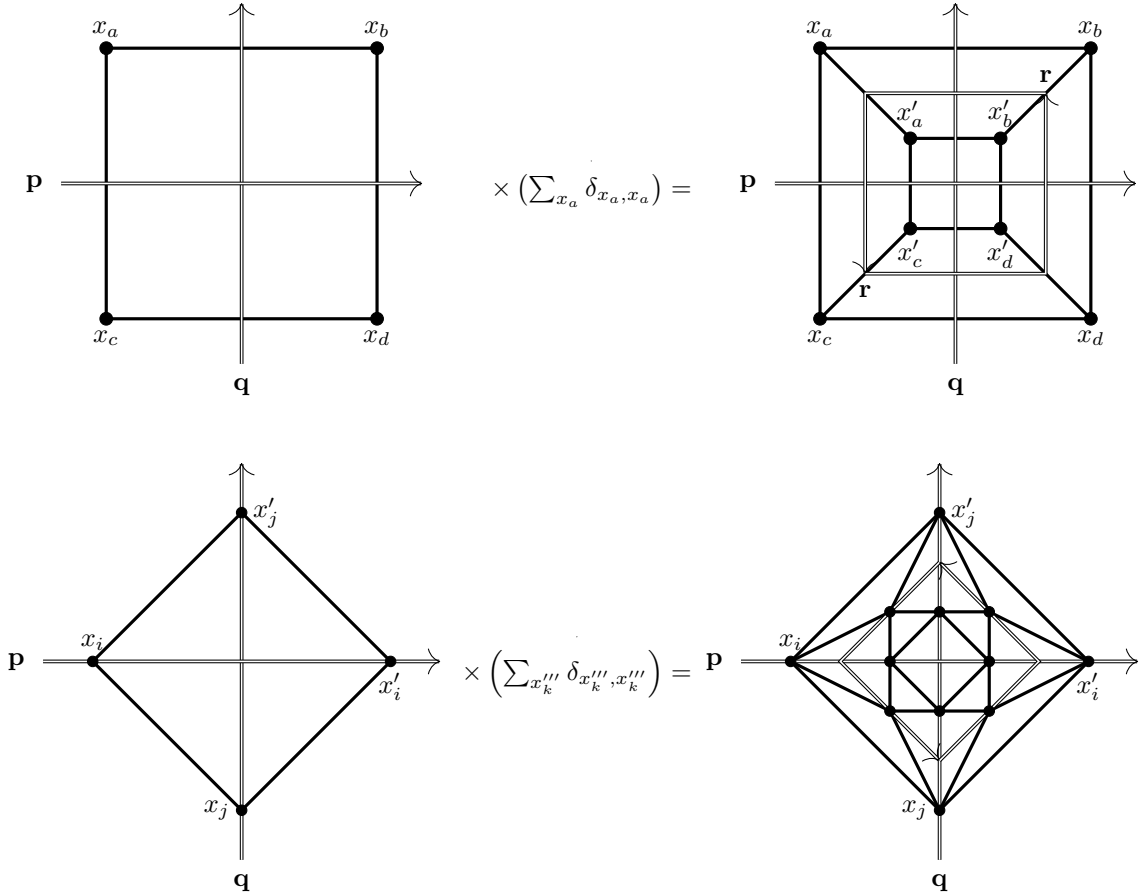

It is easy to see that the extended Z-invariance property allows for some quite interesting deformations of the usual integrable vertex and IRF models.  As an example, by repeatedly using the deformations of the type pictured in Figures \ref{Vcube} or \ref{IRFcube} respectively, a Boltzmann weight may be taken an arbitrary distance out of the plane of the model, at the cost of only adding some irrelevant constant factors to the partition function.  Such deformations are not permitted under the usual formulation of Z-invariance, because these deformations involve the introduction of rapidity lines which form closed directed loops.  It is also revealing to consider these same deformations when they are restricted to the plane, as is indicated in Figure \ref{newd}.

In Figure \ref{newd}, the rapidity lines of $\mathbf{r}$-type, form closed directed loops both with themselves, and in combination with the $\mathbf{p}$- and $\mathbf{q}$-type rapidity lines.   Repeated applications of such deformations, are then seen to add to a model an arbitrary number of IRF or vertex Boltzmann weights, which are contained inside a single IRF or vertex Boltzmann weight respectively.  These two new types of deformations for the IRF and vertex models, are central to the extended Z-invariance property that has been formulated in this paper.


\section{Quasi-classical expansion and classical discrete integrable equations}\label{sec:qcl}

In the quasi-classical limit, the fluctuating variables of the model are taken to approach a fixed ground state configuration, which is determined as the solution of a classical discrete integrable equation.  For integrable models that satisfy a star-triangle relation, this limit has been found \cite{Bazhanov:2007mh,Bazhanov:2010kz,Bazhanov:2016ajm,Kels:2017vbc,Kels:2018xge} to always lead to a classical discrete integrable equation in the Adler-Bobenko-Suris (ABS) classification \cite{ABS,ABS2}, where the latter equations are equivalent to the equations of the critical point of the star-triangle relation.  This correspondence has recently been extended to the entire ABS list, with the use of star-triangle relations that are based on hypergeometric integrals \cite{Bazhanov:2016ajm,Kels:2018xge}.  On the other hand, the analogous saddle-point equations that arise from the quasi-classical limit of the star-star relation, {\it e.g.} \cite{Bazhanov:2011mz,KelsThesis}, appear to lead to new types of discrete integrable equations.  This section will investigate those types of equations which arise in the quasi-classical limit of the IRF model on $\sigma$, and show that such equations also satisfy a classical analogue of the extended Z-invariance property.  The latter is also shown to be closely related to a closure property that is central to Lagrangian multiform formulations of discrete integrable equations \cite{LobbNijhoff}.  Note also that the IRF model will be considered here rather than the vertex model, since the IRF model has a more natural quasi-classical limit and description in terms of variational-type equations.

\subsection{Quasi-classical expansion of the star-star relation}

Recall the edge Boltzmann weights $ W_{pq}(\xv_i,\xv_j)$, $\oW_{pq}(\xv_i,\xv_j)$, as pictured in Figure \ref{fig-crosses}, which are assumed now to implicitly depend on a parameter $\hbar>0$, which may be interpreted as a temperature parameter (or Planck constant for the quantum mechanics picture) of the model. Let $f_\hbar(\xv)$ denote a scaling and translation of a variable $\xv$, that has a dependency on the parameter $\hbar$.  It is assumed here that as $\hbar\rightarrow0$, the quasi-classical expansion of Boltzmann weights $W_{pq}(\xv_i,\xv_j)$, $\oW_{pq}(\xv_i,\xv_j)$, can be written as
\begin{align}
\label{weight-exp}
\begin{split}
\ds\log W_{f_\hbar(p)f_\hbar(q)}(f_\hbar(\xv_i),f_\hbar(\xv_j))=& \ds  -\hbar^{-1}\, \Lambda_{pq}(\xv_i,\xv_j)+O(\hbar^0),\\
\ds\log \overline{ W}_{f_\hbar(p)f_\hbar(q)}(f_\hbar(\xv_i),f_\hbar(\xv_j))=&\ds -\hbar^{-1}\,\olam_{pq}(\xv_i,\xv_j) -\frac{1}{2}\log\hbar+O(\hbar^0).
\end{split}
\end{align}
This expansion is found for the majority known integrable edge-interaction models \cite{Bazhanov:2016ajm,Kels:2018xge}, under a suitable transformation $f_\hbar$, that is model dependent.  Note that depending on the model under consideration, the spin variable $\xv_i$ may take continuous real values or discrete integer values, while in the quasi-classical limit, the resulting variables\footnote{As a slight abuse of notation, both spin variables of the previous section, and classical variables that arise from the quasi-classical limit in this section, are referred to with the notation $\xv_i$.} $\xv_i$, become continuously valued \cite{Bazhanov:2016ajm}.

The Lagrangian functions are also assumed here to satisfy the following anti-symmetry relations
\begin{align}
\label{clasinversion}
\ds\Lambda_{pq}(\xv_i ,\xv_j)+\Lambda_{qp}(\xv_j,\xv_i)=0,\quad
\ds\olam_{pq}(\xv_i,\xv_j)+\olam_{qp}(\xv_j,\xv_i)=0,
\end{align}
for all $\xv_i,\xv_j$, which are the classical form of \eqref{invrels2}.

The expansion of the form \eqref{weight-exp} is the key to obtaining several types of classical discrete equations from the quasi-classical limit.  First, using the asymptotic expansion of Boltzmann weights \eqref{weight-exp}, the quasi-classical asymptotics of the IRF weight \eqref{V1}, can be determined with a saddle point method, leading to the expansion
\begin{align}
\label{IRFqce}
\begin{gathered}
\ds\log\bW_{\bp\bq}^{(1)}(\xv_a,\xv_b,\xv_c,\xv_d)=-\hbar^{-1}\, \lag^{(1)}_{\bp\bq} (\xv^{(cl)}_1\, |\, \xv_a,\xv_b,\xv_c,\xv_d) + O(\log\hbar), \\
\ds\log\bW_{\bp\bq}^{(2)}(\xv_a,\xv_b,\xv_c,\xv_d)=-\hbar^{-1}\, \lag^{(2)}_{\bp\bq} (\xv^{(cl)}_2\, |\,\xv_a,\xv_b,\xv_c,\xv_d)  + {O}(\log\hbar),
\end{gathered}
\end{align}
where the five-point Lagrangian functions are defined by
\begin{align}
\label{action1}
\begin{gathered}
\ds\lag^{(1)}_{\bp\bq}(\xv\, |\,\xv_a,\xv_b,\xv_c,\xv_d)= \olam_{pq}(\xv_c,\xv)+ \olam_{p'q'}(\xv_b,\xv)+ \Lambda_{p'q}(\xv,\xv_a)+ \Lambda_{pq'}(\xv,\xv_d), \\
\ds\lag^{(2)}_{\bp\bq}(\xv\, |\,\xv_a,\xv_b,\xv_c,\xv_d)=\olam_{pq}(\xv,\xv_b)+ \olam_{p'q'}(\xv,\xv_c)+ \Lambda_{p'q}(\xv_d,\xv)+ \Lambda_{pq'}(\xv_a,\xv).
\end{gathered}
\end{align}
Due to the anti-symmetry relations \eqref{clasinversion}, the two types of Lagrangian functions in \eqref{action1}, satisfy
\begin{align}
\label{cinv}
\lag^{(1)}_{\bq\bp}(\xv\,|\,\xv_a,\xv_c,\xv_b,\xv_d)+\lag^{(2)}_{\bp\bq}(\xv\,|\, \xv_a,\xv_b,\xv_c,\xv_d)=0.
\end{align}
The saddle points $\xv^{(cl)}_1$, and $\xv^{(cl)}_2$, in \eqref{IRFqce}, are solutions of the respective equations of motion
\begin{align}
\frac{\partial}{\partial \xv}\,\lag^{(1)}_{\bp\bq}(\xv\, |\,\xv_a,\xv_b,\xv_c,\xv_d)=0,\qquad\frac{\partial}{\partial \xv}\,\lag^{(2)}_{\bp\bq}(\xv\, |\,\xv_a,\xv_b,\xv_c,\xv_d)=0,
\end{align}
which are determined for fixed values of the variables $\xv_a,\xv_b,\xv_c,\xv_d$.  These equations may be written explicitly as
\begin{align}
\label{vareqs}
\begin{gathered}
\ds\ovphi_{pq}(\xv_c,\xv)+\ovphi_{p'q'}(\xv_b,\xv)+\varphi_{p'q}(\xv,\xv_a)+\varphi_{pq'}(\xv,\xv_d)=0, \\
\ds\ovphi_{pq}(\xv,\xv_b)+\ovphi_{p'q'}(\xv,\xv_c)+\varphi_{p'q}(\xv_d,\xv)+\varphi_{pq'}(\xv_a,\xv)=0,
\end{gathered}
\end{align}
respectively, where the functions $\varphi_{pq}(\xv_i,\xv_j)$, and $\ovphi_{pq}(\xv_i,\xv_j)$, are defined by
\begin{align}
\label{gg-def}
\begin{gathered}
\ds\varphi_{pq}(\xv_i,\xv_j) =\ds\frac{\partial}{\partial \xv_i}\,\Lambda_{pq}(\xv_i,\xv_j)=-\frac{\partial}{\partial \xv_i}\,\Lambda_{qp}(\xv_j,\xv_i),\\
\ds\ovphi_{pq}(\xv_i,\xv_j) =\ds\frac{\partial}{\partial \xv_j}\,\olam_{pq}(\xv_i,\xv_j)=-\frac{\partial}{\partial \xv_j}\,\olam_{qp}(\xv_j,\xv_i).
\end{gathered}
\end{align}
The latter functions are the analogues of ``three-leg'' functions for ABS equations which arise from the quasi-classical limit of the star-triangle relation.

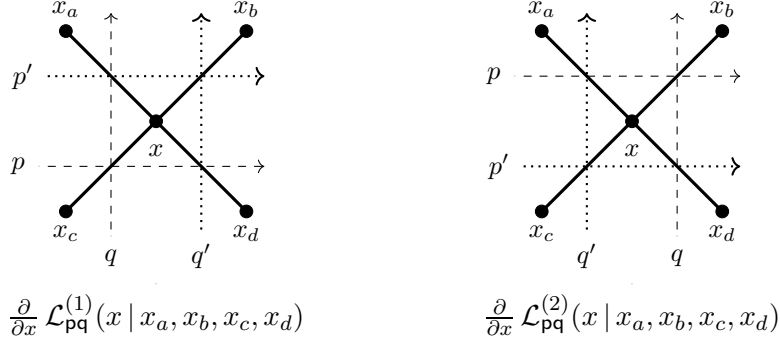
\begin{figure}[htb!]
\centering
\begin{tikzpicture}[scale=1.2]

\draw[-,very thick] (-1,-1)--(1,1);
\draw[-,very thick] (1,-1)--(-1,1);

\filldraw[fill=black!,draw=black!] (-1,-1) circle (2pt)
node[below=1.5pt]{\color{black}\small $\xv_c$};
\filldraw[fill=black!,draw=black!] (1,-1) circle (2pt)
node[below=1.5pt]{\color{black}\small $\xv_d$};
\filldraw[fill=black!,draw=black!] (-1,1) circle (2pt)
node[above=1.5pt]{\color{black}\small $\xv_a$};
\filldraw[fill=black!,draw=black!] (1,1) circle (2pt)
node[above=1.5pt]{\color{black}\small $\xv_b$};

\filldraw[fill=black!,draw=black!] (0,0) circle (2pt)
node[below=5pt]{\color{black}\small $\xv$};

\draw[->,thick,dotted] (-1.2,0.5) -- (1.2,0.5);
\draw[black!] (-1.2,0.5) circle (0.01pt)
node[left=1.5pt]{\color{black}\small $p'$};
\draw[->,dashed] (-1.2,-0.5) -- (1.2,-0.5);
\draw[black!] (-1.3,-0.5) circle (0.01pt)
node[left=1.5pt]{\color{black}\small $p$};
\draw[->,dashed] (-0.5,-1.2) -- (-0.5,1.2);
\draw[black!] (-0.5,-1.27) circle (0.01pt)
node[below=1.5pt]{\color{black}\small $q$};
\draw[->,thick,dotted] (0.5,-1.2) -- (0.5,1.2);
\draw[black!] (0.5,-1.2) circle (0.01pt)
node[below=1.5pt]{\color{black}\small $q'$};

\draw[black!] (0,-1.8) circle (0.01pt)
node[below=0.1pt]{\color{black} $\frac{\partial}{\partial \xv}\,\lag^{(1)}_{\bp\bq}(\xv\, |\,\xv_a,\xv_b,\xv_c,\xv_d)$};

\begin{scope}[xshift=150pt]

\draw[-,very thick] (-1,-1)--(1,1);
\draw[-,very thick] (1,-1)--(-1,1);

\filldraw[fill=black!,draw=black!] (-1,-1) circle (2pt)
node[below=1.5pt]{\color{black}\small $\xv_c$};
\filldraw[fill=black!,draw=black!] (1,-1) circle (2pt)
node[below=1.5pt]{\color{black}\small $\xv_d$};
\filldraw[fill=black!,draw=black!] (-1,1) circle (2pt)
node[above=1.5pt]{\color{black}\small $\xv_a$};
\filldraw[fill=black!,draw=black!] (1,1) circle (2pt)
node[above=1.5pt]{\color{black}\small $\xv_b$};

\filldraw[fill=black!,draw=black!] (0,0) circle (2pt)
node[below=5pt]{\color{black}\small $\xv$};

\draw[->,dashed] (-1.2,0.5) -- (1.2,0.5);
\draw[black!] (-1.3,0.5) circle (0.01pt)
node[left=1.5pt]{\color{black}\small $p$};
\draw[->,thick,dotted] (-1.2,-0.5) -- (1.2,-0.5);
\draw[black!] (-1.2,-0.5) circle (0.01pt)
node[left=1.5pt]{\color{black}\small $p'$};
\draw[->,thick,dotted] (-0.5,-1.2) -- (-0.5,1.2);
\draw[black!] (-0.5,-1.2) circle (0.01pt)
node[below=1.5pt]{\color{black}\small $q'$};
\draw[->,dashed] (0.5,-1.2) -- (0.5,1.2);
\draw[black!] (0.5,-1.27) circle (0.01pt)
node[below=1.5pt]{\color{black}\small $q$};

\draw[black!] (0,-1.8) circle (0.01pt)
node[below=0.1pt]{\color{black} $\frac{\partial}{\partial \xv}\,\lag^{(2)}_{\bp\bq}(\xv\, |\,\xv_a,\xv_b,\xv_c,\xv_d)$};

\end{scope}
\end{tikzpicture}

\caption{Edge configurations for classical discrete Laplace equations \eqref{vareqs}.  These equations may be interpreted as ``quad equations'' defined on elementary four-squares with vertices $a,b,c,d$, and this approach is considered in more detail in Section \ref{sec:Mcomp}.}
\label{fig:laplace}
\end{figure}

The equations \eqref{vareqs} may be interpreted as classical five-point Laplace-type equations, {\it e.g.} \cite{MR2467378,BG11}, which are respectively centered at spins $\xv_1^{(cl)}$, and $\xv_2^{(cl)}$, as in Figure \ref{fig:laplace}, where $\xv_a,\xv_b,\xv_c,\xv_d$ are taken to be fixed.  The equations depicted in Figure \ref{fig:laplace}, are a classical counterpart of the IRF Boltzmann weight configuration appearing in Figure \ref{fig-IRF}.   The equations \eqref{vareqs} may also be interpreted as ``quad equations'' \cite{BobSurQuadGraphs}, which are defined on elementary four-squares with vertices $a,b,c,d$, and this approach will be considered in some more detail (for multi-component equations) in Section \ref{sec:Mcomp}. 

Using \eqref{IRFqce}, the quasi-classical expansion of the star-star relation \eqref{star-star}, results in the following {\it classical star-star relation} \cite{Bazhanov:2011mz} at leading order $O(\hbar^{-1})$
\begin{align}
\label{cstar}
\begin{split}
&\ds\lag^{(1)}_{\bp\bq}(\xv^{(cl)}_1\,|\,\xv_a,\xv_b,\xv_c,\xv_d )+ \Lambda_{q'q}(\xv_d,\xv_c)+ \Lambda_{qq'}(\xv_a,\xv_b)\\
&\ds\phantom{xx}= \lag^{(2)}_{\bp\bq}(\xv^{(cl)}_2 \,|\,\xv_a,\xv_b,\xv_c,\xv_d) +\Lambda_{p'p}(\xv_c,\xv_a)+\Lambda_{pp'}(\xv_b,\xv_d),
\end{split}
\end{align}
that is satisfied on solutions of \eqref{vareqs}.  Note that the classical star-star relation \eqref{cstar}, also implies the following four constraints,
\begin{align}
\label{vareq_new}
\begin{split}
\ds\varphi_{p'q}(\xv^{(cl)}_1,\xv_a)+\varphi_{pq'}(\xv_a,\xv^{(cl)}_2)-\varphi_{q'q}(\xv_b,\xv_a)-\varphi_{p'p}(\xv_c,\xv_a)=0,\\
\ds\ovphi_{pq'}(\xv^{(cl)}_1,\xv_d)+\ovphi_{p'q}(\xv_d,\xv^{(cl)}_2)-\varphi_{q'q}(\xv_d,\xv_c)-\varphi_{p'p}(\xv_d,\xv_b)=0,\\
\ovphi_{p'q'}(\xv_b,\xv^{(cl)}_1)+\ovphi_{pq}(\xv^{(cl)}_2,\xv_b)-\varphi_{q'q}(\xv_b,\xv_a)-\varphi_{p'p}(\xv_d,\xv_b)=0,\\
\ovphi_{pq}(\xv_c,\xv^{(cl)}_1)+\ovphi_{p'q'}(\xv^{(cl)}_2,\xv_c)-\varphi_{q'q}(\xv_d,\xv_c)-\varphi_{p'p}(\xv_c,\xv_a)=0,
\end{split}
\end{align}
between five spins centered at $\xv_a,\xv_d,\xv_b,\xv_c$, respectively.  These equations are obtained by taking the derivative of \eqref{cstar} with respect to each of the spins $\xv_a,\xv_d,\xv_b,\xv_c$, which are stationary with respect to the saddle points $\xv^{(cl)}_1$ and $\xv^{(cl)}_2$.


The quasi-classical expansion of the Yang-Baxter equation \eqref{YBE-IRF}, results in the following {\it classical Yang-Baxter equation}
\begin{align}
\label{cybe}
\begin{split}
&\ds\lag^{(1)}_{\bp\bq}(\xv^{(cl)}_1\,|\,\xv_c,\xv^{(cl)},\xv_e,\xv_d )+\lag^{(1)}_{\bp\br}(\xv^{(cl)}_2\,|\,\xv^{(cl)},\xv_b,\xv_d,\xv_f )+\lag^{(1)}_{\bq\br}(\xv^{(cl)}_3\,|\,\xv_c,\xv_a,\xv^{(cl)},\xv_b ) \\
&\ds\phantom{xxx}+\Lambda_{qq'}(\xv_c,\xv^{(cl)})+\Lambda_{q'q}(\xv_b,\xv_a)+ \Lambda_{qq'}(\xv'^{(cl)},\xv_f)+\Lambda_{q'q}(\xv_d,\xv_e)\\[0.3cm]
&\ds\phantom{\xv}=\lag^{(1)}_{\bq\br}(\xv'^{(cl)}_3\,|\,\xv_e,\xv'^{(cl)},\xv_d,\xv_f )+\lag^{(1)}_{\bp\br}(\xv'^{(cl)}_2\,|\,\xv_c,\xv_a,\xv_e,\xv'^{(cl)} )+\lag^{(1)}_{\bp\bq}(\xv'^{(cl)}_1\,|\,\xv_a,\xv_b,\xv'^{(cl)},\xv_f ) ,
\end{split}
\end{align}
where $\xv_a,\xv_b,\xv_c,\xv_d,\xv_e,\xv_f$, are fixed, and the $\xv^{(cl)}_1$, $\xv^{(cl)}_3$, $\xv^{(cl)}_5$, $\xv'^{(cl)}_1$, $\xv'^{(cl)}_3$, $\xv'^{(cl)}_5$ are solutions of their respective equations of motion of the type appearing in \eqref{vareqs}, while the $\xv^{(cl)}$ and $\xv'^{(cl)}$ solve respectively
\begin{align}
\label{cybesp}
\begin{gathered}
\ds\ovphi_{p'q'}(\xv,\xv^{(cl)}_1)+\ovphi_{qr}(\xv,\xv^{(cl)}_5)+\varphi_{qq'}(\xv_c,\xv)+\varphi_{p'r}(\xv^{(cl)}_3,\xv)=0,\\
\ds\ovphi_{q'r'}(\xv,\xv'^{(cl)}_1)+\ovphi_{pq}(\xv,\xv'^{(cl)}_5)+\varphi_{pr'}(\xv'^{(cl)}_3,\xv)+\varphi_{q'q}(\xv_f,\xv)=0.
\end{gathered}
\end{align}
Similarly to \eqref{vareq_new}, by taking derivatives with respect to the spins $\xv_a,\xv_b,\xv_c,\xv_d,\xv_e,\xv_f$, there are six additional constraints involving $\varphi$, $\ovphi$, that are centered at each of the six latter spins respectively.  Note that as for the statistical mechanics case \eqref{YBE-IRF}, the classical Yang-Baxter equation \eqref{cybe} is already implied by the classical star-star relation \eqref{cstar} (independently of the quasi-classical expansion of \eqref{YBE-IRF}), and in this sense the equation \eqref{cstar} can be considered as more fundamental.  As seen above (and also in some previously considered specific cases \cite{Bazhanov:2011mz,KelsThesis}), starting from the expansion of Boltzmann weights \eqref{weight-exp}, one may obtain a rather rich structure of classical equations from the quasi-classical expansion of the IRF model.

\subsection{Classical discrete Laplace system of equations}

For the IRF model, taking $\hbar\rightarrow0$ corresponds to a ground state configuration, which is determined by solving the equation of motion of an action functional that arises in the leading order quasi-classical expansion of the partition function. In the following, such a quasi-classical limit will be considered for the IRF models of the previous section, which were defined on $\sigma_0$ and $\sigma$ respectively.

First for the case of the IRF model on $\sigma_0$, using the quasi-classical expansion \eqref{IRFqce}, the first expression for the partition function of the IRF model \eqref{IRFZ1} may be evaluated with a saddle point method, which leads to the expansion
\begin{align}
\label{zexp0}
\log Z_0 = -\hbar^{-1}\, \mathcal{A}_0(\xv_0^{(cl)})+O(\log\hbar).
\end{align}
Here the action functional $\mathcal{A}_0(\xv)$ is defined by
\begin{align}
\label{caction1}
\mathcal{A}_0(\x)=\sum_{\sigma^4_{ij}} \lag^{(1)}_{\bp\bq}\,(\xv^{(cl)}_i\,|\,\xv_a,\xv_b,\xv_c,\xv_d),
\end{align}
where the sum is taken over all elementary four-squares $\sigma^4_{ij}\in F^{(4)}(\sigma_0)$, of the type appearing on the right hand side of Figure \ref{e4s}.  The $\xv_i^{(cl)}$ are the solutions of the equations
\begin{align}
\label{Euler-a}
\frac{\partial}{\partial \xv_i}\, \lag^{(1)}_{\bp\bq}\, (\xv_i\,|\, \xv_a,\xv_b,\xv_c,\xv_d)=0,\quad\forall\, i\in V^{(1)}(L)\cap V_{int}(L),
\end{align}
while the $\x_0^{(cl)}$ is the saddle point of the partition function, given by the solution of
\begin{align}
\label{Euler-b}
\frac{\partial}{\partial \xv_i}\, \lag^{(2)}_{\bp\bq}\, (\xv_i\,|\, \xv_a,\xv_b,\xv_c,\xv_d)=0,\quad\forall\, i\in V^{(2)}(L)\cap V_{int}(L).
\end{align}

Together, the equations \eqref{Euler-a}, and \eqref{Euler-b}, constitute a set of constraints on each interior vertex $i\in V_{int}(L)$, under fixed boundary conditions, which may be interpreted as a general form of the classical discrete Laplace equations.  Note that the quasi-classical expansion for the second expression of the IRF partition function \eqref{IRFZ2}, leads to exactly the same system of classical Laplace equations \eqref{Euler-a}, \eqref{Euler-b}.

For the case of the IRF model on $\sigma$, there are five different Lagrangian functions associated to each of the five different types elementary four-squares in $F^{(4)}(\sigma)$.  Specifically, the Lagrangian functions are given by
\begin{align}
\begin{gathered}
\ds\lag^{(1)}_{\bp\bq}(\xv\,|\,\xv_a,\xv_b,\xv_c,\xv_d),\\
\ds\lag^{(1)}_{\bp\br}(\xv\,|\,\xv_a,\xv_b,\xv_c,\xv_d),\\
\ds\lag^{(2)}_{\br\bp}(\xv\,|\,\xv_a,\xv_b,\xv_c,\xv_d),\\
\ds\lag^{(3)}_{\bq\br}(\xv\,|\,\xv_a,\xv_b,\xv_c,\xv_d):=\lag^{(2)}_{\bq\br}(\xv\,|\,\xv_a,\xv_b,\xv_c,\xv_d)+\Lambda_{rr'}(\xv_c,\xv_d)+\Lambda_{r'r}(\xv_b,\xv_a),\\
\ds\lag^{(4)}_{\br\bq}(\xv\,|\,\xv_a,\xv_b,\xv_c,\xv_d):=\lag^{(1)}_{\br\bq}(\xv\,|\,\xv_a,\xv_b,\xv_c,\xv_d)+\Lambda_{rr'}(\xv_a,\xv_c)+\Lambda_{r'r}(\xv_d,\xv_b),
\end{gathered}
\end{align}
which are associated to elementary four-squares $\sigma^4_{ij}$, $\sigma^4_{jk}$, $\sigma^4_{kj}$, $\sigma^4_{ki}$, $\sigma^4_{ik}$ respectively, that are pictured in Figures \ref{e4s}, and \ref{4IRFBW}.  Note that due to \eqref{cinv}, pairs of Lagrangians on elementary four-squares $\sigma^4_{ik}(\n)$, and $\sigma^4_{ki}(\n)$, or respectively on $\sigma^4_{jk}(\n)$, and $\sigma^4_{kj}(\n)$, sum to zero.

Similarly to \eqref{zexp0}, the quasi-classical expansion of the partition function \eqref{zIRFsigma} of the IRF model on $\sigma$, is given by
\begin{align}
\log Z=-\hbar^{-1}\mathcal{A}(\x^{(cl)})+O(\hbar^0),
\end{align}
where the action functional $\mathcal{A}(\x)$ is given by
\begin{align}
\label{caction2}
\begin{split}
&\ds \mathcal{A}(\x)=\sum_{\sigma^4_{ij}}\lag_{\bp\bq}^{(1)}(\xv_{i_0}^{(cl)}\,|\,\xv_a,\xv_b,\xv_c,\xv_d)+\sum_{\sigma^4_{jk}}\lag_{\bp\br}^{(1)}(\xv_{i_1}^{(cl)}\,|\,\xv_a,\xv_b,\xv_c,\xv_d) \\
\ds&+\sum_{\sigma^4_{kj}}\,\lag_{\br\bp}^{(2)}(\xv_{i_2}^{(cl)}\,|\,\xv_a,\xv_b,\xv_c,\xv_d)+\sum_{\sigma^4_{ki}}\lag_{\bq\br}^{(3)}(\xv_{i_3}^{(cl)}\,|\,\xv_a,\xv_b,\xv_c,\xv_d) +\sum_{\sigma^4_{ik}}\lag_{\br\bq}^{(4)}(\xv_{i_4}^{(cl)}\,|\,\xv_a,\xv_b,\xv_c,\xv_d).
\end{split}
\end{align}
Here the sums are taken over the five different types of elementary four-squares in $F^{(4)}(\sigma)$, that are shown in Figure \ref{4IRFBW}, and on the right hand side of Figure \ref{e4s}.  The $\xv_{i_0}^{(cl)}$, $\xv_{i_1}^{(cl)}$, and $\xv_{i_4}^{(cl)}$, solve equations of motion of the first type appearing in \eqref{vareqs}, while the $\xv_{i_2}^{(cl)}$ ,$\xv_{i_3}^{(cl)}$ solve equations of motion of the second type in \eqref{vareqs}.  

The $\xv^{(cl)}$ is the saddle point of the partition function, given by the solution of the equations
\begin{align}
\label{Euler-c}
\frac{\partial}{\partial \xv_i}\mathcal{A}(\x)=0,
\end{align}
determined for all interior variables $\xv_i=\xv_1,\xv_2,\ldots,\xv_m$, that are assigned to vertices $i_1,i_2,\ldots,i_m\in V_{int}(\Gl)$, where each vertex is common to at least two different elementary four-squares of $\sigma$.  That is, the equation \eqref{Euler-c} provides a constraint on each interior variable of $\sigma$, located on the corner vertices of elementary four-squares.

Together, the equations of motion for the spin variables $\xv_{i_0}^{(cl)},\xv_{i_1}^{(cl)},\xv_{i_2}^{(cl)},\xv_{i_3}^{(cl)},\xv_{i_4}^{(cl)}$, and the saddle point equations \eqref{Euler-c}, provide a constraint on each interior vertex $i\in V_{int}(\Gl)$, under fixed boundary conditions.  The latter equations constitute the system of classical discrete Laplace equations on the surface $\sigma$.

\subsection{Z-invariance for the classical discrete Laplace systems}\label{sec:z-invar3}

The Z-invariance that was described in Section \ref{sec:z-invar2}, manifests as {\it classical Z-invariance} for the action functionals \eqref{caction1}, and \eqref{caction2}.  The statement of classical Z-invariance, is that the action functionals $\mathcal{A}_0$, and $\mathcal{A}$,, defined on graphs on $\sigma_0$, and $\sigma$ respectively, are equivalent, as a consequence of the classical Yang-Baxter equation \eqref{cybe}, and inversion relations \eqref{clasinversion}, and \eqref{cinv}.

Similarly to the case of the statistical mechanical model of the previous section, this can be straightforwardly shown, since any positively oriented rapidity lines $r^+,r'^+$, on $\sigma$, will be associated to some three elementary four-squares $\sigma^4_{ij}(\n+\khat)$, $\sigma^4_{ik}(\n)$, $\sigma^4_{jk}(\n+\ihat)$, and any negatively oriented rapidity lines $r,r'$, on $\sigma$, will be associated to some three elementary four-squares $\sigma^4_{ij}(\n)$, $\sigma^4_{ik}(\n+\jhat)$, $\sigma^4_{jk}(\n)$.  On the latter respective groups of three elementary four-squares, the classical Yang-Baxter equation \eqref{cybe} is satisfied, and can be repeatedly used on such groups of three four-squares to deform a surface $\sigma$, until all rapidity lines $r,r'$ are removed, at which stage $\sigma$ will be identical to $\sigma_0$.  As was mentioned above, the contribution to the action $\mathcal{A}$, from any pairs of elementary four-squares $\sigma^4_{ik}(\n)$, and $\sigma^4_{ki}(\n)$, or respectively $\sigma^4_{jk}(\n)$, and $\sigma^4_{kj}(\n)$, that appear through the use of the classical Yang-Baxter equation, will be zero, due to the inversion relations \eqref{clasinversion}, and \eqref{cinv}.

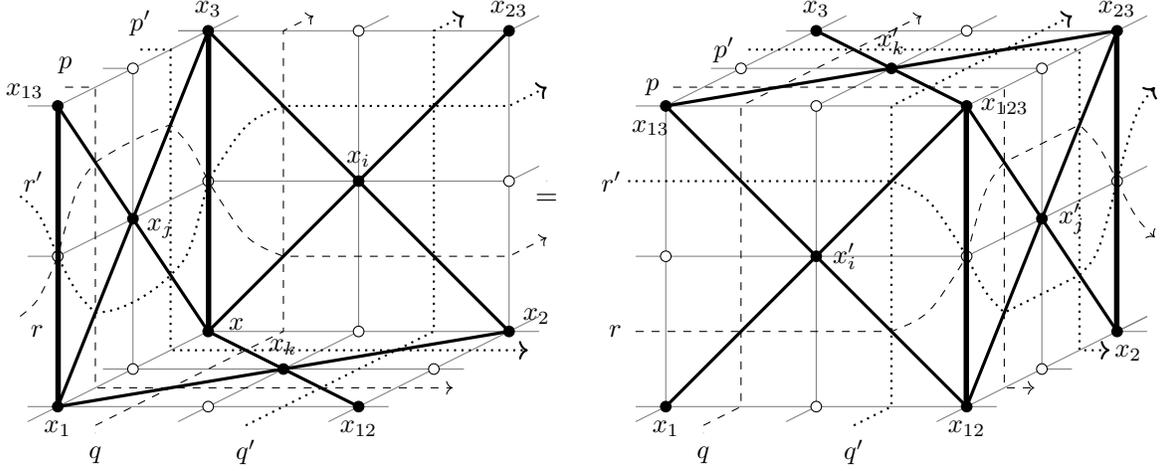
\begin{figure}[htb]
\centering
\begin{tikzpicture}

\begin{scope}[scale=1.0]
\draw[-,gray] (-3.4,-1)--(1.4,-1);\draw[-,gray] (0.6,-1.2)--(3.4,0.2);\draw[-,gray] (2.4,-0.5)--(-2.4,-0.5);\draw[-,gray] (1,0)--(-1.4,-1.2);\draw[-,gray] (-3.4,-1.2)--(-1,0)--(3.4,0);
\draw[-,gray] (-3,-1)--(-3,3)--(-1,4)--(-1,0);\draw[-,gray] (-3.4,1)--(-3,1)--(-1,2);\draw[-,gray] (-2,-0.5)--(-2,3.5)--(-2.4,3.5);\draw[-,gray] (-3.4,3)--(-3,3);\draw[-,gray] (-1.4,4)--(-1,4);
\draw[-,gray] (3,0)--(3,4)--(-1.4,4);\draw[-,gray] (3,4)--(3.4,4.2);\draw[-,gray] (1.4,4.2)--(1,4)--(1,0);\draw[-,gray] (3.4,2.2)--(3,2)--(-1,2);\draw[-,gray] (-1,4)--(-0.6,4.2);
\filldraw[fill=white,draw=black] (1,0) circle (2.0pt);
\filldraw[fill=white,draw=black] (-2.0,-0.5) circle (2.0pt);
\filldraw[fill=white,draw=black] (-1,-1) circle (2.0pt);
\filldraw[fill=white,draw=black] (2.0,-0.5) circle (2.0pt);
\filldraw[fill=black,draw=black] (1,-1) circle (2.0pt)
node[below=1.5pt]{\small $x_{12}$};
\filldraw[fill=black,draw=black] (0,-0.5) circle (2.0pt)
node[above=1.5pt]{\small $x_k$};
\filldraw[fill=black,draw=black] (-1,0) circle (2.0pt);
\fill[black!] (-1,0.1) circle (0.01pt)
node[right=4pt]{\color{black}\small $x$};
\filldraw[fill=black,draw=black] (-3,-1) circle (2.0pt)
node[below=1.5pt]{\small $x_1$};
\filldraw[fill=black,draw=black] (3,0) circle (2.0pt);
\fill[black!] (3,0.2) circle (0.01pt)
node[right=1.5pt]{\color{black}\small $x_2$};
\filldraw[fill=white,draw=black] (-3,1) circle (2.0pt);
\filldraw[fill=white,draw=black] (-1,2) circle (2.0pt);
\filldraw[fill=white,draw=black] (-2,3.5) circle (2.0pt);
\filldraw[fill=black,draw=black] (-3,3) circle (2.0pt);
\fill[black!] (-3,3.2) circle (0.01pt)
node[left=1.5pt]{\color{black}\small $x_{13}$};
\filldraw[fill=black,draw=black] (-1,4) circle (2.0pt)
node[above=1.5pt]{\small $x_3$};
\filldraw[fill=black,draw=black] (-2,1.5) circle (2.0pt);
\fill[black!] (-2,1.37) circle (0.01pt)
node[right=1.5pt]{\color{black}\small $x_j$};
\filldraw[fill=white,draw=black] (3,2) circle (2.0pt);
\filldraw[fill=white,draw=black] (1,4) circle (2.0pt);
\filldraw[fill=black,draw=black] (1,2) circle (2.0pt)
node[above=1.5pt]{\small $x_i$};
\filldraw[fill=black,draw=black] (3,4) circle (2.0pt)
node[above=1.5pt]{\small $x_{23}$};
\draw[-,very thick] (-3,-1)--(3,0);\draw[-,very thick] (-1,0)--(1,-1);
\draw[-,very thick] (-3,-1)--(-1,4);\draw[-,very thick] (-3,3)--(-1,0);
\draw[-,very thick] (-1,0)--(3,4);\draw[-,very thick] (-1,4)--(3,0);
\draw[-,very thick] (-2.98,-1)--(-2.98,3);\draw[-,very thick] (-0.98,0)--(-0.98,4);
\draw[-,very thick] (-3.01,-1)--(-3.01,3);\draw[-,very thick] (-1.01,0)--(-1.01,4);
\draw[->,black,thick,dotted] (-1.9,3.75)--(-1.5,3.75)--(-1.5,-0.25)--(3.25,-0.25);
\draw[->,black,dashed] (-2.5,-1.25)--(0,0)--(0,4)--(0.4,4.2);
\draw[->,black,dashed] (-2.9,3.25)--(-2.5,3.25)--(-2.5,-0.75)--(2.25,-0.75);
\draw[->,black,thick,dotted] (-0.5,-1.25)--(2,0)--(2,4)--(2.4,4.2);
\draw[->,black,dashed] (-3.5,0.2) .. controls (-3.25,0.4) ..(-3,1) .. controls (-2.75,1.95) .. (-2.5,2.25) .. controls (-2.2,2.45) and (-1.8,2.65) .. (-1.5,2.75) .. controls (-1.25,2.5) .. (-1,2) .. controls (-0.5,1.2) .. (0,1) -- (3,1) -- (3.5,1.25);
\draw[->,black,thick,dotted] (-3.5,1.8) .. controls (-3.25,1.6) .. (-3,1) .. controls (-2.75,0.5) .. (-2.5,0.25) .. controls (-2.2,0.25) and (-1.8,0.45) .. (-1.5,0.75) .. controls (-1.25,1) .. (-1,2) .. controls (-0.5,2.8) .. (0,3) -- (3,3) -- (3.5,3.25);
\draw[black] (-1.9,3.75) circle (0.01pt)
node[above=1.5pt]{\color{black}\small $p'$};
\draw[black] (-2.9,3.25) circle (0.01pt)
node[above=1.5pt]{\color{black}\small $p$};
\draw[black] (-2.5,-1.35) circle (0.01pt)
node[below=1.5pt]{\color{black}\small $q$};
\draw[black] (-0.5,-1.25) circle (0.01pt)
node[below=1.5pt]{\color{black}\small $q'$};
\draw[black] (-3,2) circle (0.01pt)
node[left=1.5pt]{\color{black}\small $r'$};
\draw[black] (-3,0) circle (0.01pt)
node[left=1.5pt]{\color{black}\small $r$};
\end{scope}

\draw[black] (3.5,2) circle (0.01pt)
node[below=1pt]{\color{black}$=$};

\begin{scope}[scale=1.0,xshift=230,yshift=0,rotate=0]
\draw[-,gray] (-1.4,-1.2)--(-1,-1)--(1,-1)--(1,1)--(-1,1)--(-1,-1)--(-3,-1)--(-3,1)--(-1,1)--(-1,3)--(1,4);
\draw[-,gray] (1,3)--(3,4)--(3,0)--(2,-0.5)--(2,1.5)--(1,1)--(1,3)--(-3,3)--(-3,1);
\draw[-,gray] (1.4,-1)--(1,-1)--(2,-0.5)--(2.4,-0.5);
\draw[-,gray] (-3,3)--(-1,4)--(3,4);
\draw[-,gray] (-2,3.5)--(2,3.5)--(2,1.5)--(3,2);
\draw[-,gray] (3.4,0)--(3,0)--(3.4,0.2);\draw[-,gray] (-3.4,-1)--(-3,-1)--(-3.4,-1.2);\draw[-,gray] (1,-1)--(0.6,-1.2);
\draw[-,gray] (-3,1)--(-3.4,1);\draw[-,gray] (-3,3)--(-3.4,3);\draw[-,gray] (-2,3.5)--(-2.4,3.5);\draw[-,gray] (-1,4)--(-1.4,4);
\draw[-,gray] (3,0)--(3.4,0.2);\draw[-,gray] (3,2)--(3.4,2.2);\draw[-,gray] (3,4)--(3.4,4.2);\draw[-,gray] (1,4)--(1.4,4.2);\draw[-,gray] (-1,4)--(-0.6,4.2);
\filldraw[fill=black,draw=black] (-3,3) circle (2.0pt);
\fill[black!] (-3.2,3) circle (0.01pt)
node[below=1.5pt]{\color{black}\small $x_{13}$};
\filldraw[fill=white,draw=black] (-1,3) circle (2.0pt);
\filldraw[fill=black,draw=black] (1,3) circle (2.0pt)
node[right=1.5pt]{\small $x_{123}$};
\filldraw[fill=white,draw=black] (-2,3.5) circle (2.0pt);
\filldraw[fill=black,draw=black] (0,3.5) circle (2.0pt)
node[above=1.5pt]{\small $x'_k$};
\filldraw[fill=white,draw=black] (2,3.5) circle (2.0pt);
\filldraw[fill=black,draw=black] (-1,4) circle (2.0pt)
node[above=1.5pt]{\small $x_3$};
\filldraw[fill=white,draw=black] (1,4) circle (2.0pt);
\filldraw[fill=black,draw=black] (3,4) circle (2.0pt)
node[above=1.5pt]{\small $x_{23}$};
\filldraw[fill=black,draw=black] (3,0) circle (2.0pt);
\fill[black!] (3.15,0) circle (0.01pt)
node[below=1.5pt]{\color{black}\small $x_2$};
\filldraw[fill=white,draw=black] (3,2) circle (2.0pt);
\filldraw[fill=black,draw=black] (-3,-1) circle (2.0pt)
node[below=1.5pt]{\small $x_1$};
\filldraw[fill=white,draw=black] (-3,1) circle (2.0pt);
\filldraw[fill=white,draw=black] (-1,-1) circle (2.0pt);
\filldraw[fill=white,draw=black] (1,1) circle (2.0pt);
\filldraw[fill=white,draw=black] (2,-0.5) circle (2.0pt);
\filldraw[fill=black,draw=black] (-1,1) circle (2.0pt)
node[right=2.5pt]{\small $x'_i$};
\filldraw[fill=black,draw=black] (1,-1) circle (2.0pt)
node[below=1.5pt]{\small $x_{12}$};
\filldraw[fill=black,draw=black] (2,1.5) circle (2.0pt)
node[right=2.5pt]{\small $x'_j$};
\draw[-,very thick] (-3,-1)--(1,3);\draw[-,very thick] (-3,3)--(1,-1);
\draw[-,very thick] (1,-1)--(3,4);\draw[-,very thick] (1,3)--(3,0);
\draw[-,very thick] (-3,3)--(3,4);\draw[-,very thick] (-1,4)--(1,3);
\draw[-,very thick] (0.98,-1)--(0.98,3);\draw[-,very thick] (2.98,0.0)--(2.98,4.0);
\draw[-,very thick] (1.01,-1)--(1.01,3);\draw[-,very thick] (3.01,0.0)--(3.01,4.0);
\draw[->,black,dashed] (-2.9,3.25)--(1.5,3.25)--(1.5,-0.75)--(1.9,-0.75);
\draw[->,black,thick,dotted] (-0.5,-1.25)--(0,-1)--(0,3)--(2,4)--(2.4,4.2);
\draw[->,black,dashed] (-3.4,0)--(-2,0)--(0,0) .. controls (0.5,0.2) .. (1,1) .. controls (1.25,1.95) .. (1.5,2.25) -- (2.5,2.75) .. controls (2.75,2.5) .. (3.0,2.0) .. controls (3.25,1.5) .. (3.5,1.25);  
\draw[->,black,thick,dotted] (-1.9,3.75)--(2.5,3.75)--(2.5,-0.25)--(2.9,-0.25);
\draw[->,black,dashed] (-2.5,-1.25)--(-2,-1)--(-2,3)--(0,4)--(0.4,4.2);
\draw[->,black,thick,dotted] (-3.5,2)--(-2,2)--(0,2) .. controls (0.5,1.8) .. (1,1) .. controls (1.25,0.5) .. (1.5,0.25) -- (2.5,0.75) .. controls (2.75,1.0) .. (3.0,2.0) .. controls (3.25,2.95) .. (3.5,3.25); 
\draw[black] (-1.9,3.75) circle (0.01pt)
node[left=1.5pt]{\color{black}\small $p'$};
\draw[black] (-2.9,3.25) circle (0.01pt)
node[left=1.5pt]{\color{black}\small $p$};
\draw[black] (-2.5,-1.35) circle (0.01pt)
node[below=1.5pt]{\color{black}\small $q$};
\draw[black] (-0.5,-1.25) circle (0.01pt)
node[below=1.5pt]{\color{black}\small $q'$};
\draw[black] (-3.4,2) circle (0.01pt)
node[left=1.5pt]{\color{black}\small $r'$};
\draw[black] (-3.4,0) circle (0.01pt)
node[left=1.5pt]{\color{black}\small $r$};
\end{scope}

\end{tikzpicture}
\caption{Classical Yang-Baxter equation \eqref{cybe} as ``closure relation''. }
\label{IRFclosure}
\end{figure}

The property of classical Z-invariance may also be interpreted as a classical closure property \cite{LobbNijhoff} of the action \eqref{caction1}, under deformations of elementary four-squares of $\sigma_0$.  Indeed, note that after exchanging $\bq\leftrightarrow\br$, and making the following change of variables
\begin{align}
\begin{gathered}
\ds \xv_d\to \xv_1,\; \xv_f\to \xv_{12},\; \xv^{(cl)}\to \xv,\; \xv_e\to \xv_{13},\; \xv_b\to \xv_{2},\; \xv'^{(cl)}\to \xv_{123},\; \xv_c\to \xv_{3},\; \xv_a\to \xv_{23},\\
\ds \xv_1^{(cl)}\to \xv_j,\; \xv'^{(cl)}_1\to \xv'_j,\; \xv_2^{(cl)}\to \xv_k,\; \xv'^{(cl)}_2\to \xv'_k,\; \xv_3^{(cl)}\to \xv_i,\; \xv'^{(cl)}_3\to \xv'_i,
\end{gathered}
\end{align}
the classical Yang-Baxter equation \eqref{cybe} may be written in the form
\begin{align}
\label{cybe2}
\Delta_i\lag^{(1)}_{\br\bq}(\xv_i\,|\,\xv_{3},\xv_{23},\xv,\xv_2)+\Delta_j\lag^{(1)}_{\bp\br}(\xv_j\,|\,\xv_{3},\xv,\xv_{13},\xv_{1})+\Delta_k\lag^{(1)}_{\bp\bq}(\xv_k\,|\,\xv,\xv_{2},\xv_1,\xv_{12})=0,
\end{align}
where   
\begin{align}
\begin{split}
\ds\Delta_i\lag^{(1)}_{\br\bq}(\xv_i\,|\,\xv_{3},\xv_{23},\xv,\xv_2):=&\lag^{(1)}_{\br\bq}(\xv'_i\,|\,\xv_{13},\xv_{123},\xv_1,\xv_{12})-\lag^{(1)}_{\br\bq}(\xv_i\,|\,\xv_{3},\xv_{23},\xv,\xv_2),\\
\Delta_j\lag^{(1)}_{\bp\br}(\xv_j\,|\,\xv_{3},\xv,\xv_{13},\xv_{1}):=&\lag^{(1)}_{\bp\br}(\xv'_j\,|\,\xv_{23},\xv_2,\xv_{123},\xv_{12})+\Lambda_{rr'}(x_{23},x_2)+\Lambda_{r'r}(x_{123},x_{12}) \\
&-(\lag^{(1)}_{\bp\br}(\xv_j\,|\,\xv_{3},\xv,\xv_{13},\xv_{1})+\Lambda_{rr'}(x_3,x)+\Lambda_{r'r}(x_{13},x_1)), \\
\Delta_k\lag^{(1)}_{\bp\bq}(\xv_k\,|\,\xv,\xv_{2},\xv_1,\xv_{12}):=&\lag^{(1)}_{\bp\bq}(\xv'_k\,|\,\xv_3,\xv_{23},\xv_{13},\xv_{123})-\lag^{(1)}_{\bp\bq}(\xv_k\,|\,\xv,\xv_{2},\xv_1,\xv_{12}),
\end{split}
\end{align}
and where $x$, $x_{123}$, $x_i,x'_ix_j,x'_j,x_k,x'_k$, are required to satisfy their respective saddle point equations of the form \eqref{vareqs}, \eqref{cybesp}.  The equation \eqref{cybe2} is shown graphically in Figure \ref{IRFclosure}.  In the form \eqref{cybe2}, the classical Yang-Baxter equation \eqref{cybe} more clearly represents a local closure property \cite{LobbNijhoff}, for the action \eqref{caction1} of the variational Laplace system of equations of $\sigma$.

\subsection{Quad equation interpretation}
\label{sec:Mcomp}


For the previously studied cases of the quasi-classical limit of scalar solutions of the star-triangle relation  \cite{Bazhanov:2007mh,Bazhanov:2010kz,Bazhanov:2016ajm,Kels:2017vbc,Kels:2018xge}, the equation for the critical point of the latter relation was always found to be equivalent to a 3D-consistent integrable quad equation from the ABS classification \cite{ABS,ABS2}.  This connection means that the star-triangle relation itself has a natural interpretation as being a quantum counterpart (in a path integral sense) of a discrete integrable equation.  On the other hand, there are solutions of the star-star relation \cite{Bazhanov:1990qk,Bazhanov:2011mz}, that provide multi-component variable generalisations of solutions of the star-triangle relation.  For these cases, the equations of motion for the IRF Boltzmann weights given in \eqref{vareqs}, might be expected to  correspond to some new types of multi-component discrete integrable quad equations.

As an explicit example, the following multi-component equations are hyperbolic degenerations \cite{KelsThesis} of the classical equations of motion for IRF Boltzmann weights corresponding to \eqref{vareqs}, which come from the elliptic gamma function solutions of the star-star relation \cite{Bazhanov:2011mz}
\begin{align}
\label{IRFEMO}
\begin{split}
\ds\varphi^{(1)}_{\bp\bq}((x^{(cl)}_1)_k\,|\,\x_a,\x_b,\x_c,\x_d)-\varphi^{(1)}_{\bp\bq}((x^{(cl)}_1)_{k+1}\,|\,\x_a,\x_b,\x_c,\x_d)=0,\quad k=1,\ldots,n-1,
\end{split}
\end{align}
where
\begin{align}
\label{q3d14leg}
\begin{split}
\ds\varphi^{(1)}_{\bp\bq}(x_j\,|\,\x_a,\x_b,\x_c,\x_d)=\sum_{k=1}^n&\log\frac{\sinh((x_c)_k-x_j-\ii(p-q))\sinh((x_b)_k-x_j-\ii(p'-q'))}{\cosh((x_a)_k-x_j-\ii(p'-q))\cosh((x_d)_k-x_j-\ii(p-q'))},
\end{split}
\end{align}
and the components of each variable $\x_i=((x_i)_1,\ldots,(x_i)_n)$, are subject to the constraint
\begin{align}
\label{sumconstraint}
\sum_{k=1}^n(x_i)_k=0.
\end{align}
However, besides the connection of the above equations with the Yang-Baxter equation, it is not yet clear what other integrable properties they possess.  This includes whether they satisfy a 3D-consistency property, as for the cases of the star-triangle relation, or even how this property would be formulated for these equations, which have a different form than what is usually considered for a quad equation.  The purpose of this subsection is to study the equations \eqref{IRFEMO} in a quad equation formulation, and particularly in the context of 3D-consistency.  While the equations \eqref{IRFEMO} themselves do not appear to be 3D-consistent, they will motivate the construction of some related types of multi-component quad equations which do satisfy a 3D-consistency property.

\subsubsection{Quad equation with vector variables}

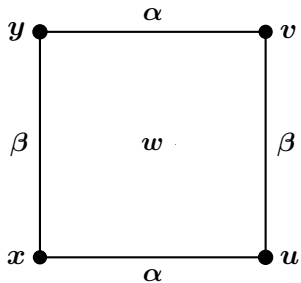
\begin{figure}[tbh]
\centering
\begin{tikzpicture}[scale=1.5]
\draw[thick,-] (3,-1)--(3,1);
\%draw[thick,-] (3,-1)--(5,1);
\draw[thick,-] (5,-1)--(3,-1);
\draw[-,thick] (3,1)--(5,1);
\draw[-,thick] (5,-1)--(5,1);
\fill (3,0) circle (0.1pt)
node[left=0.5pt]{\color{black}\small $\bt$};
\fill (5,0) circle (0.1pt)
node[right=0.5pt]{\color{black}\small $\bt$};
\fill (4,-1) circle (0.1pt)
node[below=0.5pt]{\color{black}\small $\al$};
\fill (4,1) circle (0.1pt)
node[above=0.5pt]{\color{black}\small $\al$};
\fill (4.2,0) circle (0.1pt)
node[left=0.5pt]{\color{black}\small $\w$};
\fill (3,-1) circle (1.8pt)
node[left=1.5pt]{\color{black} $\x$};
\filldraw[fill=black,draw=black] (3,1) circle (1.8pt)
node[left=1.5pt]{\color{black} $\y$};
\fill (5,1) circle (1.8pt)
node[right=1.5pt]{\color{black} $\bv$};
\filldraw[fill=black,draw=black] (5,-1) circle (1.8pt)
node[right=1.5pt]{\color{black} $\bu$};
\end{tikzpicture}
\caption{Quadrilateral associated to $n$-component quad equations considered in this sectiom.  There are $n$-component vector variables $\x$, $\bu$, $\y$, $\bv$ associated to vertices, 2-component parameters $\al$, $\bt$, associated to edges, and an $n$-component parameter $\w$ associated to the face of the quadrilateral.}
\label{figquad}
\end{figure}

 Based on the form of the equations \eqref{IRFEMO}, probably the most straightforward way to go a quad equation picture, is to simply consider equations which depend on multi-component variables on faces, edges, and vertices of a quadrilateral, of the type pictured in Figure \ref{figquad}.  The $n$-component vector variables
\begin{align}
\x=(x_1,\ldots,x_n),\quad\bu=(u_1,\ldots,u_n),\quad\y=(y_1,\ldots,y_n),\quad\bv=(v_1,\ldots,v_n),
\end{align}
are assigned to the vertices of the quadrilateral.  
The 2-component parameters
\begin{align}
\al=(\alpha_1,\alpha_2),\qquad\bt=(\beta_1,\beta_2),
\end{align}
are assigned to the edges of the quadrilateral, and the two components of each of $\al$, and $\bt$, are independent.  Finally, there is an $n$-component parameter $\w$, assigned to the face of the quadrilateral.

The quadrilateral in Figure \ref{figquad} is thus associated with a quad equation that in principle has $5n+4$ independent variables/parameters (not yet taking into account constraints of the form \eqref{sumconstraint}).  
The reason $\w$ is interpreted here as an $n$-component parameter, rather than an $n$-component variable on the same level as the $\x,\bu,\bv,\y$, is that the explicit expressions that will be given for the quad equations will be linear in the components of $\x,\bu,\bv,\y$, but not linear in the components of $\w$, $\al$, or $\bt$. In this sense, the $\w$ is more closer to the parameters $\al$, and $\bt$, rather than the corner variables $\x,\bu,\bv,\y$.



In terms of the above variables and parameters, an $n$-component vector quad equation $\qeq$, will be written here as
\begin{align}
\label{quadeq}
\qeq(\x,\bu,\y,\bv;\al,\bt;\w)=0,\qquad n=1,2,\ldots,
\end{align}
and is taken to consist of the set of $n$ individual equations
\begin{align}
\qsym_{k}(\x,\bu,\y,\bv;\al,\bt;\w)=0,\qquad k=1,\ldots,n.
\end{align}


For example, returning to the explicit example of \eqref{IRFEMO}, with the following change of variables
\begin{align}
\label{pointtrans}
y_i=\EXP^{(x_a)_i},\quad v_i=\EXP^{(x_b)_i},\quad x_i=\EXP^{(x_c)_i},\quad u_i=\EXP^{(x_d)_i},\quad w_i=\EXP^{(x^{(cl)}_1)_i},\qquad i=1,\ldots, n,
\end{align}
and
\begin{align}
\al=\{\EXP^{q},\EXP^{q'}\},\quad \bt=\{\EXP^{p},\EXP^{p'}\},
\end{align}
equation \eqref{IRFEMO} takes the form of a multi-component quad equation \eqref{quadeq}, with
\begin{align}
\label{q3m}
\begin{split}
\qsym_{k}&(\x,\bu,\y,\bv;\al,\bt;\w) \\
&=\prod_{i=1}^n(x_i+w_k(\alpha_1-\beta_1))(v_i+w_k(\alpha_2-\beta_2))(u_i+w_{k+1}(\alpha_2-\beta_1))(y_i+w_{k+1}(\alpha_1-\beta_2)) \\
&-\prod_{i=1}^n(x_i+w_{k+1}(\alpha_1-\beta_1))(v_i+w_{k+1}(\alpha_2-\beta_2))(u_i+w_k(\alpha_2-\beta_1))(y_i+w_k(\alpha_1-\beta_2)),
\end{split}
\end{align}
for $k=1,\ldots,n-1$, and where
\begin{align}
\label{prodto1}
\prod_{i=1}^nx_i=1,\qquad\prod_{i=1}^nu_i=1,\qquad\prod_{i=1}^ny_i=1,\qquad\prod_{i=1}^nv_i=1,\qquad\prod_{i=1}^nw_i=1.
\end{align}



\subsubsection{Degeneration of scalar case}

Note that the equations \eqref{q3m} (equivalent to \eqref{IRFEMO}) do not appear to satisfy a 3D-consistency property, at least not in their given form.  However a particular limit for the $n=2$ (scalar) case of the equations \eqref{IRFEMO}, may be related to a 3D-consistent linear quad equation previously studied by Atkinson \cite{Atkinson08,Atkinson09}.

First, redefining the variables in \eqref{IRFEMO} as
\begin{align}
x_a=\epsilon y^{\frac{1}{2}}+\frac{\ii\pi}{2},\quad x_b=\epsilon v^{\frac{1}{2}},\quad x_c=\epsilon x^{\frac{1}{2}},\quad x_d=\epsilon u^{\frac{1}{2}}+\frac{\ii\pi}{2},\quad x_1^{(cl)}=\epsilon w^{\frac{1}{2}},
\end{align}
and
\begin{align}
p=\epsilon^2\beta_1,\qquad p'=\epsilon^2\beta_2,\qquad q=\epsilon^2\alpha_1,\qquad q'=\epsilon^2\alpha_2,
\end{align}
then considering $\epsilon\to0$, results in the following quad equation at $O(\epsilon)$
\begin{align}
\label{scalarquadeq0}
Q(x,u,y,v;\al,\bt;w)=\frac{\alpha_1-\beta_1}{x-w}+\frac{\beta_1-\alpha_2}{u-w}+\frac{\alpha_2-\beta_2}{v-w}+\frac{\beta_2-\alpha_1}{y-w}=0.
\end{align}
The quad equation \eqref{scalarquadeq0} satisfies the 3D-consistency property (this property will be described in the following subsection). 

Equation \eqref{scalarquadeq0} may be rewritten as
\begin{align}
\label{scalarquadeq}
\begin{split}
&\left(\alpha_1(w-u)(w-v)(y-x)+\alpha_2(w-x)(w-y)(u-v)\right. \\
&\;\left.+\beta_1(w-y)(w-v)(x-u) +\beta_2(w-x)(w-u)(v-y)\right)=0,
\end{split}
\end{align}
or equivalently
\begin{align}
\label{scalarquadeqb}
\begin{split}
&\left((\alpha_1-\beta_1)(v-w)+(\alpha_2-\beta_2)(x-w)\right)(u-w)(y-w) \\
&\;+\left((\beta_1-\alpha_2)(y-w)+(\beta_2-\alpha_1)(u-w)\right)(x-w)(v-w)=0.
\end{split}
\end{align}
The parameter $w$ may also be absorbed into a redefinition of the variables and parameters, resulting in the quad equation
\begin{align}
\label{scalarquadeq2}
\begin{split}
\alpha_1uv(y-x)+\beta_1yv(x-u)+\alpha_2xy(u-v)+\beta_2xu(v-y)=0,
\end{split}
\end{align}
or equivalently
\begin{align}
\label{scalarquadeq2b}
\begin{split}
(\alpha_1-\beta_1)uyv+(\beta_1-\alpha_2)xyv+(\alpha_2-\beta_2)xuy+(\beta_2-\alpha_1)xuv=0.
\end{split}
\end{align}
Finally, inverting each of the $x,u,v,y$ results in the linear form of the quad equation \eqref{scalarquadeq0}
\begin{align}
\label{scalarquadeq2c}
\begin{split}
(\alpha_1-\beta_1)x+(\beta_1-\alpha_2)u+(\alpha_2-\beta_2)v+(\beta_2-\alpha_1)y=0.
\end{split}
\end{align}
The latter equation has also previously appeared 
in the work of Atkinson \cite{Atkinson08,Atkinson09}, via a B\"{a}cklund transformation for a 3D-consistent quad equation given by Hietarinta \cite{HietarintaCAC2004}.  Note also that the quad equation \eqref{scalarquadeq2c} arises in the limit of the parameters $\alpha_i\to\epsilon\alpha_i$, $\beta_i\to\epsilon\beta_i$, $\epsilon\to0$, of the latter equation of Hietarinta, independently of the B\"{a}cklund transformation of  \cite{Atkinson08}.

\subsubsection{Multi-component 3D-consistent equations}

The same type of limit of \eqref{IRFEMO} which resulted in \eqref{scalarquadeq0}, becomes complicated to apply for the cases of $n>2$.  However, the form of the multi-component equations \eqref{IRFEMO}, is suggestive of a multi-component generalisation of \eqref{scalarquadeq0} which takes the form 
\begin{align}
\label{q1m}
Q_k(\x,\bu,\y,\bv;\al,\bt;\w)=\sum_{i=1}^n\left\{\frac{\alpha_1-\beta_1}{x_i-w_k}+\frac{\beta_1-\alpha_2}{u_i-w_k}+\frac{\alpha_2-\beta_2}{v_i-w_k}+\frac{\beta_2-\alpha_1}{y_i-w_k}\right\}=0.
\end{align}
for\footnote{Note that the scalar case of \eqref{IRFEMO} (or \eqref{q3m}), corresponds to $n=2$, while the scalar case of \eqref{q1m} corresponds to $n=1$.  This is because the extra condition on the variables \eqref{sumconstraint} (or \eqref{prodto1}), has already been used in taking the limit to \eqref{scalarquadeq0}.} $n=1,2,\ldots$.  Remarkably, this equation can be observed to satisfy the following multi-component analogue of the 3D-consistency condition. 

\paragraph{ Multi-component 3D-consistency condition}  

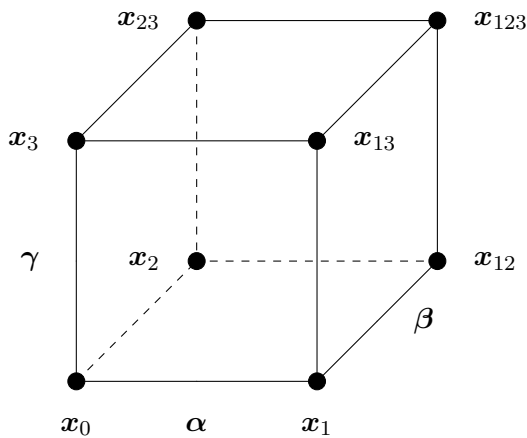
\begin{figure}[htb]
\centering
\begin{tikzpicture}[scale=1.6]
\draw[black!] (-1,0) circle (0.01pt)
node[left=10pt]{\color{black} $\gm$};
\draw[black!] (0,-1) circle (0.01pt)
node[below=10pt]{\color{black} $\al$};
\draw[black!] (1.5,-0.5) circle (0.01pt)
node[right=10pt]{\color{black} $\bt$};
\draw[-] (1,1)--(-1,1)--(-1,-1)--(1,-1)--(1,1);
\draw[-] (2,-0)--(2,2)--(-0,2);
\draw[-,dashed] (-0,2)--(-0,-0)--(2,-0);
\draw[-] (-1,1)--(0,2);
\draw[-] (1,1)--(2,2);
\draw[-] (1,-1)--(2,0);
\draw[-,dashed] (-1,-1)--(0,0);
\filldraw[fill=black,draw=black] (2,2) circle (2pt)
node[right=10pt]{$\x_{123}$};
\filldraw[fill=black,draw=black] (0,0) circle (2pt)
node[left=10pt]{$\x_{2}$};
\filldraw[fill=black,draw=black] (2,0) circle (2pt)
node[right=10pt]{$\x_{12}$};
\filldraw[fill=black,draw=black] (0,2) circle (2pt)
node[left=10pt]{$\x_{23}$};
\filldraw[fill=black,draw=black] (1,1) circle (2pt)
node[right=10pt]{$\x_{13}$};
\filldraw[fill=black,draw=black] (-1,-1) circle (2pt)
node[below=10pt]{$\x_0$};
\filldraw[fill=black,draw=black] (1,-1) circle (2pt)
node[below=10pt]{$\x_1$};
\filldraw[fill=black,draw=black] (-1,1) circle (2pt)
node[left=10pt]{$\x_3$};
\end{tikzpicture}

\caption{Cube used to formulate the property of 3D-consistency (also commonly known as ``consistency-around-a-cube'' (CAC), or ``multi-dimensional consistency''), for multi-component quad equations \eqref{quadeq}, defined on the quadrilaterals of Figure \ref{figquad}.  The $n$-component variables $\x_i$ are associated to vertices,  the two-component parameters $\al$, $\bt$, $\gm$, are associated to the edges, and an $n$-component parameter $\w$ (not shown) is associated to each face of the cube.}
\label{3D}
\end{figure}

 Consider the cube in Figure \ref{3D}, with variables and parameters assigned to the six quadrilateral faces according to Figure \ref{figquad}.   
 Edges that are parallel in Figure \ref{3D}, are always associated with the same parameter $\al$, $\bt$, or $\gm$.  The same parameter $\w$ is associated to each of the six faces.  The six multi-component quad equations
\begin{align}
\label{3dequations}
\begin{gathered}
\qeq(\x_0,\x_1,\x_2,\x_{12};\al,\bt;\w)=0, \\
\qeq(\x_0,\x_1,\x_3,\x_{13};\al,\gm;\w)=0, \\
\qeq(\x_0,\x_2,\x_3,\x_{23};\bt,\gm;\w)=0, \\
\qeq(\x_3,\x_{13},\x_{23},\x_{123};\al,\bt;\w)=0, \\
\qeq(\x_2,\x_{12},\x_{23},\x_{123};\al,\gm;\w)=0, \\
\qeq(\x_1,\x_{12},\x_{13},\x_{123};\bt,\gm;\w)=0, 
\end{gathered}
\end{align}
for $k=1,\ldots,n$, are respectively associated to the corresponding six faces of a cube, as labelled in Figure \ref{3D}.

Then consider the initial value problem, where $\x_0,\x_1,\x_2,\x_3$ and $\al$, $\bt$, $\gm$, $\w$, are known, and $\x_{12},\x_{13},\x_{23},\x_{123}$, are to be determined.  The first $3(n)$ equations in \eqref{3dequations} should provide a unique solution for the vector variables $\x_{12}$, $\x_{13}$, $\x_{23}$, respectively.  The 3D-consistency condition is that the remaining $3(n)$ equations in \eqref{3dequations}, each must agree for the solution of the remaining vector variable $\x_{123}$.
\\[-0.2cm]

The 3D-consistency of \eqref{q1m} can be checked numerically for small values of $n$ ($n\leq10$), and is expected to hold for all $n\geq1$.  However it is not known to the author how to directly prove the 3D-consistency using algebraic methods (besides the simple scalar $n=1$ case of \eqref{scalarquadeq0}), mainly because the expressions for the equations \eqref{q1m} become quite complicated for $n\geq2$.

\paragraph{Further examples}

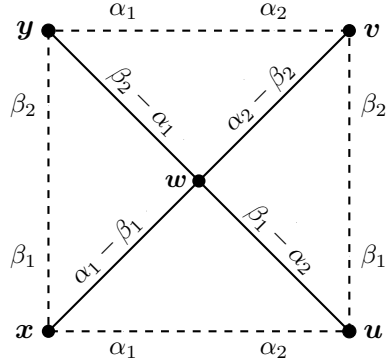
\begin{figure}[tbh]
\centering
\begin{tikzpicture}[scale=1.0]
%
\draw[-,thick,dashed] (5,-1)--(5,3)--(1,3)--(1,-1)--(5,-1);
\draw[-,thick] (5,3)--(1,-1);
\draw[-,thick] (5,-1)--(1,3);
\fill (1,0) circle (0.1pt)
node[left=0.5pt]{\color{black}\small $\beta_1$};
\fill (5,0) circle (0.1pt)
node[right=0.5pt]{\color{black}\small $\beta_1$};
\fill (2.8,1.55) circle (0.1pt)
node[left=0.5pt,rotate=-45]{\color{black}\small $\beta_2-\alpha_1$};
\fill (4,-1) circle (0.1pt)
node[below=0.5pt]{\color{black}\small $\alpha_2$};
\fill (4,3) circle (0.1pt)
node[above=0.5pt]{\color{black}\small $\alpha_2$};
\fill (1,2) circle (0.1pt)
node[left=0.5pt]{\color{black}\small $\beta_2$};
\fill (5,2) circle (0.1pt)
node[right=0.5pt]{\color{black}\small $\beta_2$};
\fill (2.3,0.6) circle (0.1pt)
node[left=0.5pt,rotate=45]{\color{black}\small $\alpha_1-\beta_1$};
\fill (2,-1) circle (0.1pt)
node[below=0.5pt]{\color{black}\small $\alpha_1$};
\fill (2,3) circle (0.1pt)
node[above=0.5pt]{\color{black}\small $\alpha_1$};
\fill (3.6,2.35) circle (0.1pt)
node[below=0.5pt,rotate=45]{\color{black}\small $\alpha_2-\beta_2$};
\fill (3,1) circle (2.5pt)
node[left=0.5pt]{\color{black}\small $\w$};
\fill (4.3,0.45) circle (0.1pt)
node[below=0.5pt,rotate=-45]{\color{black}\small $\beta_1-\alpha_2$};
\fill (1,-1) circle (2.5pt)
node[left=1.5pt]{\color{black} $\x$};
\filldraw[fill=black,draw=black] (1,3) circle (2.5pt)
node[left=1.5pt]{\color{black} $\y$};
\fill (5,3) circle (2.5pt)
node[right=1.5pt]{\color{black} $\bv$};
\filldraw[fill=black,draw=black] (5,-1) circle (2.5pt)
node[right=1.5pt]{\color{black} $\bu$};
\end{tikzpicture}
\caption{The four-leg form \eqref{4leg} of a multi-component quad equation \eqref{quadeq} associated to the quadrilateral of Figure \ref{figquad}.}
\label{fig4quad}
\end{figure}

It might be expected that there exist other multi-component equations related to \eqref{q1m}, that will also satisfy the 3D-consistency property.  For example, instead of \eqref{q1m},  a more general type of multi-component quad equation can be considered, which takes the following four-leg form
\begin{align}
\label{4leg}
\begin{split}
&\qsym_k(\x,\bu,\y,\bv;\al,\bt;\w) \\
&\;=
\sum_{i=1}^n\left(\varphi(x_i,w_k;\alpha_1,\beta_1)+\varphi(u_i,w_k;\beta_1,\alpha_2)+\varphi(v_i,w_k;\alpha_2,\beta_2)+\varphi(y_i,w_k;\beta_2,\alpha_1)\right)=0,
\end{split}
\end{align}
for $k=1,\ldots,n$, where $\varphi(x,y;\alpha,\beta)$ is a leg function, of the form
\begin{align}
\label{legdef}
\varphi(x,y;\alpha,\beta)=\frac{f(\alpha,\beta)}{x-y}+g(\alpha,\beta),
\end{align}
and the $f$, and $g$, are taken to satisfy
\begin{align}
f(\alpha,\beta)+f(\beta,\alpha)=0,\qquad g(\alpha,\beta)+g(\beta,\alpha)=0.
\end{align}
Then \eqref{q1m} would correspond to \eqref{4leg} with $f(\alpha,\beta)=\alpha-\beta$, and $g(\alpha,\beta)=0$.  Such a quad equation of the form \eqref{4leg} is pictured graphically in Figure \ref{fig4quad}. 

Note that the quad equations of the form \eqref{4leg}, are invariant under permutations of the components of a variable $\x,\bu,\y,\bv$, however are not invariant under the exchange of the components of a parameter $\alpha$, or $\beta$.  Also since the index $k$ of \eqref{4leg}, appears as the index of components of the face parameter $\w$, permuting the components of $\w$ simply results in a different permutation of the equations \eqref{4leg}.  The quad equations \eqref{4leg}, also satisfy the following square symmetries
\begin{align}
\label{quadsym1}
\qeq(\x,\bu,\y,\bv;\al,\bt;\w)=-\qeq(\x,\y,\bu,\bv;\bt,\al;\w),
\end{align}
and
\begin{align}
\label{quadsym2}
\qeq(\x,\bu,\y,\bv;\al,\bt;\w)=-\qeq(\y,\bv,\x,\bu;\al,\hat{\bt};\w)=-\qeq(\bu,\x,\bv,\y;\hat{\al},\bt;\w),
\end{align}
where $\hat{\al}$, $\hat{\bt}$, represent $\al$, $\bt$, with components exchanged, respectively.

It is not difficult to find different functions $f(\alpha,\beta)$, and $g(\alpha,\beta)$, for which the multi-component quad equation \eqref{4leg} will satisfy the 3D-consistency condition.  For example, the following choice related to \eqref{q1m} is found to be 3D-consistent
\begin{align}
\label{q1a}
\begin{array}{lll}
\qeq 1a_{(\delta)}:&f(\alpha,\beta)=\alpha-\beta,\; &g(\alpha,\beta)=\delta\alpha\beta(\alpha-\beta), 
\end{array}
\end{align}
where $\delta=0,1$.  This equation has been labelled here as $\qeq 1a_{(\delta)}$, due to the resemblence of this equation for $\delta=0$ (corresponding to \eqref{q1m}), to the discrete Laplace equation associated to the ABS equation $Q1_{(\delta=0)}$ \cite{BobSurQuadGraphs}.

Similarly, the following simple hyperbolic deformation of \eqref{q1a}
\begin{align}
\label{q1h}
\begin{array}{lll}
\qeq 1h_{(\delta)} :&f(\alpha,\beta)=\sinh(2(\alpha-\beta)),\; &g(\alpha,\beta)=\delta\sinh\alpha\sinh\beta\sinh(\alpha-\beta),
\end{array}
\end{align}
and the following simple elliptic deformation of \eqref{q1a}
\begin{align}
\label{q1e}
\begin{array}{lll}
\qeq 1e:&f(\alpha,\beta)=\sn(\alpha)^2-\sn(\beta)^2,\; &g(\alpha,\beta)=\sn(\alpha)\sn(\beta)\sn(\alpha-\beta)(1-m\sn(\alpha)^2\sn(\beta)^2), 
\end{array}
\end{align}
are both found to be 3D-consistent.   In \eqref{q1e}, $m=k^2$ (not to be confused with the index $k$ for \eqref{4leg}) is the elliptic parameter for the Jacobi elliptic function $\sn(z)$ \cite{WW}.  It is expected that there would be other equations of the form \eqref{4leg} that satisfy 3D-consistency, and it would be interesting to further investigate these types of equations, or even classify these equations along the lines of \cite{ABS,ABS2}.  There are also obviously more general types of quad equations that could be considered instead of the form \eqref{4leg}, but no other forms are known to the author that will also be 3D-consistent.

Similarly to the case of \eqref{q1m}, the 3D-consistency for the scalar $n=1$ cases of the equations \eqref{q1a}, \eqref{q1h}, \eqref{q1e}, can be proven directly, however so far it is not known how to prove the 3D-consistency for all $n\geq1$,  mainly because for $n\geq2$ the expressions for the equations become much more complicated.  The 3D-consistency of the above equations has been checked through the use of numerical calculations for up to $n=10$, and it is expected to be satisfied for all $n\geq1$.  




Finally, it is straightforward to put the equations \eqref{q1a}, \eqref{q1h}, \eqref{q1e}, in an affine linear form, where the equations are linear in each of the components of $\x,\bu,\y,\bv$.  
The affine linear form is given by
\begin{align}
\label{afflin}
\begin{split}
\qsym_k(\x,\bu,\y,\bv;\al,\bt;\w) 
&=\sum_{i=1}^n\Big\{\left[\left(f(\alpha_1,\beta_1)(w_k-v_i)+f(\alpha_2,\beta_2)(w_k-x_i)\right)(w_k-u_i)(w_k-y_i)\right. \\[-0.3cm]
&\left.\left.\hspace{0.84cm}+\left(f(\beta_1,\alpha_2)(w_k-y_i)+f(\beta_2,\alpha_1)(w_k-u_i)\right)(w_k-v_i)(w_k-x_i)\right.\right] \\[-0.1cm]
&\hspace{0.8cm}\times\prod^{n}_{\substack{j=1 \\ j\neq i}}(w_k-x_j)(w_k-u_j)(w_k-y_j)(w_k-v_j)\Big\} \\[-0.15cm]
&\hspace{0.5cm}+h(\al,\bt)\prod^{n}_{i=1}\left\{(w_k-x_i)(w_k-u_i)(w_k-y_i)(w_k-v_i)\right\},
\end{split}
\end{align}
for $k=1,\ldots,n$, where  $h(\al,\bt)$ is given in terms of $g(\alpha,\beta)$ by
\begin{align}
h(\al,\bt)=g(\alpha_1,\beta_1)+g(\beta_1,\alpha_2)+g(\alpha_2,\beta_2)+g(\beta_2,\alpha_1).
\end{align}
For the respective equations \eqref{q1a}, \eqref{q1h}, \eqref{q1e}, the functions $h(\al,\bt)$ may be factorised as
\begin{align}
\begin{array}{ll}
\qeq 1a_{(\delta)}:&h(\al,\bt)=\delta(\alpha_1-\alpha_2)(\beta_1-\beta_2)(\alpha_1+\alpha_2-\beta_1-\beta_2), \\[0.1cm]
\qeq 1h_{(\delta)}:&h(\al,\bt)=\delta\sinh\left(\alpha_1-\alpha_2\right)\sinh\left(\beta_1-\beta_2\right)\sinh\left(\alpha_1+\alpha_2-\beta_1-\beta_2\right), \\[0.1cm]
\qeq 1e:&h(\al,\bt)=\sn(\alpha_1-\alpha_2)\sn(\beta_1-\beta_2)\sn(\alpha_1+\alpha_2-\beta_1-\beta_2)(1-m(\sn(\alpha_1)\sn(\alpha_2))^2) \\[0.1cm]
&\times (1-m(\sn(\beta_1)\sn(\beta_2))^2)(1-m(\sn(\alpha_1+\alpha_2)\sn(\beta_1+\beta_2))^2)+m(\sn(\alpha_1)^2-\sn(\alpha_2)^2) \\[0.1cm]
&\times(\sn(\beta_1)^2-\sn(\beta_2)^2)(\sn(\alpha_1)\sn(\alpha_2)\sn(\alpha_1+\alpha_2)-\sn(\beta_1)\sn(\beta_2)\sn(\beta_1+\beta_2)).
\end{array}
\end{align}

Interestingly, the hyperbolic- and elliptic-type equations in \eqref{q1h}, and \eqref{q1e}, don't require a point transformation of the components of the corner variables $\x,\bu,\y,\bv$, in terms of hyperbolic or elliptic functions respectively, in order to go from the four-leg form \eqref{4leg}, to the affine-linear form \eqref{afflin}.  Particularly, such a point transformation is typically always required for the scalar cases, in order to relate a 3D-consistent quad equation to a Yang-Baxter equation \cite{Bazhanov:2016ajm,Kels:2018xge} (or even simply for a three-leg equation \cite{ABS}).  The equations \eqref{q1h}, and \eqref{q1e}, are curious in this respect, and are unlikely to arise from a counterpart multi-component Yang-Baxter equation. 

\section{Conclusion}

In this paper a new formulation of Baxter's Z-invariance property \cite{Baxter:1978xr,Baxter:1986df} is given for two-dimensional vertex and interaction-round-a-face (IRF) models of statistical mechanics that satisfy the Yang-Baxter equation.  Specifically, the models were first associated to a surface of elementary four-squares, where each elementary four-square is identified with a Boltzmann weight of the respective model.  Such elementary four-squares were used as the building blocks of more general 2-dimensional surfaces, which are extended into 3 dimensions, on which different vertex and IRF models were also defined.  The extended Z-invariance property is that the partition function for the latter deformed vertex or IRF model, is equivalent to the partition function for the original model in the plane, as a consequence of the Yang-Baxter equations and inversion relations satisfied by the respective models.  This is considered here to be an extension of Baxter's original formulation of Z-invariance \cite{Baxter:1978xr,Baxter:1986df}, because it requires the introduction of new types of rapidity lines which form closed directed loops in the rapidity graph of the model, whereas traditionally such lines are not permitted to appear.  The extended Z-invariance property was previously shown to also hold for lattice models of statistical mechanics that satisfy the star-triangle relation \cite{Kels:2017fyt}, and the results of this paper show how the Z-invariance is applicable to a wider range of integrable models of statistical mechanics, including the six- and eight-vertex models \cite{Baxter:1982zz}, and the RSOS models \cite{ABF}, as well as several different integrable lattice models based on the star-star relation \cite{Bazhanov:1990qk,Bazhanov:1992jqa,Baxter:1997tn,Bazhanov:2011mz,Bazhanov:2013bh,Yamazaki:2013nra,Gahramanov:2017idz,Kels:2017vbc}. 

Due to its association with integrability, the Z-invariance property has been previously utilised for a variety of important applications.  For example, it has been used in various studies of the Ising model \cite{AUYANG198744,REYESMARTINEZ1997203,Martı́nez1998463,Costa-Santos,Au-Yang2007,Boutillier2010,Boutillier2011}, in the derivation of the order parameter of the chiral Potts model \cite{JimboMiwaNakayashiki,Bax2,Baxter:2005jt,Au-YangPerk2011}, in relating integrable lattice models with isoradial embeddings of graphs and circle patterns \cite{Costa-Santos,Bazhanov:2007mh}, and in relating continuous spin lattice models with supersymmetric gauge theories \cite{Yamazaki:2012cp}.  It would be interesting if similar applications exist for the formulation of Z-invariance given in this paper for the vertex and IRF models.  Keep in mind that such vertex and IRF models can also be interpreted as models of physical systems \cite{Baxter:1982zz}, and another important question is whether the extended Z-invariance property of this paper (and \cite{Kels:2017fyt}) represents an observable phenomenon for the physical systems represented by these models. 

In Section \ref{sec:qcl}, the quasi-classical limit of the IRF model was also considered.  It was shown that the leading order quasi-classical expansion of the partition function, corresponds to an action functional for a system of classical discrete Laplace equations.  This action functional  was seen to be invariant under cubic deformations of the underlying surface of elementary four-squares, which is the classical manifestation of the extended Z-invariance property.  This classical Z-invariance property also has a natural interpretation as a closure property of the action functional, which has previously been studied in the context of three-point Lagrangian multiform equations \cite{LobbNijhoff}, and the related pluri-Lagrange systems \cite{BPS}.  However the latter Lagrangian systems are inherently different from the five-point equations that have been considered in this paper, and thus it would be interesting to relate the equations given here to other notions of discrete integrability.  An initial step in this direction has been taken in Section \ref{sec:Mcomp}, where some explicit new $n$-component 3D-consistent equations were proposed, which were constructed from a degeneration of elliptic/hyperbolic multi-component equations, coming from the quasi-classical limit of particular continuous spin IRF models \cite{Bazhanov:2011mz,KelsThesis}.  It will be important to further investigate the integrable properties of these new types of equations, including their relations to the Yang-Baxter equation, in future works.  

\section*{Acknowledgements}
The majority of this work was completed while the author was an overseas researcher under Postdoctoral Fellowship of Japan Society for the Promotion of Science (JSPS), at the University of Tokyo, Komaba.  The author thanks Atsuo Kuniba and Masahito Yamazaki for helpful discussions.  Some of the $n$-component equations of Section \ref{sec:Mcomp} were presented at the 13th Symmetries and Integrability of Difference Equations (SIDE) conference, in Fukuoka, Japan, on November, 2018. The author thanks the organisers for the opportunity to give a talk at this conference, and also thanks the audience for their interest and feedback.

\begin{appendices}
\numberwithin{equation}{section}

\section{Deformations for vertex and IRF formulations}\label{app:IRF}
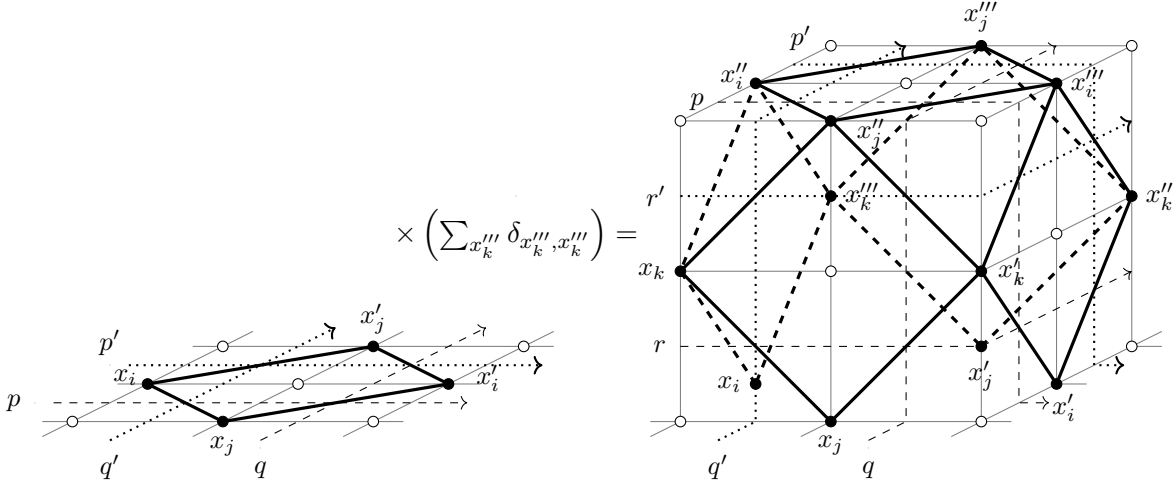
\begin{figure}[hbt!]
\centering
\begin{tikzpicture}

\begin{scope}[scale=1.0]
\draw[-,gray] (-3.4,-1)--(1.4,-1);\draw[-,gray] (0.6,-1.2)--(3.4,0.2);\draw[-,gray] (2.4,-0.5)--(-2.4,-0.5);\draw[-,gray] (1.4,0.2)--(-1.4,-1.2);\draw[-,gray] (-3.4,-1.2)--(-0.6,0.2);\draw[-,gray] (-1.4,0)--(3.4,0);
\filldraw[fill=black,draw=black] (1,0) circle (2.0pt)
node[above=1pt]{\small $x'_j$};
\filldraw[fill=black,draw=black] (-2.0,-0.5) circle (2.0pt);
\fill[black!] (-2,-0.41) circle (0.01pt)
node[left=0pt]{\color{black}\small $x_i$};
\filldraw[fill=black,draw=black] (-1,-1) circle (2.0pt)
node[below=2pt]{\small $x_j$};
\filldraw[fill=black,draw=black] (2.0,-0.5) circle (2.0pt);
\fill[black] (2.2,-0.42) circle (0.01pt)
node[right=1pt]{\color{black}\small $x'_i$};
\filldraw[fill=white,draw=black] (1,-1) circle (2.0pt);
\filldraw[fill=white,draw=black] (0,-0.5) circle (2.0pt);
\filldraw[fill=white,draw=black] (-1,0) circle (2.0pt);
\filldraw[fill=white,draw=black] (-3,-1) circle (2.0pt);
\filldraw[fill=white,draw=black] (3,0) circle (2.0pt);
\draw[-,very thick] (-1,-1)--(2,-0.5)--(1,0)--(-2,-0.5)--(-1,-1);
\draw[->,black,thick,dotted] (-2.25,-0.25)--(3.25,-0.25);
\draw[->,black,thick,dotted] (-2.5,-1.25)--(0.5,0.25);
\draw[->,black,dashed] (-3.25,-0.75)--(2.25,-0.75);
\draw[->,black,dashed] (-0.5,-1.25)--(2.5,0.25);
\draw[black] (-2.5,-0.25) circle (0.01pt)
node[above=0pt]{\color{black}\small $p'$};
\draw[black] (-2.5,-1.25) circle (0.01pt)
node[below=1.5pt]{\color{black}\small $q'$};
\draw[black] (-3.5,-0.75) circle (0.01pt)
node[left=1.5pt]{\color{black}\small $p$};
\draw[black] (-0.5,-1.35) circle (0.01pt)
node[below=1.5pt]{\color{black}\small $q$};
\end{scope}

\draw[black] (2.9,2) circle (0.01pt)
node[below=1pt]{\color{black}$\times\left(\sum_{x'''_k}\delta_{x'''_k,x'''_k}\right)=$};

\begin{scope}[scale=1.0,xshift=230,yshift=0,rotate=0]
\draw[-,gray] (-1.4,-1.2)--(-1,-1)--(1,-1)--(1,1)--(-1,1)--(-1,-1)--(-3,-1)--(-3,1)--(-1,1)--(-1,3)--(1,4);
\draw[-,gray] (1,3)--(3,4)--(3,0)--(2,-0.5)--(2,1.5)--(1,1)--(1,3)--(-3,3)--(-3,1);
\draw[-,gray] (1.4,-1)--(1,-1)--(2,-0.5)--(2.4,-0.5);
\draw[-,gray] (-3,3)--(-1,4)--(3,4);
\draw[-,gray] (-2,3.5)--(2,3.5)--(2,1.5)--(3,2);
\draw[-,gray] (3.4,0)--(3,0)--(3.4,0.2);\draw[-,gray] (-3.4,-1)--(-3,-1)--(-3.4,-1.2);\draw[-,gray] (1,-1)--(0.6,-1.2);
\filldraw[fill=white,draw=black] (-3,3) circle (2.0pt);
\filldraw[fill=black,draw=black] (-1,3) circle (2.0pt);
\fill[black!] (-1,2.83) circle (0.01pt)
node[right=6pt]{\color{black}\small $x''_j$};
\filldraw[fill=white,draw=black] (1,3) circle (2.0pt);
\filldraw[fill=black,draw=black] (-2,3.5) circle (2.0pt);
\fill[black!] (-1.9,3.6) circle (0.01pt)
node[left=1.5pt]{\color{black}\small $x''_i$};
\filldraw[fill=white,draw=black] (0,3.5) circle (2.0pt);
\filldraw[fill=black,draw=black] (2,3.5) circle (2.0pt)
node[right=1.5pt]{\color{black}\small $x'''_i$};
\filldraw[fill=white,draw=black] (-1,4) circle (2.0pt);
\filldraw[fill=black,draw=black] (1,4) circle (2.0pt)
node[above=1.5pt]{\color{black}\small $x'''_j$};
\filldraw[fill=white,draw=black] (3,4) circle (2.0pt);
\filldraw[fill=white,draw=black] (3,0) circle (2.0pt);
\filldraw[fill=black,draw=black] (3,2) circle (2.0pt)
node[right=1.5pt]{\color{black}\small $x''_k$};
\filldraw[fill=black,draw=black] (-1,2) circle (2.0pt)
node[right=1.5pt]{\color{black}\small $x'''_k$};
\filldraw[fill=white,draw=black] (-3,-1) circle (2.0pt);
\filldraw[fill=black,draw=black] (-3,1) circle (2.0pt)
node[left=1.5pt]{\color{black}\small $x_k$};
\filldraw[fill=black,draw=black] (-2,-0.5) circle (2.0pt)
node[left=1.5pt]{\color{black}\small $x_i$};
\filldraw[fill=black,draw=black] (-1,-1) circle (2.0pt)
node[below=1.5pt]{\color{black}\small $x_j$};
\filldraw[fill=black,draw=black] (1,1) circle (2.0pt)
node[right=2.5pt]{\color{black}\small $x'_k$};
\filldraw[fill=black,draw=black] (2,-0.5) circle (2.0pt);
\fill[black!] (2.1,-0.5) circle (0.01pt)
node[below=0pt]{\color{black}\small $x'_i$};
\filldraw[fill=white,draw=black] (-1,1) circle (2.0pt);
\filldraw[fill=white,draw=black] (1,-1) circle (2.0pt);
\filldraw[fill=white,draw=black] (2,1.5) circle (2.0pt);
\filldraw[fill=black,draw=black] (1,0) circle (2.0pt)
node[below=1.5pt]{\color{black}\small $x'_j$};
\draw[-,very thick] (-1,-1)--(1,1)--(-1,3)--(-3,1)--(-1,-1);
\draw[-,dashed,very thick] (1,0)--(3,2)--(1,4)--(-1,2)--(1,0);
\draw[-,very thick] (2,-0.5)--(3,2)--(2,3.5)--(1,1)--(2,-0.5);
\draw[-,dashed,very thick] (-2,-0.5)--(-3,1)--(-2,3.5)--(-1,2)--(-2,-0.5);
\draw[-,very thick] (-1,3)--(2,3.5)--(1,4)--(-2,3.5)--(-1,3);
\draw[->,black,dashed] (-2.5,3.25)--(1.5,3.25)--(1.5,-0.75)--(1.9,-0.75);
\draw[->,black,dashed] (-0.5,-1.25)--(0,-1)--(0,3)--(2,4);
\draw[->,black,dashed] (-3,0)--(1,0)--(3,1);
\draw[->,black,thick,dotted] (-1.5,3.75)--(2.5,3.75)--(2.5,-0.25)--(2.9,-0.25);
\draw[->,black,thick,dotted] (-2.5,-1.25)--(-2,-1)--(-2,3)--(0,4);
\draw[->,black,thick,dotted] (-3,2)--(1,2)--(3,3);
\draw[black] (-1.4,3.75) circle (0.01pt)
node[above=1.5pt]{\color{black}\small $p'$};
\draw[black] (-2.5,3.25) circle (0.01pt)
node[left=1.5pt]{\color{black}\small $p$};
\draw[black] (-2.5,-1.25) circle (0.01pt)
node[below=1.5pt]{\color{black}\small $q'$};
\draw[black] (-0.5,-1.35) circle (0.01pt)
node[below=1.5pt]{\color{black}\small $q$};
\draw[black] (-3,2) circle (0.01pt)
node[left=1.5pt]{\color{black}\small $r'$};
\draw[black] (-3,0) circle (0.01pt)
node[left=1.5pt]{\color{black}\small $r$};
\end{scope}

\end{tikzpicture}
\caption{A deformation in the vertex formulation corresponding to \eqref{veq1}.  Note that the rapidity lines which are not shown for two hidden four-squares on the right hand side are assigned according to Figure \ref{4VBW}. }
\label{Vcube}
\end{figure}

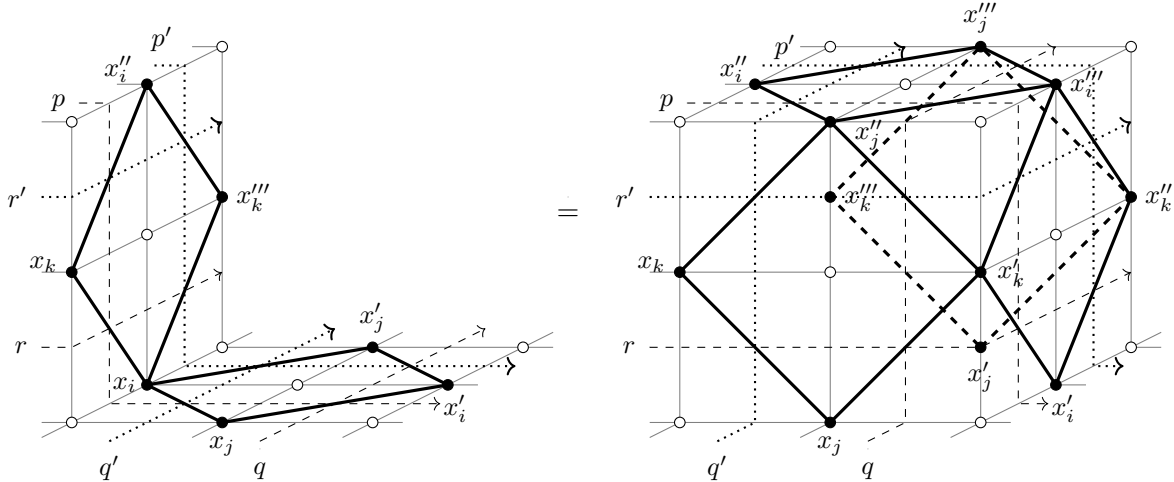
\begin{figure}[hbt!]
\centering
\begin{tikzpicture}

\begin{scope}[scale=1.0]
\fill[white] (0,6.5) circle (0.01pt);

\draw[-,gray] (-3.4,-1)--(1.4,-1);\draw[-,gray] (0.6,-1.2)--(3.4,0.2);\draw[-,gray] (2.4,-0.5)--(-2,-0.5);\draw[-,gray] (1.4,0.2)--(-1.4,-1.2);\draw[-,gray] (-3.4,-1.2)--(-0.6,0.2);\draw[-,gray] (-1,0)--(3.4,0);
\draw[-,gray] (-3,-1)--(-3,3)--(-1,4)--(-1,0); \draw[-,gray] (-2,-0.5)--(-2,3.5)--(-2.4,3.5); \draw[-,gray] (-1,2)--(-3,1)--(-3.4,1); \draw[-,gray] (-3,3)--(-3.4,3); \draw[-,gray] (-1,4)--(-1.4,4);
\filldraw[fill=black,draw=black] (1,0) circle (2.0pt)
node[above=1.5pt]{\color{black}\small $x'_j$};
\filldraw[fill=black,draw=black] (-2,-0.5) circle (2.0pt)
node[left=0pt]{\color{black}\small $x_i$};
\filldraw[fill=black,draw=black] (-1,-1) circle (2.0pt)
node[below=1.5pt]{\color{black}\small $x_j$};
\filldraw[fill=black,draw=black] (2,-0.5) circle (2.0pt);
\fill[black!] (2.1,-0.5) circle (0.01pt)
node[below=0pt]{\color{black}\small $x'_i$};
\filldraw[fill=white,draw=black] (1,-1) circle (2.0pt);
\filldraw[fill=white,draw=black] (0,-0.5) circle (2.0pt);
\filldraw[fill=white,draw=black] (-1,0) circle (2.0pt);
\filldraw[fill=white,draw=black] (-3,-1) circle (2.0pt);
\filldraw[fill=white,draw=black] (3,0) circle (2.0pt);
\filldraw[fill=black,draw=black] (-2,3.5) circle (2.0pt);
\fill[black!] (-2,3.7) circle (0.01pt)
node[left=1.5pt]{\color{black}\small $x''_i$};
\filldraw[fill=black,draw=black] (-3,1) circle (2.0pt);
\fill[black!] (-3,1.1) circle (0.01pt)
node[left=1.5pt]{\color{black}\small $x_k$};
\filldraw[fill=black,draw=black] (-1,2) circle (2.0pt)
node[right=1.5pt]{\color{black}\small $x'''_k$};
\filldraw[fill=white,draw=black] (-2,1.5) circle (2.0pt);
\filldraw[fill=white,draw=black] (-3,3) circle (2.0pt);
\filldraw[fill=white,draw=black] (-1,4) circle (2.0pt);
\draw[-,very thick] (-1,-1)--(2,-0.5)--(1,0)--(-2,-0.5)--(-1,2)--(-2,3.5)--(-3,1)--(-2,-0.5)--(-1,-1);
\draw[->,black,thick,dotted] (-1.9,3.75)--(-1.5,3.75)--(-1.5,-0.25)--(2.9,-0.25);
\draw[->,black,thick,dotted] (-2.5,-1.25)--(0.5,0.25);
\draw[->,black,dashed] (-2.9,3.25)--(-2.5,3.25)--(-2.5,-0.75)--(1.9,-0.75);
\draw[->,black,dashed] (-0.5,-1.25)--(2.5,0.25);
\draw[->,black,dashed] (-3.4,0)--(-3,0)--(-1,1);
\draw[->,black,thick,dotted] (-3.4,2)--(-3,2)--(-1,3);
\draw[black] (-1.8,3.75) circle (0.01pt)
node[above=1.5pt]{\color{black}\small $p'$};
\draw[black] (-2.5,-1.25) circle (0.01pt)
node[below=1.5pt]{\color{black}\small $q'$};
\draw[black] (-2.9,3.25) circle (0.01pt)
node[left=1.5pt]{\color{black}\small $p$};
\draw[black] (-0.5,-1.35) circle (0.01pt)
node[below=1.5pt]{\color{black}\small $q$};
\draw[black] (-3.4,0) circle (0.01pt)
node[left=1.5pt]{\color{black}\small $r$};
\draw[black] (-3.4,2) circle (0.01pt)
node[left=1.5pt]{\color{black}\small $r'$};

\end{scope}

\draw[black] (3.6,2) circle (0.01pt)
node[below=1pt]{\color{black}$=$};

\begin{scope}[scale=1.0,xshift=230,yshift=0,rotate=0]
\draw[-,gray] (-1.4,-1.2)--(-1,-1)--(1,-1)--(1,1)--(-1,1)--(-1,-1)--(-3,-1)--(-3,1)--(-1,1)--(-1,3)--(1,4);
\draw[-,gray] (1,3)--(3,4)--(3,0)--(2,-0.5)--(2,1.5)--(1,1)--(1,3)--(-3,3)--(-3,1);
\draw[-,gray] (1.4,-1)--(1,-1)--(2,-0.5)--(2.4,-0.5);
\draw[-,gray] (-3,3)--(-1,4)--(3,4);
\draw[-,gray] (-2,3.5)--(2,3.5)--(2,1.5)--(3,2);
\draw[-,gray] (3.4,0)--(3,0)--(3.4,0.2);\draw[-,gray] (-3.4,-1)--(-3,-1)--(-3.4,-1.2);\draw[-,gray] (1,-1)--(0.6,-1.2);
\draw[-,gray] (-3,1)--(-3.4,1);\draw[-,gray] (-3,3)--(-3.4,3);\draw[-,gray] (-2,3.5)--(-2.4,3.5);\draw[-,gray] (-1,4)--(-1.4,4);
\filldraw[fill=white,draw=black] (-3,3) circle (2.0pt);
\filldraw[fill=black,draw=black] (-1,3) circle (2.0pt);
\fill[black!] (-1,2.83) circle (0.01pt)
node[right=5.5pt]{\color{black}\small $x''_j$};
\filldraw[fill=white,draw=black] (1,3) circle (2.0pt);
\filldraw[fill=black,draw=black] (-2,3.5) circle (2.0pt);
\fill[black!] (-1.9,3.7) circle (0.01pt)
node[left=1.5pt]{\color{black}\small $x''_i$};
\filldraw[fill=white,draw=black] (0,3.5) circle (2.0pt);
\filldraw[fill=black,draw=black] (2,3.5) circle (2.0pt)
node[right=1.5pt]{\color{black}\small $x'''_i$};
\filldraw[fill=white,draw=black] (-1,4) circle (2.0pt);
\filldraw[fill=black,draw=black] (1,4) circle (2.0pt)
node[above=1.5pt]{\color{black}\small $x'''_j$};
\filldraw[fill=white,draw=black] (3,4) circle (2.0pt);
\filldraw[fill=white,draw=black] (3,0) circle (2.0pt);
\filldraw[fill=black,draw=black] (3,2) circle (2.0pt)
node[right=1.5pt]{\color{black}\small $x''_k$};
\filldraw[fill=black,draw=black] (-1,2) circle (2.0pt)
node[right=1.5pt]{\color{black}\small $x'''_k$};
\filldraw[fill=white,draw=black] (-3,-1) circle (2.0pt);
\filldraw[fill=black,draw=black] (-3,1) circle (2.0pt);
\fill[black!] (-3,1.1) circle (0.01pt)
node[left=1.5pt]{\color{black}\small $x_k$};
\filldraw[fill=black,draw=black] (-1,-1) circle (2.0pt)
node[below=1.5pt]{\color{black}\small $x_j$};
\filldraw[fill=black,draw=black] (1,1) circle (2.0pt)
node[right=2.5pt]{\color{black}\small $x'_k$};
\filldraw[fill=black,draw=black] (2,-0.5) circle (2.0pt);
\fill[black!] (2.1,-0.5) circle (0.01pt)
node[below=0pt]{\color{black}\small $x'_i$};
\filldraw[fill=white,draw=black] (-1,1) circle (2.0pt);
\filldraw[fill=white,draw=black] (1,-1) circle (2.0pt);
\filldraw[fill=white,draw=black] (2,1.5) circle (2.0pt);
\filldraw[fill=black,draw=black] (1,0) circle (2.0pt)
node[below=1.5pt]{\color{black}\small $x'_j$};
\draw[-,very thick] (-1,-1)--(1,1)--(-1,3)--(-3,1)--(-1,-1);
\draw[-,dashed,very thick] (1,0)--(3,2)--(1,4)--(-1,2)--(1,0);
\draw[-,very thick] (2,-0.5)--(3,2)--(2,3.5)--(1,1)--(2,-0.5);
\draw[-,very thick] (-1,3)--(2,3.5)--(1,4)--(-2,3.5)--(-1,3);
\draw[->,black,dashed] (-2.9,3.25)--(1.5,3.25)--(1.5,-0.75)--(1.9,-0.75);
\draw[->,black,dashed] (-0.5,-1.25)--(0,-1)--(0,3)--(2,4);
\draw[->,black,dashed] (-3.4,0)--(1,0)--(3,1);
\draw[->,black,thick,dotted] (-1.9,3.75)--(2.5,3.75)--(2.5,-0.25)--(2.9,-0.25);
\draw[->,black,thick,dotted] (-2.5,-1.25)--(-2,-1)--(-2,3)--(0,4);
\draw[->,black,thick,dotted] (-3.4,2)--(1,2)--(3,3);
\draw[black] (-1.6,3.75) circle (0.01pt)
node[above=-1.5pt]{\color{black}\small $p'$};
\draw[black] (-2.9,3.25) circle (0.01pt)
node[left=1.5pt]{\color{black}\small $p$};
\draw[black] (-2.5,-1.25) circle (0.01pt)
node[below=1.5pt]{\color{black}\small $q'$};
\draw[black] (-0.5,-1.35) circle (0.01pt)
node[below=1.5pt]{\color{black}\small $q$};
\draw[black] (-3.4,2) circle (0.01pt)
node[left=1.5pt]{\color{black}\small $r'$};
\draw[black] (-3.4,0) circle (0.01pt)
node[left=1.5pt]{\color{black}\small $r$};
\end{scope}

\end{tikzpicture}
\caption{A deformation in the vertex formulation.  Note that the rapidity lines which are not shown for the hidden four-square on the right hand side are assigned according to Figure \ref{4VBW}.}
\label{Vcube2}
\end{figure}

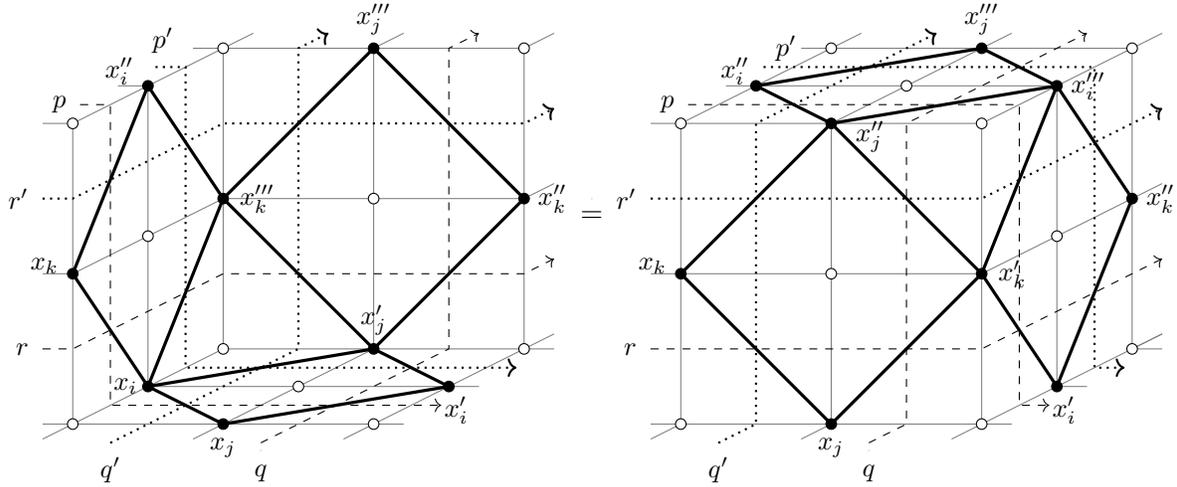
\begin{figure}[hbt!]
\centering
\begin{tikzpicture}

\begin{scope}[scale=1.0]
\draw[-,gray] (-3.4,-1)--(1.4,-1);\draw[-,gray] (0.6,-1.2)--(3.4,0.2);\draw[-,gray] (2.4,-0.5)--(-2,-0.5);\draw[-,gray] (1,0)--(-1.4,-1.2);\draw[-,gray] (-3.4,-1.2)--(-1,0)--(3.4,0);
\draw[-,gray] (-3,-1)--(-3,3)--(-1,4)--(-1,0); \draw[-,gray] (-2,-0.5)--(-2,3.5)--(-2.4,3.5); \draw[-,gray] (-1,2)--(-3,1)--(-3.4,1); \draw[-,gray] (-3,3)--(-3.4,3); \draw[-,gray] (-1,4)--(-1.4,4);
\draw[-,gray] (3,0)--(3,4)--(-1.4,4);\draw[-,gray] (3,4)--(3.4,4.2);\draw[-,gray] (1.4,4.2)--(1,4)--(1,0);\draw[-,gray] (3.4,2.2)--(3,2)--(-1,2);\draw[-,gray] (-1,4)--(-0.6,4.2);
\filldraw[fill=black,draw=black] (1,0) circle (2.0pt)
node[above=1.5pt]{\color{black}\small $x'_j$};
\filldraw[fill=black,draw=black] (-2,-0.5) circle (2.0pt)
node[left=0pt]{\color{black}\small $x_i$};
\filldraw[fill=black,draw=black] (-1,-1) circle (2.0pt)
node[below=1.5pt]{\color{black}\small $x_j$};
\filldraw[fill=black,draw=black] (2,-0.5) circle (2.0pt);
\fill[black!] (2.1,-0.5) circle (0.01pt)
node[below=0pt]{\color{black}\small $x'_i$};
\filldraw[fill=white,draw=black] (1,-1) circle (2.0pt);
\filldraw[fill=white,draw=black] (0,-0.5) circle (2.0pt);
\filldraw[fill=white,draw=black] (-1,0) circle (2.0pt);
\filldraw[fill=white,draw=black] (-3,-1) circle (2.0pt);
\filldraw[fill=white,draw=black] (3,0) circle (2.0pt);
\filldraw[fill=black,draw=black] (-2,3.5) circle (2.0pt);
\fill[black!] (-2,3.7) circle (0.01pt)
node[left=1.5pt]{\color{black}\small $x''_i$};
\filldraw[fill=black,draw=black] (-3,1) circle (2.0pt);
\fill[black!] (-3,1.1) circle (0.01pt)
node[left=1.5pt]{\color{black}\small $x_k$};
\filldraw[fill=black,draw=black] (-1,2) circle (2.0pt)
node[right=2.5pt]{\color{black}\small $x'''_k$};
\filldraw[fill=white,draw=black] (-2,1.5) circle (2.0pt);
\filldraw[fill=white,draw=black] (-3,3) circle (2.0pt);
\filldraw[fill=white,draw=black] (-1,4) circle (2.0pt);
\filldraw[fill=black,draw=black] (3,2) circle (2.0pt)
node[right=1.5pt]{\small $x''_k$};
\filldraw[fill=black,draw=black] (1,4) circle (2.0pt)
node[above=1.5pt]{\small $x'''_j$};
\filldraw[fill=white,draw=black] (1,2) circle (2.0pt);
\filldraw[fill=white,draw=black] (3,4) circle (2.0pt);
\draw[-,very thick] (-1,-1)--(2,-0.5)--(1,0)--(-2,-0.5)--(-1,2)--(-2,3.5)--(-3,1)--(-2,-0.5)--(-1,-1);
\draw[-,very thick] (1,0)--(-1,2)--(1,4)--(3,2)--(1,0);
\draw[->,black,thick,dotted] (-1.9,3.75)--(-1.5,3.75)--(-1.5,-0.25)--(2.9,-0.25);
\draw[->,black,thick,dotted] (-2.5,-1.25)--(0,0)--(0,4)--(0.4,4.2);
\draw[->,black,dashed] (-2.9,3.25)--(-2.5,3.25)--(-2.5,-0.75)--(1.9,-0.75);
\draw[->,black,dashed] (-0.5,-1.25)--(2,0)--(2,4)--(2.4,4.2);
\draw[->,black,dashed] (-3.4,0)--(-3,0)--(-1,1)--(3,1)--(3.4,1.2);
\draw[->,black,thick,dotted] (-3.4,2)--(-3,2)--(-1,3)--(3,3)--(3.4,3.2);
\draw[black] (-1.8,3.75) circle (0.01pt)
node[above=1.5pt]{\color{black}\small $p'$};
\draw[black] (-2.5,-1.25) circle (0.01pt)
node[below=1.5pt]{\color{black}\small $q'$};
\draw[black] (-2.9,3.25) circle (0.01pt)
node[left=1.5pt]{\color{black}\small $p$};
\draw[black] (-0.5,-1.35) circle (0.01pt)
node[below=1.5pt]{\color{black}\small $q$};
\draw[black] (-3.4,0) circle (0.01pt)
node[left=1.5pt]{\color{black}\small $r$};
\draw[black] (-3.4,2) circle (0.01pt)
node[left=1.5pt]{\color{black}\small $r'$};
\end{scope}

\draw[black] (3.9,2) circle (0.01pt)
node[below=1pt]{\color{black}$=$};

\begin{scope}[scale=1.0,xshift=230,yshift=0,rotate=0]
\draw[-,gray] (-1.4,-1.2)--(-1,-1)--(1,-1)--(1,1)--(-1,1)--(-1,-1)--(-3,-1)--(-3,1)--(-1,1)--(-1,3)--(1,4);
\draw[-,gray] (1,3)--(3,4)--(3,0)--(2,-0.5)--(2,1.5)--(1,1)--(1,3)--(-3,3)--(-3,1);
\draw[-,gray] (1.4,-1)--(1,-1)--(2,-0.5)--(2.4,-0.5);
\draw[-,gray] (-3,3)--(-1,4)--(3,4);
\draw[-,gray] (-2,3.5)--(2,3.5)--(2,1.5)--(3,2);
\draw[-,gray] (3.4,0)--(3,0)--(3.4,0.2);\draw[-,gray] (-3.4,-1)--(-3,-1)--(-3.4,-1.2);\draw[-,gray] (1,-1)--(0.6,-1.2);
\draw[-,gray] (-3,1)--(-3.4,1);\draw[-,gray] (-3,3)--(-3.4,3);\draw[-,gray] (-2,3.5)--(-2.4,3.5);\draw[-,gray] (-1,4)--(-1.4,4);
\draw[-,gray] (3,0)--(3.4,0.2);\draw[-,gray] (3,2)--(3.4,2.2);\draw[-,gray] (3,4)--(3.4,4.2);\draw[-,gray] (1,4)--(1.4,4.2);\draw[-,gray] (-1,4)--(-0.6,4.2);
\filldraw[fill=white,draw=black] (-3,3) circle (2.0pt);
\filldraw[fill=black,draw=black] (-1,3) circle (2.0pt);
\fill[black!] (-1,2.83) circle (0.01pt)
node[right=5.5pt]{\color{black}\small $x''_j$};
\filldraw[fill=white,draw=black] (1,3) circle (2.0pt);
\filldraw[fill=black,draw=black] (-2,3.5) circle (2.0pt);
\fill[black!] (-1.9,3.7) circle (0.01pt)
node[left=1.5pt]{\color{black}\small $x''_i$};
\filldraw[fill=white,draw=black] (0,3.5) circle (2.0pt);
\filldraw[fill=black,draw=black] (2,3.5) circle (2.0pt)
node[right=1.5pt]{\color{black}\small $x'''_i$};
\filldraw[fill=white,draw=black] (-1,4) circle (2.0pt);
\filldraw[fill=black,draw=black] (1,4) circle (2.0pt)
node[above=1.5pt]{\color{black}\small $x'''_j$};
\filldraw[fill=white,draw=black] (3,4) circle (2.0pt);
\filldraw[fill=white,draw=black] (3,0) circle (2.0pt);
\filldraw[fill=black,draw=black] (3,2) circle (2.0pt)
node[right=1.5pt]{\color{black}\small $x''_k$};
\filldraw[fill=white,draw=black] (-3,-1) circle (2.0pt);
\filldraw[fill=black,draw=black] (-3,1) circle (2.0pt);
\fill[black!] (-3,1.1) circle (0.01pt)
node[left=1.5pt]{\color{black}\small $x_k$};
\filldraw[fill=black,draw=black] (-1,-1) circle (2.0pt)
node[below=1.5pt]{\color{black}\small $x_j$};
\filldraw[fill=black,draw=black] (1,1) circle (2.0pt)
node[right=2.5pt]{\color{black}\small $x'_k$};
\filldraw[fill=black,draw=black] (2,-0.5) circle (2.0pt);
\fill[black!] (2.1,-0.5) circle (0.01pt)
node[below=0pt]{\color{black}\small $x'_i$};
\filldraw[fill=white,draw=black] (-1,1) circle (2.0pt);
\filldraw[fill=white,draw=black] (1,-1) circle (2.0pt);
\filldraw[fill=white,draw=black] (2,1.5) circle (2.0pt);
\draw[-,very thick] (-1,-1)--(1,1)--(-1,3)--(-3,1)--(-1,-1);
\draw[-,very thick] (2,-0.5)--(3,2)--(2,3.5)--(1,1)--(2,-0.5);
\draw[-,very thick] (-1,3)--(2,3.5)--(1,4)--(-2,3.5)--(-1,3);
\draw[->,black,dashed] (-2.9,3.25)--(1.5,3.25)--(1.5,-0.75)--(1.9,-0.75);
\draw[->,black,dashed] (-0.5,-1.25)--(0,-1)--(0,3)--(2.4,4.2);
\draw[->,black,dashed] (-3.4,0)--(1,0)--(3.4,1.2);
\draw[->,black,thick,dotted] (-1.9,3.75)--(2.5,3.75)--(2.5,-0.25)--(2.9,-0.25);
\draw[->,black,thick,dotted] (-2.5,-1.25)--(-2,-1)--(-2,3)--(0.4,4.2);
\draw[->,black,thick,dotted] (-3.4,2)--(1,2)--(3.4,3.2);
\draw[black] (-1.6,3.75) circle (0.01pt)
node[above=-1.5pt]{\color{black}\small $p'$};
\draw[black] (-2.9,3.25) circle (0.01pt)
node[left=1.5pt]{\color{black}\small $p$};
\draw[black] (-2.5,-1.25) circle (0.01pt)
node[below=1.5pt]{\color{black}\small $q'$};
\draw[black] (-0.5,-1.35) circle (0.01pt)
node[below=1.5pt]{\color{black}\small $q$};
\draw[black] (-3.4,2) circle (0.01pt)
node[left=1.5pt]{\color{black}\small $r'$};
\draw[black] (-3.4,0) circle (0.01pt)
node[left=1.5pt]{\color{black}\small $r$};
\end{scope}

\end{tikzpicture}
\caption{A deformation in the vertex formulation that is equivalent to the Yang-Baxter equation \eqref{YBE-vertex}.}
\label{Vcube3}
\end{figure}

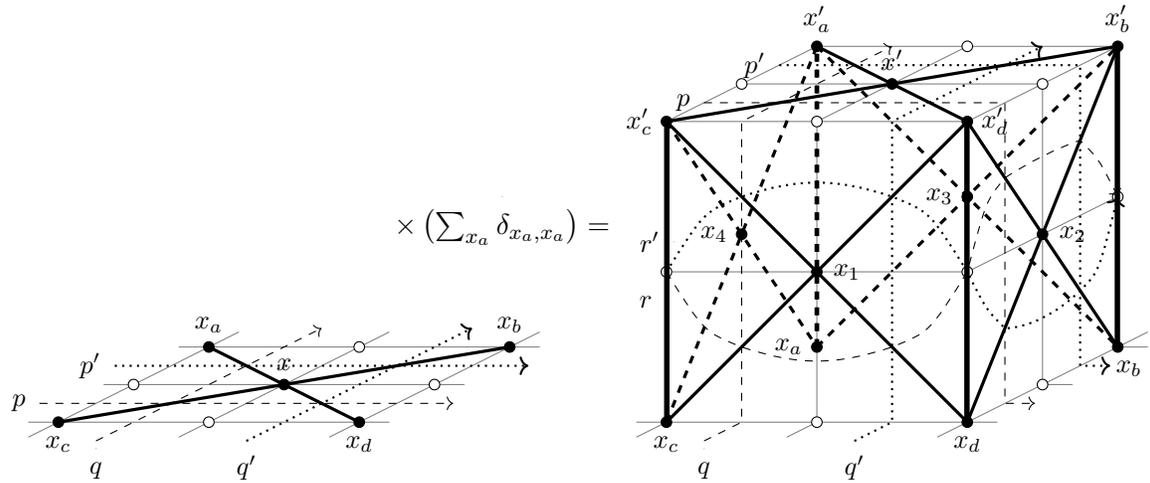
\begin{figure}[hbt!]
\centering
\begin{tikzpicture}

\begin{scope}[scale=1.0]
\draw[-,gray] (-3.4,-1)--(1.4,-1);\draw[-,gray] (0.6,-1.2)--(3.4,0.2);\draw[-,gray] (2.4,-0.5)--(-2.4,-0.5);\draw[-,gray] (1.4,0.2)--(-1.4,-1.2);\draw[-,gray] (-3.4,-1.2)--(-0.6,0.2);\draw[-,gray] (-1.4,0)--(3.4,0);
\filldraw[fill=white,draw=black] (1,0) circle (2.0pt);
\filldraw[fill=white,draw=black] (-2.0,-0.5) circle (2.0pt);
\filldraw[fill=white,draw=black] (-1,-1) circle (2.0pt);
\filldraw[fill=white,draw=black] (2.0,-0.5) circle (2.0pt);
\filldraw[fill=black,draw=black] (1,-1) circle (2.0pt)
node[below=1.5pt]{\small $x_d$};
\filldraw[fill=black,draw=black] (0,-0.5) circle (2.0pt)
node[above=1.5pt]{\small $x$};
\filldraw[fill=black,draw=black] (-1,0) circle (2.0pt)
node[above=1.5pt]{\small $x_a$};
\filldraw[fill=black,draw=black] (-3,-1) circle (2.0pt)
node[below=1.5pt]{\small $x_c$};
\filldraw[fill=black,draw=black] (3,0) circle (2.0pt)
node[above=1.5pt]{\small $x_b$};
\draw[-,very thick] (-3,-1)--(3,0);\draw[-,very thick] (-1,0)--(1,-1);
\draw[->,black,thick,dotted] (-2.25,-0.25)--(3.25,-0.25);
\draw[->,black,dashed] (-2.5,-1.25)--(0.5,0.25);
\draw[->,black,dashed] (-3.25,-0.75)--(2.25,-0.75);
\draw[->,black,thick,dotted] (-0.5,-1.25)--(2.5,0.25);
\draw[black] (-2.25,-0.25) circle (0.01pt)
node[left=1.5pt]{\color{black}\small $p'$};
\draw[black] (-2.5,-1.35) circle (0.01pt)
node[below=1.5pt]{\color{black}\small $q$};
\draw[black] (-3.25,-0.75) circle (0.01pt)
node[left=1.5pt]{\color{black}\small $p$};
\draw[black] (-0.5,-1.25) circle (0.01pt)
node[below=1.5pt]{\color{black}\small $q'$};
\end{scope}

\draw[black] (2.9,2) circle (0.01pt)
node[below=1pt]{\color{black}$\times\left(\sum_{x_a}\delta_{x_a,x_a}\right)=$};

\begin{scope}[scale=1.0,xshift=230,yshift=0,rotate=0]
\draw[-,gray] (-1.4,-1.2)--(-1,-1)--(1,-1)--(1,1)--(-1,1)--(-1,-1)--(-3,-1)--(-3,1)--(-1,1)--(-1,3)--(1,4);
\draw[-,gray] (1,3)--(3,4)--(3,0)--(2,-0.5)--(2,1.5)--(1,1)--(1,3)--(-3,3)--(-3,1);
\draw[-,gray] (1.4,-1)--(1,-1)--(2,-0.5)--(2.4,-0.5);
\draw[-,gray] (-3,3)--(-1,4)--(3,4);
\draw[-,gray] (-2,3.5)--(2,3.5)--(2,1.5)--(3,2);
\draw[-,gray] (3.4,0)--(3,0)--(3.4,0.2);\draw[-,gray] (-3.4,-1)--(-3,-1)--(-3.4,-1.2);\draw[-,gray] (1,-1)--(0.6,-1.2);
\draw[-,dashed,very thick] (-1.01,0)--(-1.01,4);\draw[-,dashed,very thick] (-0.99,0)--(-0.99,4);
\filldraw[fill=black,draw=black] (-3,3) circle (2.0pt)
node[left=1.5pt]{\small $x'_c$};
\filldraw[fill=white,draw=black] (-1,3) circle (2.0pt);
\filldraw[fill=black,draw=black] (1,3) circle (2.0pt)
node[right=1.5pt]{\small $x'_d$};
\filldraw[fill=white,draw=black] (-2,3.5) circle (2.0pt);
\filldraw[fill=black,draw=black] (0,3.5) circle (2.0pt)
node[above=1.5pt]{\small $x'$};
\filldraw[fill=white,draw=black] (2,3.5) circle (2.0pt);
\filldraw[fill=black,draw=black] (-1,4) circle (2.0pt)
node[above=1.5pt]{\small $x'_a$};
\filldraw[fill=white,draw=black] (1,4) circle (2.0pt);
\filldraw[fill=black,draw=black] (3,4) circle (2.0pt)
node[above=1.5pt]{\small $x'_b$};
\filldraw[fill=black,draw=black] (3,0) circle (2.0pt);
\fill[black!] (3.15,0) circle (0.01pt)
node[below=1.5pt]{\color{black}\small $x_b$};
\filldraw[fill=white,draw=black] (3,2) circle (2.0pt);
\filldraw[fill=black,draw=black] (-1,0) circle (2.0pt)
node[left=1.5pt]{\small $x_a$};
\filldraw[fill=black,draw=black] (-3,-1) circle (2.0pt)
node[below=1.5pt]{\small $x_c$};
\filldraw[fill=white,draw=black] (-3,1) circle (2.0pt);
\filldraw[fill=black,draw=black] (-2,1.5) circle (2.0pt)
node[left=1.5pt]{\small $x_4$};
\filldraw[fill=white,draw=black] (-1,-1) circle (2.0pt);
\filldraw[fill=white,draw=black] (1,1) circle (2.0pt);
\filldraw[fill=white,draw=black] (2,-0.5) circle (2.0pt);
\filldraw[fill=black,draw=black] (-1,1) circle (2.0pt)
node[right=2.5pt]{\small $x_1$};
\filldraw[fill=black,draw=black] (1,-1) circle (2.0pt)
node[below=1.5pt]{\small $x_d$};
\filldraw[fill=black,draw=black] (2,1.5) circle (2.0pt)
node[right=2.5pt]{\small $x_2$};
\filldraw[fill=black,draw=black] (1,2) circle (2.0pt)
node[left=1.5pt]{\small $x_3$};
\draw[-,very thick] (-3,-1)--(1,3);\draw[-,very thick] (-3,3)--(1,-1);
\draw[-,dashed,very thick] (-1,0)--(3,4);\draw[-,dashed,very thick] (-1,4)--(3,0);
\draw[-,very thick] (1,-1)--(3,4);\draw[-,very thick] (1,3)--(3,0);
\draw[-,dashed,very thick] (-3,-1)--(-1,4);\draw[-,dashed,very thick] (-3,3)--(-1,0);
\draw[-,very thick] (-3,3)--(3,4);\draw[-,very thick] (-1,4)--(1,3);
\draw[-,very thick] (-2.98,-1)--(-2.98,3);\draw[-,very thick] (0.98,-1)--(0.98,3);\draw[-,very thick] (2.98,0)--(2.98,4);\draw[-,very thick] (-3.01,-1)--(-3.01,3);\draw[-,very thick] (1.01,-1)--(1.01,3);\draw[-,very thick] (3.01,0)--(3.01,4);
\draw[->,black,dashed] (-2.5,3.25)--(1.5,3.25)--(1.5,-0.75)--(1.9,-0.75);
\draw[->,black,thick,dotted] (-0.5,-1.25)--(0,-1)--(0,3)--(2,4);
\draw[->,black,dashed] (-3,1) .. controls (-2.5,0.2) .. (-2,0) .. controls (-1.25,-0.25) and (-0.75,-0.25) .. (0,0) .. controls (0.5,0.2) .. (1,1) .. controls (1.25,1.95) .. (1.5,2.25) .. controls (1.8,2.45) and (2.2,2.65) .. (2.5,2.75) .. controls (2.75,2.5) .. (3,2);
\draw[->,black,thick,dotted] (-1.5,3.75)--(2.5,3.75)--(2.5,-0.25)--(2.9,-0.25);
\draw[->,black,dashed] (-2.5,-1.25)--(-2,-1)--(-2,3)--(0,4);
\draw[->,black,thick,dotted] (-3,1) .. controls (-2.5,1.8) .. (-2,2) .. controls (-1.25,2.25) and (-0.75,2.25) .. (0,2) .. controls (0.5,1.8) .. (1,1) .. controls (1.25,0.5) .. (1.5,0.25) .. controls (1.8,0.25) and (2.2,0.45) .. (2.5,0.75) .. controls (2.75,1) .. (3,2);
\draw[black] (-1.5,3.75) circle (0.01pt)
node[left=1.5pt]{\color{black}\small $p'$};
\draw[black] (-2.5,3.25) circle (0.01pt)
node[left=1.5pt]{\color{black}\small $p$};
\draw[black] (-2.5,-1.35) circle (0.01pt)
node[below=1.5pt]{\color{black}\small $q$};
\draw[black] (-0.5,-1.25) circle (0.01pt)
node[below=1.5pt]{\color{black}\small $q'$};
\draw[black] (-2.9,1.4) circle (0.01pt)
node[left=1.5pt]{\color{black}\small $r'$};
\draw[black] (-3,0.6) circle (0.01pt)
node[left=1.5pt]{\color{black}\small $r$};
\end{scope}

\end{tikzpicture}
\caption{A deformation in the IRF formulation corresponding to \eqref{irfeq1}.  Note that the rapidity lines which are not shown for two hidden four-squares on the right hand side are assigned according to Figure \ref{4IRFBW}. }
\label{IRFcube}
\end{figure}

\begin{figure}[hbt!]
\centering
\begin{tikzpicture}

\begin{scope}[scale=1.0]
\draw[-,gray] (-3.4,-1)--(1.4,-1);\draw[-,gray] (0.6,-1.2)--(3.4,0.2);\draw[-,gray] (2.4,-0.5)--(-2.4,-0.5);\draw[-,gray] (1.4,0.2)--(-1.4,-1.2);\draw[-,gray] (-3.4,-1.2)--(-0.6,0.2);\draw[-,gray] (-1,0)--(3.4,0);
\draw[-,gray] (-3,-1)--(-3,3)--(-1,4)--(-1,0);\draw[-,gray] (-3.4,1)--(-3,1)--(-1,2);\draw[-,gray] (-2,-0.5)--(-2,3.5)--(-2.4,3.5);\draw[-,gray] (-3.4,3)--(-3,3);\draw[-,gray] (-1.4,4)--(-1,4);
\filldraw[fill=white,draw=black] (1,0) circle (2.0pt);
\filldraw[fill=white,draw=black] (-2.0,-0.5) circle (2.0pt);
\filldraw[fill=white,draw=black] (-1,-1) circle (2.0pt);
\filldraw[fill=white,draw=black] (2.0,-0.5) circle (2.0pt);
\filldraw[fill=black,draw=black] (1,-1) circle (2.0pt)
node[below=1.5pt]{\color{black}\small $x_d$};
\filldraw[fill=black,draw=black] (0,-0.5) circle (2.0pt)
node[above=1.5pt]{\small $x$};
\filldraw[fill=black,draw=black] (-1,0) circle (2.0pt);
\fill[black!] (-0.75,0.05) circle (0.01pt)
node[above=1.5pt]{\color{black}\small $x_a$};
\filldraw[fill=black,draw=black] (-3,-1) circle (2.0pt)
node[below=1.5pt]{\small $x_c$};
\filldraw[fill=black,draw=black] (3,0) circle (2.0pt)
node[above=1.5pt]{\small $x_b$};
\filldraw[fill=white,draw=black] (-3,1) circle (2.0pt);
\filldraw[fill=white,draw=black] (-1,2) circle (2.0pt);
\filldraw[fill=white,draw=black] (-2,3.5) circle (2.0pt);
\filldraw[fill=black,draw=black] (-3,3) circle (2.0pt);
\fill[black!] (-3,3.2) circle (0.01pt)
node[left=1.5pt]{\color{black}\small $x'_c$};
\filldraw[fill=black,draw=black] (-1,4) circle (2.0pt)
node[above=1.5pt]{\small $x'_a$};
\filldraw[fill=black,draw=black] (-2,1.5) circle (2.0pt);
\fill[black!] (-2,1.37) circle (0.01pt)
node[right=1.5pt]{\color{black}\small $x_4$};
\draw[-,very thick] (-3,-1)--(3,0);\draw[-,very thick] (-1,0)--(1,-1);
\draw[-,very thick] (-3,-1)--(-1,4);\draw[-,very thick] (-3,3)--(-1,0);
\draw[-,very thick] (-2.98,-1)--(-2.98,3);\draw[-,very thick] (-0.98,0)--(-0.98,4);
\draw[-,very thick] (-3.01,-1)--(-3.01,3);\draw[-,very thick] (-1.01,0)--(-1.01,4);
\draw[->,black,thick,dotted] (-1.9,3.75)--(-1.5,3.75)--(-1.5,-0.25)--(3.25,-0.25);
\draw[->,black,dashed] (-2.5,-1.25)--(0.5,0.25);
\draw[->,black,dashed] (-2.9,3.25)--(-2.5,3.25)--(-2.5,-0.75)--(2.25,-0.75);
\draw[->,black,thick,dotted] (-0.5,-1.25)--(2.5,0.25);
\draw[->,black,dashed] (-3.5,0.2) .. controls (-3.25,0.4) ..(-3,1) .. controls (-2.75,1.95) .. (-2.5,2.25) .. controls (-2.2,2.45) and (-1.8,2.65) .. (-1.5,2.75) .. controls (-1.25,2.5) .. (-1,2);
\draw[->,black,thick,dotted] (-3.5,1.8) .. controls (-3.25,1.6) .. (-3,1) .. controls (-2.75,0.5) .. (-2.5,0.25) .. controls (-2.2,0.25) and (-1.8,0.45) .. (-1.5,0.75) .. controls (-1.25,1) .. (-1,2);
\draw[black] (-1.9,3.75) circle (0.01pt)
node[above=1.5pt]{\color{black}\small $p'$};
\draw[black] (-2.5,-1.35) circle (0.01pt)
node[below=1.5pt]{\color{black}\small $q$};
\draw[black] (-2.9,3.25) circle (0.01pt)
node[above=1.5pt]{\color{black}\small $p$};
\draw[black] (-0.5,-1.25) circle (0.01pt)
node[below=1.5pt]{\color{black}\small $q'$};
\draw[black] (-3,2) circle (0.01pt)
node[left=1.5pt]{\color{black}\small $r'$};
\draw[black] (-3,0) circle (0.01pt)
node[left=1.5pt]{\color{black}\small $r$};
\end{scope}

\draw[black] (3.5,2) circle (0.01pt)
node[below=1pt]{\color{black}$=$};

\begin{scope}[scale=1.0,xshift=230,yshift=0,rotate=0]
\draw[-,gray] (-1.4,-1.2)--(-1,-1)--(1,-1)--(1,1)--(-1,1)--(-1,-1)--(-3,-1)--(-3,1)--(-1,1)--(-1,3)--(1,4);
\draw[-,gray] (1,3)--(3,4)--(3,0)--(2,-0.5)--(2,1.5)--(1,1)--(1,3)--(-3,3)--(-3,1);
\draw[-,gray] (1.4,-1)--(1,-1)--(2,-0.5)--(2.4,-0.5);
\draw[-,gray] (-3,3)--(-1,4)--(3,4);
\draw[-,gray] (-2,3.5)--(2,3.5)--(2,1.5)--(3,2);
\draw[-,gray] (3.4,0)--(3,0)--(3.4,0.2);\draw[-,gray] (-3.4,-1)--(-3,-1)--(-3.4,-1.2);\draw[-,gray] (1,-1)--(0.6,-1.2);
\draw[-,gray] (-3,1)--(-3.4,1);\draw[-,gray] (-3,3)--(-3.4,3);\draw[-,gray] (-2,3.5)--(-2.4,3.5);\draw[-,gray] (-1,4)--(-1.4,4);
\filldraw[fill=black,draw=black] (-3,3) circle (2.0pt);
\fill[black!] (-3.2,3) circle (0.01pt)
node[below=1.5pt]{\color{black}\small $x'_c$};
\filldraw[fill=white,draw=black] (-1,3) circle (2.0pt);
\filldraw[fill=black,draw=black] (1,3) circle (2.0pt)
node[right=1.5pt]{\small $x'_d$};
\filldraw[fill=white,draw=black] (-2,3.5) circle (2.0pt);
\filldraw[fill=black,draw=black] (0,3.5) circle (2.0pt)
node[above=1.5pt]{\small $x'$};
\filldraw[fill=white,draw=black] (2,3.5) circle (2.0pt);
\filldraw[fill=black,draw=black] (-1,4) circle (2.0pt)
node[above=1.5pt]{\small $x'_a$};
\filldraw[fill=white,draw=black] (1,4) circle (2.0pt);
\filldraw[fill=black,draw=black] (3,4) circle (2.0pt)
node[above=1.5pt]{\small $x'_b$};
\filldraw[fill=black,draw=black] (3,0) circle (2.0pt);
\fill[black!] (3.15,0) circle (0.01pt)
node[below=1.5pt]{\color{black}\small $x_b$};
\filldraw[fill=white,draw=black] (3,2) circle (2.0pt);
\filldraw[fill=black,draw=black] (-1,0) circle (2.0pt);
\fill[black!] (-1,0.1) circle (0.01pt)
node[left=1.5pt]{\color{black}\small $x_a$};
\filldraw[fill=black,draw=black] (-3,-1) circle (2.0pt)
node[below=1.5pt]{\small $x_c$};
\filldraw[fill=white,draw=black] (-3,1) circle (2.0pt);
\filldraw[fill=white,draw=black] (-1,-1) circle (2.0pt);
\filldraw[fill=white,draw=black] (1,1) circle (2.0pt);
\filldraw[fill=white,draw=black] (2,-0.5) circle (2.0pt);
\filldraw[fill=black,draw=black] (-1,1) circle (2.0pt)
node[right=2.5pt]{\small $x_1$};
\filldraw[fill=black,draw=black] (1,-1) circle (2.0pt)
node[below=1.5pt]{\small $x_d$};
\filldraw[fill=black,draw=black] (2,1.5) circle (2.0pt)
node[right=2.5pt]{\small $x_2$};
\filldraw[fill=black,draw=black] (1,2) circle (2.0pt)
node[left=1.5pt]{\small $x_3$};
\draw[-,very thick] (-3,-1)--(1,3);\draw[-,very thick] (-3,3)--(1,-1);
\draw[-,dashed,very thick] (-1,0)--(3,4);\draw[-,dashed,very thick] (-1,4)--(3,0);
\draw[-,very thick] (1,-1)--(3,4);\draw[-,very thick] (1,3)--(3,0);
\draw[-,very thick] (-3,3)--(3,4);\draw[-,very thick] (-1,4)--(1,3);
\draw[-,very thick] (0.98,-1)--(0.98,3);\draw[-,very thick] (2.98,0)--(2.98,4);
\draw[-,very thick] (1.01,-1)--(1.01,3);\draw[-,very thick] (3.01,0)--(3.01,4);
\draw[->,black,dashed] (-2.9,3.25)--(1.5,3.25)--(1.5,-0.75)--(1.9,-0.75);
\draw[->,black,thick,dotted] (-0.5,-1.25)--(0,-1)--(0,3)--(2,4);
\draw[->,black,dashed] (-3.4,0)--(-2,0)--(0,0) .. controls (0.5,0.2) .. (1,1) .. controls (1.25,1.95) .. (1.5,2.25) .. controls (1.8,2.45) and (2.2,2.65) .. (2.5,2.75) .. controls (2.75,2.5) .. (3,2);
\draw[->,black,thick,dotted] (-1.9,3.75)--(2.5,3.75)--(2.5,-0.25)--(2.9,-0.25);
\draw[->,black,dashed] (-2.5,-1.25)--(-2,-1)--(-2,3)--(0,4);
\draw[->,black,thick,dotted] (-3.5,2)--(-2,2)--(0,2) .. controls (0.5,1.8) .. (1,1) .. controls (1.25,0.5) .. (1.5,0.25) .. controls (1.8,0.25) and (2.2,0.45) .. (2.5,0.75) .. controls (2.75,1) .. (3,2);
\draw[black] (-1.9,3.75) circle (0.01pt)
node[left=1.5pt]{\color{black}\small $p'$};
\draw[black] (-2.9,3.25) circle (0.01pt)
node[left=1.5pt]{\color{black}\small $p$};
\draw[black] (-2.5,-1.35) circle (0.01pt)
node[below=1.5pt]{\color{black}\small $q$};
\draw[black] (-0.5,-1.25) circle (0.01pt)
node[below=1.5pt]{\color{black}\small $q'$};
\draw[black] (-3.4,2) circle (0.01pt)
node[left=1.5pt]{\color{black}\small $r$};
\draw[black] (-3.3,0) circle (0.01pt)
node[left=1.5pt]{\color{black}\small $r'$};
\end{scope}

\end{tikzpicture}
\caption{A deformation in the IRF formulation.  Note that the rapidity lines which are not shown for the hidden four-square on the right hand side are assigned according to Figure \ref{4IRFBW}.}
\label{IRFcube2}
\end{figure}
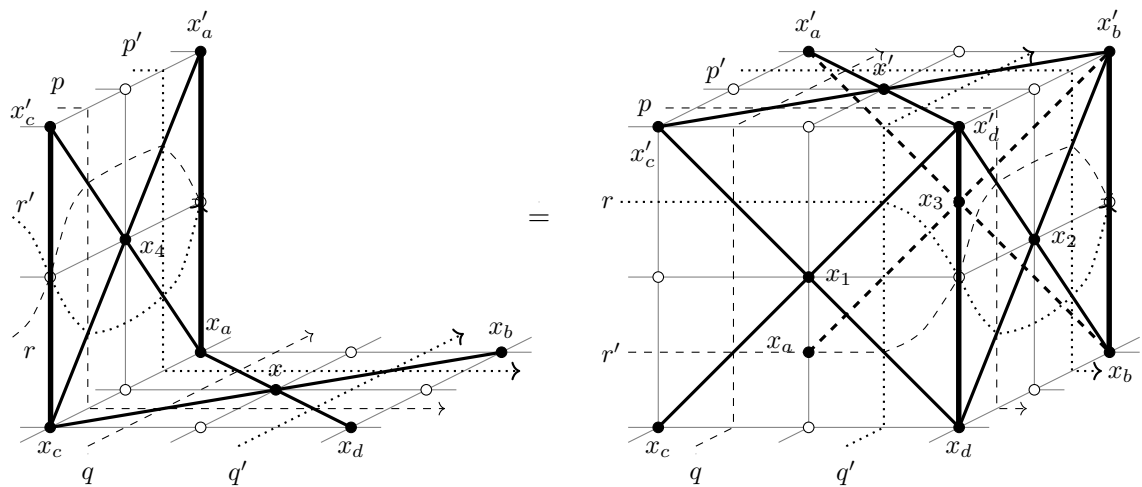

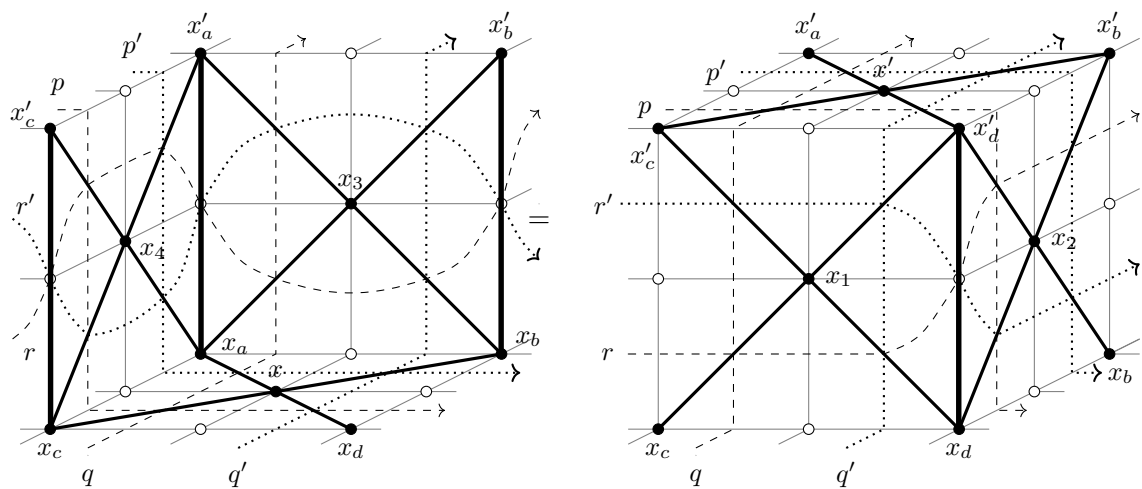
\begin{figure}[hbt!]
\centering
\begin{tikzpicture}

\begin{scope}[scale=1.0]
\draw[-,gray] (-3.4,-1)--(1.4,-1);\draw[-,gray] (0.6,-1.2)--(3.4,0.2);\draw[-,gray] (2.4,-0.5)--(-2.4,-0.5);\draw[-,gray] (1,0)--(-1.4,-1.2);\draw[-,gray] (-3.4,-1.2)--(-1,0)--(3.4,0);
\draw[-,gray] (-3,-1)--(-3,3)--(-1,4)--(-1,0);\draw[-,gray] (-3.4,1)--(-3,1)--(-1,2);\draw[-,gray] (-2,-0.5)--(-2,3.5)--(-2.4,3.5);\draw[-,gray] (-3.4,3)--(-3,3);\draw[-,gray] (-1.4,4)--(-1,4);
\draw[-,gray] (3,0)--(3,4)--(-1.4,4);\draw[-,gray] (3,4)--(3.4,4.2);\draw[-,gray] (1.4,4.2)--(1,4)--(1,0);\draw[-,gray] (3.4,2.2)--(3,2)--(-1,2);\draw[-,gray] (-1,4)--(-0.6,4.2);
\filldraw[fill=white,draw=black] (1,0) circle (2.0pt);
\filldraw[fill=white,draw=black] (-2.0,-0.5) circle (2.0pt);
\filldraw[fill=white,draw=black] (-1,-1) circle (2.0pt);
\filldraw[fill=white,draw=black] (2.0,-0.5) circle (2.0pt);
\filldraw[fill=black,draw=black] (1,-1) circle (2.0pt)
node[below=1.5pt]{\small $x_d$};
\filldraw[fill=black,draw=black] (0,-0.5) circle (2.0pt)
node[above=1.5pt]{\small $x$};
\filldraw[fill=black,draw=black] (-1,0) circle (2.0pt);
\fill[black!] (-1,0.1) circle (0.01pt)
node[right=4pt]{\color{black}\small $x_a$};
\filldraw[fill=black,draw=black] (-3,-1) circle (2.0pt)
node[below=1.5pt]{\small $x_c$};
\filldraw[fill=black,draw=black] (3,0) circle (2.0pt);
\fill[black!] (3,0.2) circle (0.01pt)
node[right=1.5pt]{\color{black}\small $x_b$};
\filldraw[fill=white,draw=black] (-3,1) circle (2.0pt);
\filldraw[fill=white,draw=black] (-1,2) circle (2.0pt);
\filldraw[fill=white,draw=black] (-2,3.5) circle (2.0pt);
\filldraw[fill=black,draw=black] (-3,3) circle (2.0pt);
\fill[black!] (-3,3.2) circle (0.01pt)
node[left=1.5pt]{\color{black}\small $x'_c$};
\filldraw[fill=black,draw=black] (-1,4) circle (2.0pt)
node[above=1.5pt]{\small $x'_a$};
\filldraw[fill=black,draw=black] (-2,1.5) circle (2.0pt);
\fill[black!] (-2,1.37) circle (0.01pt)
node[right=1.5pt]{\color{black}\small $x_4$};
\filldraw[fill=white,draw=black] (3,2) circle (2.0pt);
\filldraw[fill=white,draw=black] (1,4) circle (2.0pt);
\filldraw[fill=black,draw=black] (1,2) circle (2.0pt)
node[above=1.5pt]{\small $x_3$};
\filldraw[fill=black,draw=black] (3,4) circle (2.0pt)
node[above=1.5pt]{\small $x'_b$};
\draw[-,very thick] (-3,-1)--(3,0);\draw[-,very thick] (-1,0)--(1,-1);
\draw[-,very thick] (-3,-1)--(-1,4);\draw[-,very thick] (-3,3)--(-1,0);
\draw[-,very thick] (-1,0)--(3,4);\draw[-,very thick] (-1,4)--(3,0);
\draw[-,very thick] (-2.98,-1)--(-2.98,3);\draw[-,very thick] (-0.98,0)--(-0.98,4);\draw[-,very thick] (2.98,0)--(2.98,4);
\draw[-,very thick] (-3.01,-1)--(-3.01,3);\draw[-,very thick] (-1.01,0)--(-1.01,4);\draw[-,very thick] (3.01,0)--(3.01,4);
\draw[->,black,thick,dotted] (-1.9,3.75)--(-1.5,3.75)--(-1.5,-0.25)--(3.25,-0.25);
\draw[->,black,dashed] (-2.5,-1.25)--(0,0)--(0,4)--(0.4,4.2);
\draw[->,black,dashed] (-2.9,3.25)--(-2.5,3.25)--(-2.5,-0.75)--(2.25,-0.75);
\draw[->,black,thick,dotted] (-0.5,-1.25)--(2,0)--(2,4)--(2.4,4.2);
\draw[->,black,dashed] (-3.5,0.2) .. controls (-3.25,0.4) ..(-3,1) .. controls (-2.75,1.95) .. (-2.5,2.25) .. controls (-2.2,2.45) and (-1.8,2.65) .. (-1.5,2.75) .. controls (-1.25,2.5) .. (-1,2) .. controls (-0.5,1.2) .. (0,1) .. controls (0.75,0.75) and (1.25,0.75) .. (2,1) .. controls (2.5,1.2) .. (3,2) .. controls (3.25,2.95) .. (3.5,3.25);
\draw[->,black,thick,dotted] (-3.5,1.8) .. controls (-3.25,1.6) .. (-3,1) .. controls (-2.75,0.5) .. (-2.5,0.25) .. controls (-2.2,0.25) and (-1.8,0.45) .. (-1.5,0.75) .. controls (-1.25,1) .. (-1,2) .. controls (-0.5,2.8) .. (0,3) .. controls (0.75,3.25) and (1.25,3.25) .. (2,3) .. controls (2.5,2.8) .. (3,2) .. controls (3.25,1.5) .. (3.5,1.25);
\draw[black] (-1.9,3.75) circle (0.01pt)
node[above=1.5pt]{\color{black}\small $p'$};
\draw[black] (-2.9,3.25) circle (0.01pt)
node[above=1.5pt]{\color{black}\small $p$};
\draw[black] (-2.5,-1.35) circle (0.01pt)
node[below=1.5pt]{\color{black}\small $q$};
\draw[black] (-0.5,-1.25) circle (0.01pt)
node[below=1.5pt]{\color{black}\small $q'$};
\draw[black] (-3,2) circle (0.01pt)
node[left=1.5pt]{\color{black}\small $r'$};
\draw[black] (-3,0) circle (0.01pt)
node[left=1.5pt]{\color{black}\small $r$};
\end{scope}

\draw[black] (3.5,2) circle (0.01pt)
node[below=1pt]{\color{black}$=$};

\begin{scope}[scale=1.0,xshift=230,yshift=0,rotate=0]
\draw[-,gray] (-1.4,-1.2)--(-1,-1)--(1,-1)--(1,1)--(-1,1)--(-1,-1)--(-3,-1)--(-3,1)--(-1,1)--(-1,3)--(1,4);
\draw[-,gray] (1,3)--(3,4)--(3,0)--(2,-0.5)--(2,1.5)--(1,1)--(1,3)--(-3,3)--(-3,1);
\draw[-,gray] (1.4,-1)--(1,-1)--(2,-0.5)--(2.4,-0.5);
\draw[-,gray] (-3,3)--(-1,4)--(3,4);
\draw[-,gray] (-2,3.5)--(2,3.5)--(2,1.5)--(3,2);
\draw[-,gray] (3.4,0)--(3,0)--(3.4,0.2);\draw[-,gray] (-3.4,-1)--(-3,-1)--(-3.4,-1.2);\draw[-,gray] (1,-1)--(0.6,-1.2);
\draw[-,gray] (-3,1)--(-3.4,1);\draw[-,gray] (-3,3)--(-3.4,3);\draw[-,gray] (-2,3.5)--(-2.4,3.5);\draw[-,gray] (-1,4)--(-1.4,4);
\draw[-,gray] (3,0)--(3.4,0.2);\draw[-,gray] (3,2)--(3.4,2.2);\draw[-,gray] (3,4)--(3.4,4.2);\draw[-,gray] (1,4)--(1.4,4.2);\draw[-,gray] (-1,4)--(-0.6,4.2);
\filldraw[fill=black,draw=black] (-3,3) circle (2.0pt);
\fill[black!] (-3.2,3) circle (0.01pt)
node[below=1.5pt]{\color{black}\small $x'_c$};
\filldraw[fill=white,draw=black] (-1,3) circle (2.0pt);
\filldraw[fill=black,draw=black] (1,3) circle (2.0pt)
node[right=1.5pt]{\small $x'_d$};
\filldraw[fill=white,draw=black] (-2,3.5) circle (2.0pt);
\filldraw[fill=black,draw=black] (0,3.5) circle (2.0pt)
node[above=1.5pt]{\small $x'$};
\filldraw[fill=white,draw=black] (2,3.5) circle (2.0pt);
\filldraw[fill=black,draw=black] (-1,4) circle (2.0pt)
node[above=1.5pt]{\small $x'_a$};
\filldraw[fill=white,draw=black] (1,4) circle (2.0pt);
\filldraw[fill=black,draw=black] (3,4) circle (2.0pt)
node[above=1.5pt]{\small $x'_b$};
\filldraw[fill=black,draw=black] (3,0) circle (2.0pt);
\fill[black!] (3.15,0) circle (0.01pt)
node[below=1.5pt]{\color{black}\small $x_b$};
\filldraw[fill=white,draw=black] (3,2) circle (2.0pt);
\filldraw[fill=black,draw=black] (-3,-1) circle (2.0pt)
node[below=1.5pt]{\small $x_c$};
\filldraw[fill=white,draw=black] (-3,1) circle (2.0pt);
\filldraw[fill=white,draw=black] (-1,-1) circle (2.0pt);
\filldraw[fill=white,draw=black] (1,1) circle (2.0pt);
\filldraw[fill=white,draw=black] (2,-0.5) circle (2.0pt);
\filldraw[fill=black,draw=black] (-1,1) circle (2.0pt)
node[right=2.5pt]{\small $x_1$};
\filldraw[fill=black,draw=black] (1,-1) circle (2.0pt)
node[below=1.5pt]{\small $x_d$};
\filldraw[fill=black,draw=black] (2,1.5) circle (2.0pt)
node[right=2.5pt]{\small $x_2$};
\draw[-,very thick] (-3,-1)--(1,3);\draw[-,very thick] (-3,3)--(1,-1);
\draw[-,very thick] (1,-1)--(3,4);\draw[-,very thick] (1,3)--(3,0);
\draw[-,very thick] (-3,3)--(3,4);\draw[-,very thick] (-1,4)--(1,3);
\draw[-,very thick] (0.98,-1)--(0.98,3);
\draw[-,very thick] (1.01,-1)--(1.01,3);
\draw[->,black,dashed] (-2.9,3.25)--(1.5,3.25)--(1.5,-0.75)--(1.9,-0.75);
\draw[->,black,thick,dotted] (-0.5,-1.25)--(0,-1)--(0,3)--(2,4)--(2.4,4.2);
\draw[->,black,dashed] (-3.4,0)--(-2,0)--(0,0) .. controls (0.5,0.2) .. (1,1) .. controls (1.25,1.95) .. (1.5,2.25)--(3.4,3.2);
\draw[->,black,thick,dotted] (-1.9,3.75)--(2.5,3.75)--(2.5,-0.25)--(2.9,-0.25);
\draw[->,black,dashed] (-2.5,-1.25)--(-2,-1)--(-2,3)--(0,4)--(0.4,4.2);
\draw[->,black,thick,dotted] (-3.5,2)--(-2,2)--(0,2) .. controls (0.5,1.8) .. (1,1) .. controls (1.25,0.5) .. (1.5,0.25)--(3.4,1.2);
\draw[black] (-1.9,3.75) circle (0.01pt)
node[left=1.5pt]{\color{black}\small $p'$};
\draw[black] (-2.9,3.25) circle (0.01pt)
node[left=1.5pt]{\color{black}\small $p$};
\draw[black] (-2.5,-1.35) circle (0.01pt)
node[below=1.5pt]{\color{black}\small $q$};
\draw[black] (-0.5,-1.25) circle (0.01pt)
node[below=1.5pt]{\color{black}\small $q'$};
\draw[black] (-3.4,2) circle (0.01pt)
node[left=1.5pt]{\color{black}\small $r'$};
\draw[black] (-3.4,0) circle (0.01pt)
node[left=1.5pt]{\color{black}\small $r$};
\end{scope}

\end{tikzpicture}
\caption{A deformation in the IRF formulation that is equivalent to the Yang-Baxter equation \eqref{YBE-IRF}. }
\label{IRFcube3}
\end{figure}

The cubic deformations in Figures \ref{Vcube}-\ref{Vcube3}, and Figures \ref{IRFcube}-\ref{IRFcube3}, are for the vertex and IRF models respectively, as defined in Section \ref{sec:ssr}.  These deformations are derived with use of the respective Yang-Baxter equations \eqref{YBE-vertex}, and \eqref{YBE-IRF}, and inversion relations \eqref{Vinv}, \eqref{IRFinv}, and are used to show the extended Z-invariance property described in Section \ref{sec:z-invar2} for the vertex and IRF models.

For the case of the vertex formulation, the most complicated case is when changing one four-square into five four-squares, where two positively oriented rapidity lines $r$ and $r'$ are added/removed, as is shown in Figure \ref{Vcube}.  Using the Yang-Baxter equation \eqref{YBE-vertex} and inversion relation \eqref{Vinv}, the contribution to the partition function of the right hand side of Figure \ref{Vcube} is given by

\begin{align}
\label{veq1}
\begin{split}
&\ds\sum_{x''_i,x'''_i,x''_j,x'''_j,x_k,x'_k,x''_k,x'''_k}\left<x''_i,x''_j\,|\,\mathbb{R}_{\bp\bq}\,|\,x'''_i,x'''_j\right>\,\left<x_k,x_j\,|\,\mathbb{R}_{\br\bq}\,|\,x'_k,x'_j\right>\,\left<x'''_i,x'_k\,|\,\mathbb{R}_{\bp\br}\,|\,x'_i,x''_k\right> \\
&\ds\phantom{\sum_{x''_i,x'''_i,x''_j,x'''_j,x_k,x'_k,x''_k,x'''_k}}\times\left<x'''_j,x''_k\,|\,\mathbb{R}_{\bq\br}\,|\,x'_j,x'''_k\right>\,\left<x'''_k,x_i\,|\,\mathbb{R}_{\br\bp}\,|\,x_k,x''_i\right> \\
&\ds\phantom{xxxx}=\sum_{x''_i,x'''_j,x_k,x''_k,x'''_k}\;\sum_{\hat{x}_i,\hat{x}'_j,\hat{x}'''_k}\left<x'''_j,x''_k\,|\,\mathbb{R}_{\bq\br}\,|\,x'_j,x'''_k\right>\,\left<x'''_k,x_i\,|\,\mathbb{R}_{\br\bp}\,|\,x_k,x''_i\right> \\[0.2cm]
&\ds\hspace{3cm}\times\left<\hat{x}'''_k,\hat{x}'_j\,|\,\mathbb{R}_{\br\bq}\,|\,x''_k,x'''_j\right>\,\left<x''_i,x_k\,|\,\mathbb{R}_{\bp\br}\,|\,\hat{x}_i,\hat{x}'''_k\right>\,\left<\hat{x}_i,x_j\,|\,\mathbb{R}_{\bp\bq}\,|\,x'_i,\hat{x}'_j\right> \\[0.2cm]
&\ds\phantom{xxxx}=\left<x_i,x_j\,|\,\mathbb{R}_{\bp\bq}\,|\,x'_i,x'_j\right>\sum_{x'''_k}\delta_{x'''_k,x'''_k}.
\end{split}
\end{align}

This shows that the contributions to the partition function of the left and right hand sides of Figure \ref{Vcube} are equal, up to the constant factor $\sum_{x'''_k}\delta_{x'''_k,x'''_k}$ (this is left here as a $\delta$-function, because when considering models with continuous valued spins, this becomes an infinite constant).  A similar type of deformation not pictured here, that instead adds two negatively oriented rapidity lines $r$ and $r'$, may be shown to hold with an analogous calculation to \eqref{veq1}.  Similar calculations involving the Yang-Baxter equation \eqref{YBE-vertex} and inversion relation \eqref{Vinv}, can be used to show the equalities of Figures \ref{Vcube2} and \ref{Vcube3}, with the latter Figure only requiring a simple use of \eqref{YBE-vertex}.

For the case of the IRF formulation, the most complicated case again involves changing one four-square into five four-squares, where the deformation adds two positively oriented rapidity lines $r$ and $r'$, as is depicted in Figure \ref{IRFcube}.  Using the Yang-Baxter equation \eqref{YBE-IRF} and inversion relation \eqref{IRFinv}, the contribution to the partition function of the right hand side of Figure \ref{IRFcube} is given by

\begin{align}
\label{irfeq1}
\begin{split}
&\ds\sum_{x'_a,x'_b,x'_c,x'_d}\V^{(1)}_{\bp\bq}(x'_a,x'_b,x'_c,x'_d)\,\V^{(1)}_{\br\bq}(x'_c,x'_d,x_c,x_d)\,\V^{(1)}_{\bp\br}(x'_b,x_b,x'_d,x_d)\,\V^{(2)}_{\bq\br}(x'_a,x_a,x'_b,x_b)\\
&\ds\phantom{\sum_{x'_a,x'_b,x'_c,x'_d}}\times\V^{(2)}_{\br\bp}(x'_a,x'_c,x_a,x_c)\, W_{qq'}(x_a,x'_a)\, W_{q'q}(x'_b,x_b)\, W_{q'q}(x'_c,x_c)\, W_{qq'}(x_d,x'_d)\\
&\ds\phantom{xxxx}=\sum_{x'_a,x'_b,x'_c}\sum_{\hat{x}_a}\V^{(2)}_{\bq\br}(x'_a,x_a,x'_b,x_b)\,\V^{(2)}_{\br\bp}(x'_a,x'_c,x_a,x_c)\, W_{qq'}(x_a,x'_a)\, W_{q'q}(x'_a,\hat{x}_a)\\
&\ds\phantom{xxxx=\sum_{x'_a,x'_b,x'_c}\sum_{\hat{x}_a}}\times\V^{(1)}_{\br\bq}(x'_a,x'_b,\hat{x}_a,x_b)\,\V^{(1)}_{\bp\br}(x'_a,\hat{x}_a,x'_c,x_c)\,\V^{(1)}_{\bp\bq}(\hat{x}_a,x_b,x_c,x_d)\\
&\ds\phantom{xxxx}=\V^{(1)}_{\bp\bq}(x_a,x_b,x_c,x_d)\sum_{x_a}\delta_{x_a,x_a}.
\end{split}
\end{align}

This shows that the contributions to the partition function of the left and right hand sides of Figure \ref{IRFcube} are equal, up to the constant factor $\left(\sum_{x_a}\delta_{x_a,x_a}\right)$. A similar type of deformation not pictured here, that instead adds two negatively oriented rapidity lines $r$ and $r'$, may be shown with an analogous calculation to \eqref{irfeq1}.  Similar calculations involving the Yang-Baxter equation \eqref{YBE-IRF} and inversion relation \eqref{IRFinv}, can be used to show the equalities of Figures \ref{IRFcube2} and \ref{IRFcube3}, with the latter Figure only requiring a simple use of \eqref{YBE-IRF}.

\clearpage

\end{appendices}

\bibliography{MComp}

\providecommand{\href}[2]{#2}\begingroup\raggedright\begin{thebibliography}{10}

\bibitem{Kels:2017fyt}
A.~P. Kels, ``{Exactly solved models on planar graphs with vertices in
  $\mathbb{Z}^3$},'' \href{http://dx.doi.org/10.1088/1751-8121/aa8f68}{{\em J.
  Phys.} {\bfseries A50} no.~49, (2017) 495202},
\href{http://arxiv.org/abs/1705.06528}{{\ttfamily arXiv:1705.06528 [math-ph]}}.

\bibitem{Baxter:1978xr}
R.~J. Baxter, ``{Solvable eight vertex model on an arbitrary planar lattice},''
\href{http://dx.doi.org/10.1098/rsta.1978.0062}{{\em Phil. Trans. Roy. Soc.
  Lond.} {\bfseries 289} (1978) 315--346}.

\bibitem{Baxter:1986df}
R.~J. Baxter, ``{Free-fermion, checkerboard and Z-invariant lattice models in
  statistical mechanics},''
\href{http://dx.doi.org/10.1098/rspa.1986.0016}{{\em Proc. Roy. Soc. Lond.}
  {\bfseries A404} (1986) 1--33}.

\bibitem{Baxter:1982zz}
R.~J. Baxter, {\em Exactly {S}olved {M}odels in {S}tatistical {M}echanics}.
\newblock Academic, London,
1982.
\newblock

\bibitem{PerkYBEs}
J.~Perk and H.~Au-Yang,
  \href{http://dx.doi.org/https://doi.org/10.1016/B0-12-512666-2/00191-7}{``Yang–baxter
  equations,''} in {\em Encyclopedia of Mathematical Physics}, J.-P.
  Françoise, G.~L. Naber, and T.~S. Tsun, eds., pp.~465 -- 473.
\newblock Academic Press, Oxford, 2006.

\bibitem{LobbNijhoff}
S.~Lobb and F.~Nijhoff, ``Lagrangian multiforms and multidimensional
  consistency,'' \href{http://dx.doi.org/10.1088/1751-8113/42/45/454013}{{\em
  J. Phys.} {\bfseries A42} no.~45, (2009) 454013}.

\bibitem{ABS}
V.~Adler, A.~Bobenko, and Y.~Suris, ``Classification of integrable equations on
  quad-graphs. the consistency approach,''
  \href{http://dx.doi.org/10.1007/s00220-002-0762-8}{{\em Commun. Math. Phys.}
  {\bfseries 233} no.~3, (2003) 513--543}.

\bibitem{Bazhanov:1990qk}
V.~V. Bazhanov, R.~M. Kashaev, V.~V. Mangazeev, and Y.~G. Stroganov,
  ``${Z}_n^{\otimes(n-1)}$ generalization of the chiral {P}otts model,''
{\em Commun. Math. Phys.} {\bfseries 138} (1991) 393--408.

\bibitem{Bazhanov:1992jqa}
V.~V. Bazhanov and R.~J. Baxter, ``{New solvable lattice models in
  three-dimensions},''
\href{http://dx.doi.org/10.1007/BF01050423}{{\em J. Statist. Phys.} {\bfseries
  69} (1992) 453--585}.

\bibitem{Baxter:1997tn}
R.~J. Baxter, ``{Star-triangle and star-star relations in statistical
  mechanics},''
\href{http://dx.doi.org/10.1142/S0217979297000058}{{\em Int. J. Mod. Phys.}
  {\bfseries B11} (1997) 27--37}.

\bibitem{Bazhanov:2011mz}
V.~V. Bazhanov and S.~M. Sergeev, ``{Elliptic gamma-function and multi-spin
  solutions of the Yang-Baxter equation},''
  \href{http://dx.doi.org/10.1016/j.nuclphysb.2011.10.032}{{\em Nucl. Phys.}
  {\bfseries B856} (2012) 475--496},
\href{http://arxiv.org/abs/1106.5874}{{\ttfamily arXiv:1106.5874 [math-ph]}}.

\bibitem{Bazhanov:2013bh}
V.~V. Bazhanov, A.~P. Kels, and S.~M. Sergeev, ``{Comment on star-star
  relations in statistical mechanics and elliptic gamma-function identities},''
  \href{http://dx.doi.org/10.1088/1751-8113/46/15/152001}{{\em J. Phys.}
  {\bfseries A46} (2013) 152001},
\href{http://arxiv.org/abs/1301.5775}{{\ttfamily arXiv:1301.5775 [math-ph]}}.

\bibitem{Yamazaki:2013nra}
M.~Yamazaki, ``{New Integrable Models from the Gauge/YBE Correspondence},''
  \href{http://dx.doi.org/10.1007/s10955-013-0884-8}{{\em J. Statist. Phys.}
  {\bfseries 154} (2014) 895},
\href{http://arxiv.org/abs/1307.1128}{{\ttfamily arXiv:1307.1128 [hep-th]}}.

\bibitem{Gahramanov:2017idz}
I.~Gahramanov and S.~Jafarzade, ``{Comments on the multi-spin solution to the
  Yang-Baxter equation and basic hypergeometric sum/integral identity},'' in
  {\em {Proceedings, ISQS-25: Prague, Czech Republic, June 6-10}}.
\newblock 2017.
\newblock
\href{http://arxiv.org/abs/1710.09106}{{\ttfamily arXiv:1710.09106 [math-ph]}}.
\newblock

\bibitem{Kels:2017vbc}
A.~P. Kels and M.~Yamazaki, ``{Lens elliptic gamma function solution of the
  Yang–Baxter equation at roots of unity},''
  \href{http://dx.doi.org/10.1088/1742-5468/aaa8fd}{{\em J. Stat. Mech.}
  {\bfseries 1802} no.~2, (2018) 023108},
\href{http://arxiv.org/abs/1709.07148}{{\ttfamily arXiv:1709.07148 [math-ph]}}.

\bibitem{Bazhanov:2007mh}
V.~V. Bazhanov, V.~V. Mangazeev, and S.~M. Sergeev, ``{Faddeev-Volkov solution
  of the Yang-Baxter equation and discrete conformal symmetry},''
  \href{http://dx.doi.org/10.1016/j.nuclphysb.2007.05.013}{{\em Nucl. Phys.}
  {\bfseries B784} (2007) 234--258},
\href{http://arxiv.org/abs/hep-th/0703041}{{\ttfamily arXiv:hep-th/0703041
  [hep-th]}}.

\bibitem{Bazhanov:2010kz}
V.~V. Bazhanov and S.~M. Sergeev, ``{A Master solution of the quantum
  Yang-Baxter equation and classical discrete integrable equations},''
  \href{http://dx.doi.org/10.4310/ATMP.2012.v16.n1.a3}{{\em Adv. Theor. Math.
  Phys.} {\bfseries 16} no.~1, (2012) 65--95},
\href{http://arxiv.org/abs/1006.0651}{{\ttfamily arXiv:1006.0651 [math-ph]}}.

\bibitem{Bazhanov:2016ajm}
V.~V. Bazhanov, A.~P. Kels, and S.~M. Sergeev, ``{Quasi-classical expansion of
  the star-triangle relation and integrable systems on quad-graphs},''
  \href{http://dx.doi.org/10.1088/1751-8113/49/46/464001}{{\em J. Phys.}
  {\bfseries A49} (2016) 464001},
\href{http://arxiv.org/abs/1602.07076}{{\ttfamily arXiv:1602.07076 [math-ph]}}.

\bibitem{Kels:2018xge}
A.~P. Kels, ``{Integrable quad equations derived from the quantum Yang-Baxter
  equation},''
\href{http://arxiv.org/abs/1803.03219}{{\ttfamily arXiv:1803.03219 [math-ph]}}.

\bibitem{nijhoff_walker_2001}
F.~W. Nijhoff and A.~J. Walker, ``The discrete and continuous {P}ainlev\'{e}
  {VI} hierarchy and the {G}arnier systems,''
  \href{http://dx.doi.org/10.1017/S0017089501000106}{{\em Glasgow Math. J.}
  {\bfseries 43} no.~A, (2001) 109--123}.

\bibitem{BobSurQuadGraphs}
A.~I. Bobenko and Y.~B. Suris, ``{I}ntegrable systems on quad-graphs,''
  \href{http://dx.doi.org/10.1155/S1073792802110075}{{\em Int. Math. Res. Not.}
  {\bfseries 2002} no.~11, (2002) 573--611}.

\bibitem{MR2467378}
A.~I. Bobenko and Y.~B. Suris, {\em Discrete differential geometry: Integrable
  structure}, vol.~98 of {\em Graduate Studies in Mathematics}.
\newblock American Mathematical Society, Providence, RI, 2008.

\bibitem{BG11}
A.~Bobenko and F.~G\"{u}nther, ``On discrete integrable equations with convex
  variational principles,''
  \href{http://dx.doi.org/10.1007/s11005-012-0583-4}{{\em Lett. Math. Phys.}
  {\bfseries 102} no.~2, (2012) 181--202}.

\bibitem{Atkinson08}
J.~Atkinson, ``B\"{a}cklund transformations for integrable lattice equations,''
  \href{http://dx.doi.org/http://stacks.iop.org/1751-8121/41/i=13/a=135202}{{\em
  J. Phys. A: Math. Theor.} {\bfseries 41} no.~13, (2008) 135202}.

\bibitem{Atkinson09}
J.~Atkinson, ``Linear quadrilateral lattice equations and multidimensional
  consistency,''
  \href{http://dx.doi.org/http://stacks.iop.org/1751-8121/42/i=45/a=454005}{{\em
  J. Phys. A: Math. Theor.} {\bfseries 42} no.~45, (2009) 454005}.

\bibitem{Baxter:1972hz}
R.~J. Baxter, ``{Partition function of the eight vertex lattice model},''
  \href{http://dx.doi.org/10.1016/0003-4916(72)90335-1}{{\em Annals Phys.}
  {\bfseries 70} (1972) 193--228}.
[Annals Phys.281,187(2000)].

\bibitem{ABF}
G.~E. Andrews, R.~J. Baxter, and P.~J. Forrester, ``Eight-vertex {SOS} model
  and generalized {R}ogers-{R}amanujan-type identities,''
  \href{http://dx.doi.org/10.1007/BF01014383}{{\em J. Stat. Phys.} {\bfseries
  35} no.~3, (May, 1984) 193--266}.

\bibitem{ABS2}
V.~E. Adler, A.~I. Bobenko, and Y.~B. Suris, ``Discrete nonlinear hyperbolic
  equations. classification of integrable cases,''
  \href{http://dx.doi.org/10.1007/s10688-009-0002-5}{{\em Funct. Anal. Appl.}
  {\bfseries 43} no.~1, (Mar, 2009) 3--17}.

\bibitem{KelsThesis}
A.~P. Kels, {\em Analytic and numerical investigation of lattice models}.
\newblock PhD thesis, Australian National University, 2013.

\bibitem{HietarintaCAC2004}
J.~Hietarinta, ``A new two-dimensional lattice model that is 'consistent around
  a cube',''
  \href{http://dx.doi.org/http://stacks.iop.org/0305-4470/37/i=6/a=L01}{{\em J.
  Phys. A: Math. Gen.} {\bfseries 37} no.~6, (2004) L67}.

\bibitem{WW}
E.~T. Whittaker and G.~N. Watson,
  \href{http://dx.doi.org/https://doi.org/10.1017/CBO9780511608759}{{\em A
  course of modern analysis}}.
\newblock Cambridge Mathematical Library. Cambridge University Press,
  Cambridge, 1996.
\newblock Reprint of the fourth (1927) edition.

\bibitem{AUYANG198744}
H.~Au-Yang and J.~H. Perk, ``Critical correlations in a {Z}-invariant
  inhomogeneous {I}sing model,''
  \href{http://dx.doi.org/http://dx.doi.org/10.1016/0378-4371(87)90145-2}{{\em
  Physica A} {\bfseries 144} no.~1, (1987) 44 -- 104}.

\bibitem{REYESMARTINEZ1997203}
J.~R. Martínez, ``Correlation functions for the {Z}-invariant {I}sing model,''
  \href{http://dx.doi.org/http://dx.doi.org/10.1016/S0375-9601(97)00057-1}{{\em
  Phys. Lett. A} {\bfseries 227} no.~3, (1997) 203 -- 208}.

\bibitem{Martı́nez1998463}
J.~Martı́nez, ``Multi-spin correlation functions for the {Z}-{I}nvariant
  {I}sing model,''
  \href{http://dx.doi.org/https://doi.org/10.1016/S0378-4371(98)00106-X}{{\em
  Physica A} {\bfseries 256} no.~3–4, (1998) 463 -- 484}.

\bibitem{Costa-Santos}
R.~Costa-Santos, ``Geometrical aspects of the {Z}-invariant {I}sing model,''
  \href{http://dx.doi.org/10.1140/epjb/e2006-00336-1}{{\em Eur. Phys. J. B}
  {\bfseries 53} no.~1, (2006) 85--90}.

\bibitem{Au-Yang2007}
H.~Au-Yang and J.~H.~H. Perk, ``Q-dependent susceptibilities in ferromagnetic
  quasiperiodic {Z}-invariant {I}sing models,''
  \href{http://dx.doi.org/10.1007/s10955-006-9213-9}{{\em J. Statist. Phys.}
  {\bfseries 127} no.~2, (2007) 265--286}.

\bibitem{Boutillier2010}
C.~Boutillier and B.~de~Tili{\`e}re, ``The critical {Z}-invariant {I}sing model
  via dimers: the periodic case,''
  \href{http://dx.doi.org/10.1007/s00440-009-0210-1}{{\em Probab. Theory
  Related Fields} {\bfseries 147} no.~3, (2010) 379--413}.

\bibitem{Boutillier2011}
C.~Boutillier and B.~de~Tili{\`e}re, ``The critical {Z}-invariant {I}sing model
  via dimers: locality property.,''
  \href{http://dx.doi.org/10.1007/s00440-009-0210-1}{{\em Commun. Math. Phys.}
  {\bfseries 301} (2011) 473--516}.

\bibitem{JimboMiwaNakayashiki}
M.~Jimbo, T.~Miwa, and A.~Nakayashiki, ``Difference equations for the
  correlation functions of the eight-vertex model,''
  \href{http://dx.doi.org/http://stacks.iop.org/0305-4470/26/i=9/a=015}{{\em J.
  Phys. A} {\bfseries 26} no.~9, (1993) 2199}.

\bibitem{Bax2}
R.~J. Baxter, ``Functional relations for the order parameters of the chiral
  {P}otts model,'' \href{http://dx.doi.org/10.1023/A:1023096408679}{{\em J.
  Statist. Phys.} {\bfseries 91} no.~3-4, (1998) 499--524}.

\bibitem{Baxter:2005jt}
R.~J. Baxter, ``Derivation of the order parameter of the chiral {P}otts
  model,'' {\em Phys. Rev. Lett.} {\bfseries 94} (2005) 130602,
\href{http://arxiv.org/abs/cond-mat/0501227}{{\ttfamily cond-mat/0501227}}.

\bibitem{Au-YangPerk2011}
H.~Au-Yang and J.~H.~H. Perk, ``Spontaneous magnetization of the integrable
  chiral {P}otts model,''
  \href{http://dx.doi.org/10.1088/1751-8113/44/44/445005}{{\em J. Phys. A:
  Math. Theor.} {\bfseries 44} no.~44, (2011) 445005}.

\bibitem{Yamazaki:2012cp}
M.~Yamazaki, ``{Quivers, YBE and 3-manifolds},''
  \href{http://dx.doi.org/10.1007/JHEP05(2012)147}{{\em JHEP} {\bfseries 05}
  (2012) 147},
\href{http://arxiv.org/abs/1203.5784}{{\ttfamily arXiv:1203.5784 [hep-th]}}.

\bibitem{BPS}
R.~Boll, M.~Petrera, and Y.~B. Suris, ``What is integrability of discrete
  variational systems?,'' \href{http://dx.doi.org/10.1098/rspa.2013.0550}{{\em
  P. Roy. Soc. A} {\bfseries 470} (2014) 20130550}.

\end{thebibliography}\endgroup
\bibliographystyle{utphys}

\end{document}